\documentclass[prb,aps,floatfix,twocolumn,superscriptaddress,amsmath,amssymb]{revtex4-2}
\usepackage{graphicx}
\usepackage{xcolor}
\usepackage{bm}
\usepackage{hyperref}
\hypersetup{colorlinks=true,citecolor=blue,linkcolor=blue,urlcolor=blue}
\usepackage[T1]{fontenc}
\usepackage{newtxtext,newtxmath}
\usepackage[normalem]{ulem}

\allowdisplaybreaks

\begin{document}

\title{Seeding neural network quantum states with tensor network states}

\author{Ryui Kaneko}
\email{ryuikaneko@sophia.ac.jp}
\affiliation{Physics Division, Sophia University, Chiyoda, Tokyo 102-8554, Japan}

\author{Shimpei Goto}
\email{shimpei.goto@phys.s.u-tokyo.ac.jp}
\affiliation{Department of Physics, University of Tokyo, Hongo, Tokyo 113-0033, Japan}

\date{\today}

\begin{abstract}
We find an efficient approach to approximately convert matrix product states (MPSs) into
restricted Boltzmann machine wave functions
consisting of a multinomial hidden unit
through a canonical polyadic (CP) decomposition of the MPSs.
This method allows us to generate well-behaved initial neural network
quantum states for quantum many-body ground-state calculations
in polynomial time of the number of variational parameters
and systematically shorten the distance between
the initial states and the ground states
while increasing the rank of the CP decomposition.
We demonstrate the efficiency of our method by taking the
transverse-field Ising model as an example
and discuss possible applications of our method to
more general quantum many-body systems
in which the ground-state wave functions possess complex nodal structures.
\end{abstract}

\maketitle

\section{Introduction}
\label{sec:intro}

Solving quantum many-body problems is one of the most challenging tasks
in modern physics, and
tensor network states have been widely used
to efficiently represent quantum many-body states
in recent years.
Matrix product states (MPSs)~\cite{baxter1968,white1992,
vidal2007,
schollwock2011,oseledets2011,
hauschild2018,hauschild2024,fishman2022,
verstraete2008,orus2014},
often specialized in one spatial dimension,
and their generalizations to higher dimensions,
such as projected entangled pair states
(PEPSs)~\cite{nishino2001,verstraete2004a_arxiv,verstraete2008,corboz2011,
orus2014,orus2019}
and tree tensor networks~\cite{fannes1992,shi2006,murg2010,nakatani2013},
have been successfully applied to
various quantum many-body problems
in low-dimensional quantum systems
by keeping the entanglement entropy of the wave function
as low as possible.
The number of variational parameters in tensor network states
remains relatively small and grows only polynomially
with the number of sites in most of quantum many-body systems.

Recently, neural network quantum states
(NNQSs) have been proposed as a new class of
variational wave functions for quantum many-body
systems~\cite{carleo2017,jeswal2018,jia2019,carleo2019,
carrasquilla2021,medvidovi2024,lange2024,nomura2025}.
One of the basic NNQSs is the restricted Boltzmann machine
(RBM) wave function~\cite{carleo2017,nomura2017,
torlai2018,sehayek2019,huang2021,nomura2021,
rrapaj2021,nomura2022,golubeva2022,
kaubruegger2018,pastori2019,lu2019,wu2023}.
In contrast to tensor network states,
NNQSs try to represent quantum many-body states
by overparameterizing variational wave functions
with neural networks.
Because of the flexibility of the network structure
and the large number of variational parameters,
NNQSs can, in principle, represent arbitrary quantum many-body states
even with the volume-law entanglement
entropy~\cite{carleo2018,levine2019,
schmitt2020,medina2021,
sharir2022,passetti2023,wurst2025,
denis2025,
deng2017a,
pei2021b,
chen2018,
gao2017},
irrespective of the spatial dimensionality of quantum
systems~\cite{astrakhantsev2021,pohle2023_arxiv,machaczek2025}.

Despite the advantages in representing
quantum many-body states with NNQSs,
the large number of variational parameters in
NNQSs often complicates the optimization of the wave function.
Optimized NNQSs may become trapped in local minima
within the energy landscape, leading to inaccurate energy estimations.
Such difficulties are sometimes mitigated by the use of
initial states that are sufficiently
close to the quantum states of interest.
For best performance, NNQSs ought to be seeded with
a well-prepared initial configuration.
However, there have been few studies on
generating initial states for NNQSs
that are well-suited for optimization toward ground states.

\begin{figure}[!t]
\centering
\includegraphics[width=.75\linewidth]{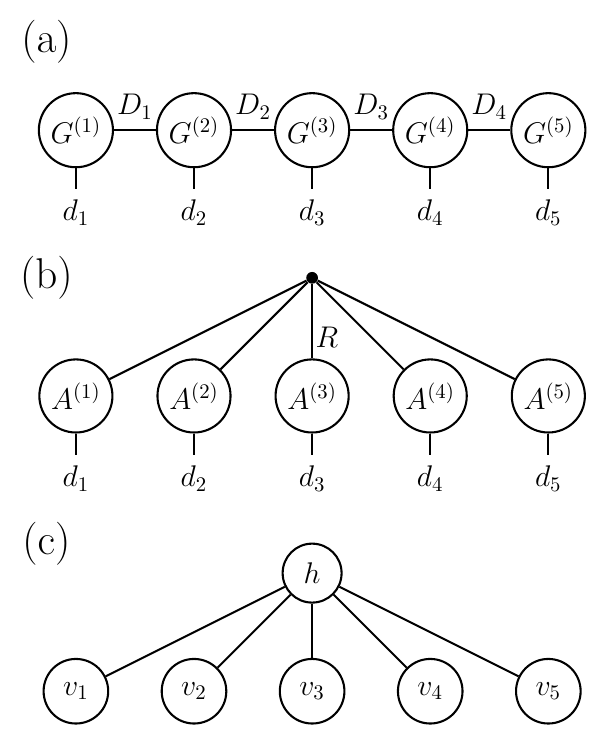}
\caption{%
Schematic figure of the conversion from MPSs to the RBM wave function
through the CP decomposition.
(a) MPS representation of tensor $T$.
(b) CP decomposition of tensor $T$,
which is efficiently computed from the MPS representation.
(c) RBM wave function with a multinomial hidden unit,
which is equivalent to the CP decomposed tensor.
}
\label{fig:mps_to_rbm}
\end{figure}

In this paper, we find that suitable initial states of
NNQSs can efficiently be prepared by tensor network states.
Previously, there have been numerous studies on constructing
tensor network states, MPSs in particular,
from RBM wave
functions~\cite{deng2017a,deng2017b,
zheng2019,
pei2021a,pei2021b,
chen2018,
gao2017,
harney2021,zhang2022,
glasser2018,he2019_arxiv,
clark2018,glasser2020,
sharir2020,li2021}.
On the contrary,
although it is widely believed that
NNQSs have the same or higher expressive power
than tensor network states~\cite{sharir2022,lange2024},
examples of generating RBM wave functions from
MPSs have been limited to special
cases~\cite{deng2017a,deng2017b,
zheng2019,
pei2021a,pei2021b,
harney2021,zhang2022,
chen2018,
glasser2018,he2019_arxiv,
clark2018}.
Such examples include
RBM wave functions from
primitive tensor network states,
essentially described
by the so-called stabilizer state~\cite{aaronson2004}
such as the Greenberger-Horne-Zeilinger (GHZ) state
(e.g., $|\mathrm{GHZ}\rangle
\propto |000\rangle + |111\rangle
$ for three qubits)~\cite{greenberger1989,greenberger1990}
or more general (in a sense that it is a non-stabilizer state)
but still a simple one
such as the W state
(e.g., $|\mathrm{W}\rangle
\propto |001\rangle + |010\rangle + |100\rangle
$ for three qubits)~\cite{duer2000};
otherwise,
resulting Boltzmann machine wave functions require
more than one hidden
layer~\cite{gao2017,
clark2018,chen2018,
glasser2018,he2019_arxiv,
glasser2020,
sharir2020,li2021},
thereby losing the structural simplicity inherent to shallow network
configurations.

We explore the possibility of generating
RBM wave functions from MPSs in a more general setting.
To this end,
instead of directly transforming MPSs into RBM wave functions,
we introduce intermediate tensor network states,
which can efficiently be obtained by the canonical polyadic (CP)
decomposition~\cite{hitchcock1927,hitchcock1928,
cattell1944,cattell1952,carroll1970,harshman1970,
mocks1988,kiers2000,kolda2009}
of arbitrary MPSs for a given rank of the
decomposition (see Fig.~\ref{fig:mps_to_rbm}).
The CP decomposed tensor network state is found to be
equivalent to the RBM wave function having a single multinomial hidden
unit.
Therefore,
when we are able to prepare accurate MPSs of a given quantum system
using conventional tensor network methods,
we can easily generate equivalent RBM wave functions
as initial states of NNQSs.

This paper is organized as follows:
In Sec.~\ref{sec:methods},
we introduce our method for approximately converting
MPSs into RBM wave functions consisting of multinomial hidden units
by the CP decomposition of the MPSs.
The computational complexity of our method
scales polynomially with the number of variational parameters
when the practical rank of the CP decomposition is
assumed to be known and is fixed by the user.
In Sec.~\ref{sec:applications},
we demonstrate the efficiency of our method by taking
the transverse-field Ising model as an example.
We first examine systems with open boundary conditions
and further test systems with periodic boundary conditions
using the initial states generated by our method
under open boundary conditions.
Finally, in Sec.~\ref{sec:summary},
we summarize our results and
discuss possible applications of our method
to more general quantum many-body systems
in which the ground states possess complex nodal structures.

\section{Methods}
\label{sec:methods}

In this section, we introduce our method for approximately converting
MPSs into RBM wave functions consisting of multinomial hidden units
by the CP decomposition of the MPSs.

\subsection{MPS representation of tensors}
\label{subsec:mps_tensor}

When the order of a tensor $T$ increases,
the number of elements in $T$ grows exponentially large;
eventually,
it is not possible to store the full tensor $T$ in memory.
Such a tensor having a large number of axes often appears in calculating
the ground state of quantum many-body systems with a large number of
sites $n$ using the exact diagonalization method.
The wave function is represented as
\begin{align}
 |\Psi\rangle
 =
 \sum_{\{s_i\}} T_{s_1, s_2, \dots, s_n} |s_1, s_2, \dots, s_n\rangle,
\end{align}
where $s_i$ denotes the physical index at site $i$.
In such cases, we can use the MPS
representation~\cite{baxter1968,white1992,verstraete2008,
vidal2007,
schollwock2011,oseledets2011,orus2014,hauschild2018,hauschild2024,fishman2022}
of the tensor,
which can be more efficiently obtained by the density matrix
renormalization group (DMRG) method~\cite{white1992}
than the exact diagonalization method.

The MPS representation of tensor $T$ only requires
parameters of the form
$G^{(i)} \in \mathbb{C}^{D_{i-1}\times d_i\times D_{i}}$,
where $d_i$ denotes the physical bond dimension at site $i$
and $D_i$ ($D_0=D_n=1$) are the virtual bond dimensions.
For a spin-$1/2$ system,
a physical spin index $s_i$ takes the value
$\uparrow$ or $\downarrow$, corresponding to $d_i=2$.
This structure allows for an efficient approximation of high-dimensional tensors,
particularly when the entanglement entropy in the system is low.
By contracting these matrices $G^{(i)}$ for each ${s_i}$ sequentially,
the full tensor $T_{s_1, s_2, \dots, s_n} \in
\mathbb{C}^{d_1\times d_2\times \cdots \times d_n}$
can be reconstructed as
\begin{align}
\label{eq:mps_tensor}
 T_{s_1, s_2, \dots, s_n}
 =
 \sum_{\{\alpha_i\}}
 G^{(1)}_{\alpha_0,s_1,\alpha_1}
 G^{(2)}_{\alpha_1,s_2,\alpha_2}
 \cdots
 G^{(n)}_{\alpha_{n-1},s_n,\alpha_n},
\end{align}
where the internal indices $\alpha_i(=1,2,\dots,D_i)$ are summed over
according to the bond dimensions.

\subsection{CP decomposition of tensors}
\label{subsec:cpd_tensor}

We first review the CP decomposition of an order-$n$ tensor
$X_{i_1, i_2, \dots, i_n} \in
\mathbb{C}^{d_1\times d_2\times \cdots \times d_n}$,
where $d_i$ is the dimension of the $i$th
axis~\cite{hitchcock1927,hitchcock1928,
cattell1944,cattell1952,carroll1970,harshman1970,
mocks1988,kiers2000,kolda2009}.
The tensor elements are expressed by
\begin{align}
\label{eq:cpd_tensor}
 X_{i_1, i_2, \dots, i_n}
 =
 \sum_{r=1}^{R} A^{(1)}_{i_1,r} A^{(2)}_{i_2,r} \cdots A^{(n)}_{i_n,r},
\end{align}
where $R$ is the rank of the CP decomposition
and $A^{(j)}_{i_j,r} \in \mathbb{C}^{d_i\times R}$
is the $r$th component of the vector
corresponding to the $j$th axis.
The CP decomposition becomes more accurate as the rank $R$ increases.
We discuss the case where the tensor is represented
as MPSs in the next section.

One of the most famous algorithms to compute the CP decomposition
is the alternating least squares (ALS)
method~\cite{comon2009,kolda2009,erichson2020}.
For simplicity, let us consider the case with $n=3$.
Our goal is to find matrices $A$, $B$, and $C$
that minimize the following loss function:
\begin{align}
 L(A, B, C)
 &=
 \sum_{i,j,k}
 \left|
 X_{i,j,k}
 -
 \sum_{r=1}^{R} A_{i,r} B_{j,r} C_{k,r}
 \right|^2.
\end{align}
The ALS method iteratively updates matrices
$A$, $B$, and $C$.
The initial values of $A$, $B$, and $C$ are often
randomly chosen.
In the first step, we fix $B$ and $C$ and
minimize the loss function with respect to $A$.
By taking the derivative of the loss function with respect to $A_{i,r}$
and setting the derivative to zero,
we see that the optimal element $A_{i,r}$ is given by
\begin{align}
\label{eq:als_update_a_element}
 A_{i,r}
 =
 \sum_{j,k} X_{i,j,k}
 \sum_{r'} B^{*}_{j,r'} C^{*}_{k,r'}
 \left(
 \sum_{j',k'} B^{*}_{j',r'} B_{j',r} C^{*}_{k',r'} C_{k',r}
 \right)^{-1}.
\end{align}
The corresponding matrix form is
\begin{align}
\label{eq:als_update_a}
 A
 =
 X_{(1)}
 \left(C^{*} \odot B^{*}\right)
 \left(C^{\dagger}C \circledast B^{\dagger}B\right)^{-1},
\end{align}
where $X_{(i)}$ is matricized tensor $X$ with respect to the $i$th
axis, $*$ is the complex conjugate,
$\dagger$ is the complex conjugate transpose,
$\odot$ is the Khatri-Rao product, and
$\circledast$ is the element-wise (Hadamard) product.
We then fix $A$ and $C$ and minimize the loss function $L(A, B, C)$.
We can similarly obtain the optimal element $B_{j,r}$ as
\begin{align}
\label{eq:als_update_b_element}
 B_{j,r}
 &=
 \sum_{i,k} X_{i,j,k}
 \sum_{r'} A^{*}_{i,r'} C^{*}_{k,r'}
 \left(
 \sum_{i',k'} A^{*}_{i',r'} A_{i',r} C^{*}_{k',r'} C_{k',r}
 \right)^{-1}
\end{align}
and the matrix form as
\begin{align}
\label{eq:als_update_b}
 B
 &=
 X_{(2)}
 \left(C^{*} \odot A^{*}\right)
 \left(C^{\dagger}C \circledast A^{\dagger}A\right)^{-1}.
\end{align}
Finally, we fix $A$ and $B$ and minimize the loss function $L(A, B, C)$.
The optimal element $C_{k,r}$ is given by
\begin{align}
\label{eq:als_update_c_element}
 C_{k,r}
 &=
 \sum_{i,j} X_{i,j,k}
 \sum_{r'} A^{*}_{i,r'} B^{*}_{j,r'}
 \left(
 \sum_{i',j'} A^{*}_{i',r'} A_{i',r} B^{*}_{j',r'} B_{j',r}
 \right)^{-1},
\end{align}
and the matrix form is given by
\begin{align}
\label{eq:als_update_c}
 C
 &=
 X_{(3)}
 \left(B^{*} \odot A^{*}\right)
 \left(B^{\dagger}B \circledast A^{\dagger}A\right)^{-1}.
\end{align}
We repeat these three steps until the loss function becomes sufficiently
small.

The ALS method can be generalized to the case with $n>3$.
The loss function is given by
\begin{align}
 L(\{A^{(i)}\})
 &=
 \sum_{i_1,i_2,\dots,i_n}
 \left|
 X_{i_1, i_2, \dots, i_n}
 -
 \sum_{r=1}^{R} A^{(1)}_{i_1,r} A^{(2)}_{i_2,r} \cdots A^{(n)}_{i_n,r}
 \right|^2.
\end{align}
The ALS method iteratively updates matrices
$A^{(1)}$, $A^{(2)}$, $\dots$, $A^{(n)}$
by fixing all but one of the matrices
and minimizing the loss function with respect to the remaining matrix.
The update rule for matrix $A^{(j)}$ is given by
\begin{align}
 A^{(j)}
 &=
 X_{(j)}
 \left[
 A^{(n)} \odot
 \cdots \odot
 A^{(j+1)} \odot
 A^{(j-1)} \odot
 \cdots \odot
 A^{(1)}
 \right]
\nonumber
\\
 &~\phantom{=}
 \cdot
 \bigl[
 {A^{(n)}}^{\dagger} A^{(n)} \circledast
 \cdots \circledast
 {A^{(j+1)}}^{\dagger} A^{(j+1)}
\nonumber
\\
 &~\phantom{=}
 \circledast
 {A^{(j-1)}}^{\dagger} A^{(j-1)} \circledast
 \cdots
 \circledast
 {A^{(1)}}^{\dagger} A^{(1)}
 \bigr]^{-1}.
\end{align}
Here, the centered dot $\cdot$ is the conventional matrix product.

In the ALS method, the rank $R$ is a hyperparameter that determines the
accuracy of the CP decomposition.
The larger the rank $R$ is, the more accurate the CP decomposition becomes.
For an order-$3$ tensor,
the sufficient number of the rank in the CP decomposition
to completely reproduce the original tensor is
$\min(d_1d_2, d_1d_3, d_2d_3)$~\cite{kolda2009},
which is smaller than the number of elements
$d_1d_2d_3$ of the tensor.
However, there is no algorithm to determine the rank of a given tensor
in general.
This problem is known to be NP-hard~\cite{hastad1990}.
Furthermore,
the ALS method is not guaranteed to converge to the global minimum
because of the nonconvexity of the loss function~\cite{auddy2025}.
In realistic applications,
we can try several initial values of factors $A$, $B$, and $C$ and
numerically determine the practical rank of
the tensor by fitting various CP decomposition results with different trials
of rank $R$.

When the order $n$ of the tensor is large,
the most time-consuming part of the ALS method
is the computation of the products between
the matricized tensor and the matrix obtained by the Khatri-Rao product.
The dimension of the matricized tensor is
$d_i \times \prod_{j=1,j\not=i}^{n} d_j
\ge d_i \times (\min_{j\not=i} d_j)^{n-1}$,
which results in the exponential
cost of evaluating the products.
As we will see in the next section,
the MPS representation is more efficient
than the original tensor representation
in terms of memory and computational cost of the ALS method.

\subsection{CP decomposition of MPSs}
\label{subsec:cpd_mps}

Even when the number of elements in a tensor increases,
the memory for the MPS representation remains manageable.
Despite the efficiency of the MPS representation,
to the best of our knowledge,
there are very few studies on the CP decomposition of MPSs.
Several algorithms have been proposed to compute the CP decomposition
and its variant, such as the Tucker decomposition,
of MPSs very
recently~\cite{diniz2019,zniyed2020,giraud2023,prvost2025};
however, the rank of the CP decomposition
is often smaller than the bond dimension of the MPS.
This limitation makes it difficult to obtain accurate
CP decompositions of MPSs, especially for MPSs that
are highly entangled and require large bond dimensions to represent
the wave function accurately.

\begin{figure}[!t]
\centering
\includegraphics[width=.75\linewidth]{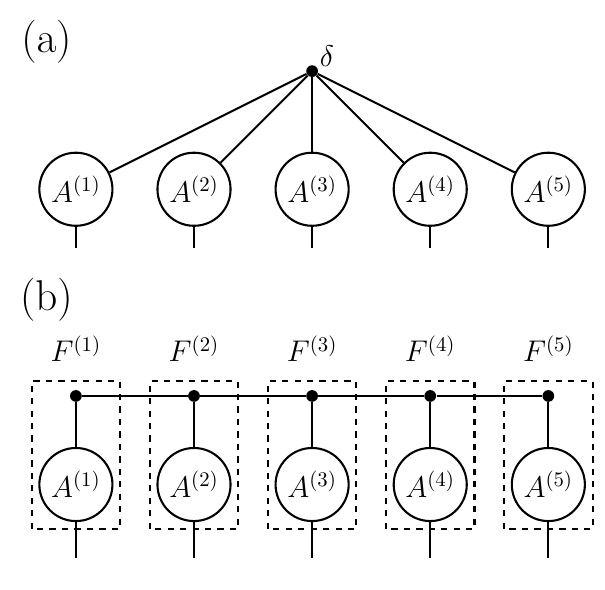}
\caption{%
(a) CP decomposition of a tensor.
A black dot represent the Kronecker delta tensor $\delta$.
Each factor $A^{(i)}$ is a $d_i \times R$ matrix
with $R$ being the rank of the CP decomposition.
(b) MPS representation of the CP decomposed tensor.
The black dot in (a) can be rewritten as
a product of Kronecker delta tensors
with smaller orders.
The product of the Kronecker delta tensor
and factor $A^{(i)}$ can be rewritten as MPS $F^{(i)}$.
}
\label{fig:cpd_is_mps}
\end{figure}

\begin{figure}[!t]
\centering
\includegraphics[width=\linewidth]{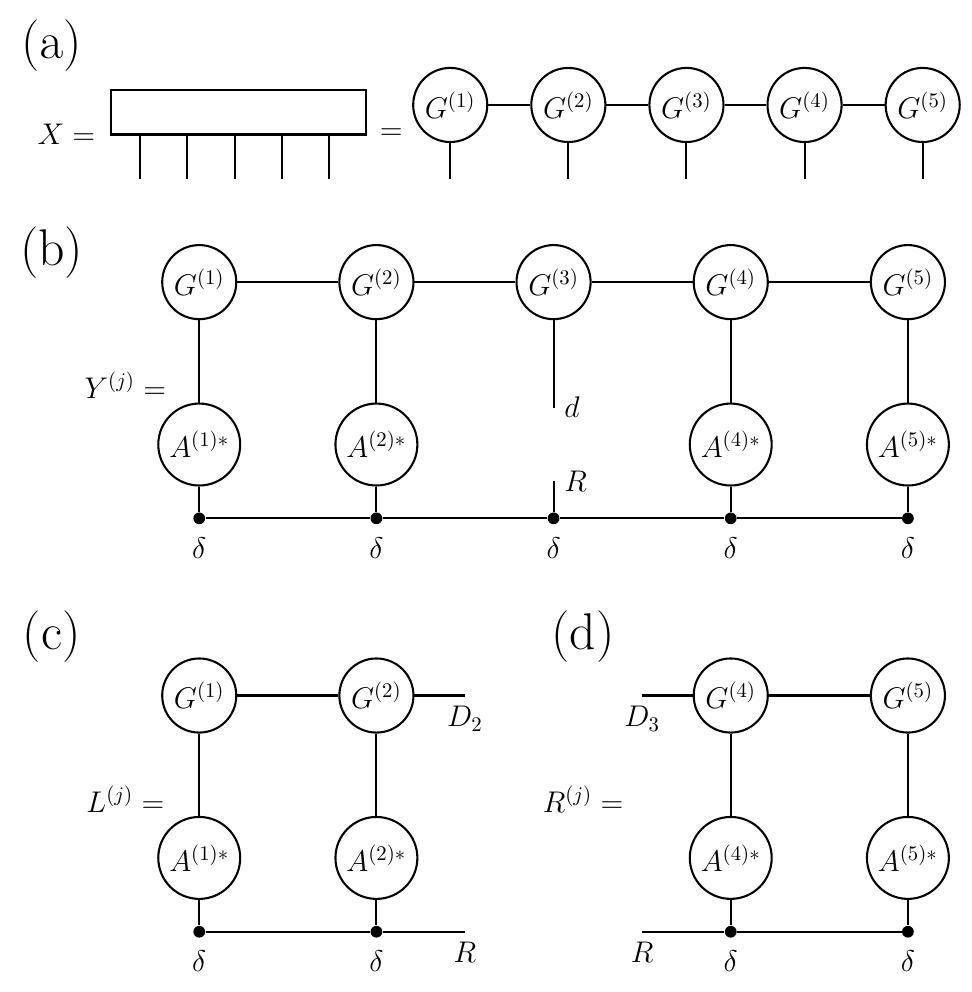}
\caption{%
Tensors needed for the ALS method
for the CP decomposition of MPSs.
We show an example of a tensor with the order $n=5$
when updating the $j=3$ element.
(a) MPS representation of tensor $X$.
(b) Matrix $Y^{(j)}$ obtained by the product between
the matricized tensor, written in MPS form $G^{(i)}$,
and the matrix obtained by the Khatri-Rao product,
written by the Kronecker delta tensors $\delta$
and matrix $A^{(i\not=j)}$.
Matrix $Y^{(j)}$ is used to update matrix $A^{(j)}$
during the ALS method.
(c) Left matrix $L^{(j)}$ at the $j$th step of the ALS method.
(d) Right matrix $R^{(j)}$ at the $j$th step of the ALS method.
Matrices $L^{(j)}$ and $R^{(j)}$ are used to construct
matrix $Y^{(j)}$.
}
\label{fig:als_mps}
\end{figure}

We propose a straightforward method for computing the CP decomposition
of MPSs by using the ALS method.
Hereafter, we assume that the rank $R$ of the CP decomposition
is fixed and given as a hyperparameter.
The input is the MPS representation $G^{(i)}$ of tensor $T$,
rather than the full tensor $T$ itself,
and the output is matrices $A^{(i)}$ that represent the CP
decomposition of the MPSs.
Here, the MPS representation $G^{(i)}$
is a $D_{i-1} \times d_i \times D_{i}$ tensor
for $i=1,2,\dots,n$, whereas
the factor $A^{(i)}$ is a $d_i \times R$ matrix
for $i=1,2,\dots,n$
($d_i=2$ for a spin-$1/2$ system).
This procedure can be effectively accomplished because
the CP decomposed tensor is already expressed in the form of MPSs.
To be more precise,
utilizing the Kronecker delta tensor
\begin{align}
 \delta_{r_1,r_2,\dots,r_n}
 &=
 \begin{cases}
 1 & (r_1=r_2=\cdots=r_n) \\
 0 & (\text{otherwise}),
 \end{cases}
\end{align}
with $r_i=1,2,\dots,R$ for all $i$,
we can rewrite the definition of the CP decomposed tensor as
\begin{align}
 X_{i_1, i_2, \dots, i_n}
 =
 \sum_{r_1=1}^{R} \sum_{r_2=1}^{R} \cdots \sum_{r_n=1}^{R}
 \delta_{r_1,r_2,\dots,r_n}
 A^{(1)}_{i_1,r_1} A^{(2)}_{i_2,r_2} \cdots A^{(n)}_{i_n,r_n}.
\end{align}
Because the Kronecker delta tensor $\delta_{r_1,r_2,\dots,r_n}$
can be expressed as a product of order-$3$ Kronecker delta
tensors~\cite{biamonte2011,denny2011,clark2018},
e.g., $\delta_{r_1,r_2,r_3,r_4,r_5} =
\delta_{r_1,r_2,r_3} \delta_{r_3,r_4,r_5}$,
we can rewrite the above equation as
\begin{align}
 X_{i_1, i_2, \dots, i_n}
 &=
 \sum_{\{\alpha_i\}}
 F^{(1)}_{\alpha_0,i_1,\alpha_1}
 F^{(2)}_{\alpha_1,i_2,\alpha_2}
 \cdots
 F^{(n)}_{\alpha_{n-1},i_n,\alpha_n},
\\
 F^{(i)}_{\alpha_{i-1},s_i,\alpha_i}
 &=
 \delta_{\alpha_{i-1},r,\alpha_i} A^{(i)}_{s_i,r}
 \quad (i=1,2,\dots,n),
\end{align}
where $F^{(i)}$ is the corresponding MPS representation,
as shown in Fig.~\ref{fig:cpd_is_mps}.
Thus, tensor multiplications in the ALS method can efficiently be
performed by the contraction of MPSs.

Let us describe the ALS method specifically for the case of MPSs.
We assume that the original tensor $X$ is represented as MPSs $G^{(i)}$,
although tensor $X$ is not explicitly given because it is too large
to store in memory
[see Fig.~\ref{fig:als_mps}(a)].
We need to evaluate the counterparts of
the $d_i\times R$ matrix
$X_{(1)} (C^{*} \odot B^{*})$
and the $R\times R$ matrix
$(C^{\dagger}C \circledast B^{\dagger}B)^{-1}$
in Eq.~\eqref{eq:als_update_a}
by contracting the MPSs $G^{(i)}$
and the MPS representations $F^{(i)}$
of factors in the CP decomposition.

The counterpart of the $d_i\times R$ matrix
$X_{(1)} (C^{*} \odot B^{*})$,
as shown in Fig.~\ref{fig:als_mps}(b),
can be evaluated in the following manner:
For factors $A^{(j)}$ ($2\leq j \leq n-1$)
having the MPS representations $F^{(i)}$,
we construct the left and right matrices
$L^{(j)}$ and $R^{(j)}$, defined as
\begin{align}
 &
 L^{(j)}_{\alpha_{j-1}, \tilde{\alpha}_{j-1}}
 =
 \sum_{\{\alpha_i\},\{s_i\},\{\tilde{\alpha_i}\}}
 G^{(1)}_{\alpha_0=1,s_1,\alpha_1}
 F^{(1)*}_{\tilde{\alpha}_0=1,s_1,\tilde{\alpha}_1}
\nonumber
\\
 &~
 \cdot
 G^{(2)}_{\alpha_1,s_2,\alpha_2}
 F^{(2)*}_{\tilde{\alpha}_1,s_2,\tilde{\alpha}_2}
 \cdots
 G^{(j-1)}_{\alpha_{j-2},s_{j-1},\alpha_{j-1}}
 F^{(j-1)*}_{\tilde{\alpha}_{j-2},s_{j-1},\tilde{\alpha}_{j-1}},
\\
 &
 R^{(j)}_{\alpha_{j}, \tilde{\alpha}_{j}}
 =
 \sum_{\{\alpha_i\},\{s_i\},\{\tilde{\alpha_i}\}}
 G^{(n)}_{\alpha_{n-1},s_n,\alpha_n=1}
 F^{(n)*}_{\tilde{\alpha}_{n-1},s_n,\tilde{\alpha}_n=1}
\nonumber
\\
 &~
 \cdot
 G^{(n-1)}_{\alpha_{n-2},s_{n-1},\alpha_{n-1}}
 F^{(n-1)*}_{\tilde{\alpha}_{n-2},s_{n-1},\tilde{\alpha}_{n-1}}
 \cdots
 G^{(j+1)}_{\alpha_{j},s_{j+1},\alpha_{j+1}}
 F^{(j+1)*}_{\tilde{\alpha}_{j},s_{j+1},\tilde{\alpha}_{j+1}},
\end{align}
respectively, as shown in Figs.~\ref{fig:als_mps}(c) and \ref{fig:als_mps}(d).
Then, we can evaluate the $d_i\times R$ matrix $Y^{(j)}$ as
\begin{align}
 Y^{(j)}_{s_j,r}
 &=
 \sum_{\alpha_{j-1},\tilde{\alpha}_{j-1},\alpha_j,\tilde{\alpha}_{j}}
 \delta_{\tilde{\alpha}_{j-1},r,\tilde{\alpha}_{j-1}}
 L^{(j)}_{\alpha_{j-1}, \tilde{\alpha}_{j-1}}
 G^{(j)}_{\alpha_{j-1},s_j,\alpha_j}
 R^{(j)}_{\alpha_{j}, \tilde{\alpha}_{j}}.
\end{align}
For factor $A^{(1)}$ [$A^{(n)}$],
we only need the right (left) matrix to construct matrix $Y^{(j)}$.

The counterpart of the $R\times R$ matrix
$(C^{\dagger}C \circledast B^{\dagger}B)$
can be evaluated just as in the original ALS method.
We have to calculate the element-wise product of the $R\times R$
matrices $n-2$ times. The corresponding matrix $Z^{(j)}$ is given by
\begin{align}
 Z^{(j)}_{r,r'}
 =
 \prod_{k=1, k\neq j}^{n}
 \sum_{i_k}
 A^{(k)*}_{i_k,r} A^{(k)}_{i_k,r'}.
\end{align}

Finally, we can iteratively update matrix $A^{(j)}$ as
\begin{align}
 A^{(j)}_{s_j,r}
 &=
 \sum_{r'}
 Y_{s_j,r'} \left[Z^{(j)}\right]^{-1}_{r',r}
\end{align}
for all $j=1,2,\dots,n$.
To safely compute the inverse of matrix $Z^{(j)}$,
we add a small constant to the diagonal elements of the
matrix $Z^{(j)}$ in practical calculations.
We initialize matrices $A^{(j)}$ with random values
taken from the standard normal distribution
and normalize matrix $A^{(j)}$ in each iteration
to avoid numerical instabilities.
When the fidelity
($|\langle\mathrm{Orig}|\mathrm{CP}\rangle|^2
\langle\mathrm{Orig}|\mathrm{Orig}\rangle^{-1}
\langle\mathrm{CP}|\mathrm{CP}\rangle^{-1}$)
between the MPS representation of the original tensor
($|\mathrm{Orig}\rangle$)
and the MPS representation of the CP decomposed tensor
($|\mathrm{CP}\rangle$)
is sufficiently high after several iterations,
the ALS method gives a good approximation of the CP decomposition.

In general, the most time-consuming part of the ALS method is
the computation of the products between
the matricized tensor and the matrix obtained by the Khatri-Rao product.
When using the MPS representation,
this time-consuming part
is reduced to the computation of the products among MPSs.
The computational cost grows only linearly in the tensor order $n$.
This is in contrast to the original ALS method using the full tensor,
which requires exponential computational cost in the tensor order $n$.

\subsection{RBM wave functions with multinomial hidden units
derived from the CP decomposed MPSs}
\label{subsec:rbm}

We will see that the CP decomposed MPSs are equivalent to the RBM wave
function. Before that, we briefly review the RBM wave function
that is used in quantum many-body problems.

In general,
the RBM wave function for a spin-$1/2$ system~\cite{carleo2017},
\begin{align}
 |\Psi\rangle
 =
 \sum_{\{s_i\}} \Psi(\{s_i\}) |s_1, s_2, \dots, s_n\rangle,
\end{align}
is defined with its wave function amplitude,
\begin{align}
 \Psi(\{v_i\})
 &=
 \sum_{\{h_j\}} \exp[-E(\{v_i\}, \{h_j\})],
\\
 E(\{v_i\}, \{h_j\})
 &=
 - \sum_{i=1}^{n_v} a_i v_i
 - \sum_{j=1}^{n_h} b_j h_j
 - \sum_{i=1}^{n_v} \sum_{j=1}^{n_h} W_{i,j} v_i h_j.
\end{align}
Here, $a_i$, $b_j$, and $W_{i,j}$ are the parameters of the RBM.
The symbols $n_v$ and $n_h$ represent the number of
the visible and hidden units, respectively.
The variable $v_i(=s_i)$ is the index of the $i$th spin
and takes the value $+1$ or $-1$, depending on the state of the spin
$\uparrow$ or $\downarrow$.
For binomial hidden units,
the variable $h_j$ also takes the value $+1$ or $-1$.
Note that, for simplicity,
we ignore the normalization factor of the wave function amplitude, which
is given by
$\sqrt{\langle\Psi|\Psi\rangle}
= \sqrt{ \sum_{\{s_i\}} |\Psi(\{s_i\})|^2 }$.

For the transformation from MPSs to RBM wave functions,
we specifically consider
the multinomial hidden units that take the states from $1$ to
$n_r$~\cite{salakhutdinov2007,guo2016_arxiv,vlugt2020,pei2021b,pei2024}.
Using the one-hot encoding of the hidden units,
we can express the hidden unit $h_j$ as an $n_r$ component vector,
\begin{align}
 h_{j,k} = (0, 0, \dots, 0, 1, 0, \dots, 0),
\end{align}
where only the $k$th component is $1$ and all other components are $0$.
Then, the virtual energy function is given by
\begin{align}
 E(\{v_i\}, \{h_j\})
 &=
 - \sum_{i=1}^{n_v} a_i v_i
 - \sum_{j=1}^{n_h} \sum_{k=1}^{n_r} b_{j,k} h_{j,k}
\nonumber
\\
 &~\phantom{=}
 - \sum_{i=1}^{n_v} \sum_{j=1}^{n_h} \sum_{k=1}^{n_r} W_{i,j,k} v_i h_{j,k},
\end{align}
where the sum over $k$ is taken over the states of the hidden unit $h_j$.
We also add the index $k$ to the parameters $b_{j,k}$ and $W_{i,j,k}$.
Defining the parameter
\begin{align}
 \theta_{j,k} = b_{j,k} + \sum_{i=1}^{n_v} W_{i,j,k} v_i,
\end{align}
we obtain the amplitude in the RBM wave function as
\begin{align}
 \Psi(\{v_i\})
 &=
 \sum_{\{h_j\}}
 \exp\left(
 \sum_{i=1}^{n_v} a_i v_i + \sum_{j=1}^{n_h} \sum_{k=1}^{n_r} h_{j,k}
 \theta_{j,k}
 \right)
\\
 &=
 \exp\left( \sum_{i=1}^{n_v} a_i v_i \right)
 \prod_{j=1}^{n_h}
 \left[ \sum_{k=1}^{n_r} \exp(\theta_{j,k}) \right].
\end{align}

We wish to convert the CP decomposed MPSs
for $n=n_v$ sites, given by
\begin{align}
 \Psi(\{s_i\})
 &=
 T_{s_1, s_2, \dots, s_n}
 =
 \sum_{r=1}^{R}
 A^{(1)}_{i_1,r} A^{(2)}_{i_2,r} \cdots A^{(n)}_{i_n,r},
\\
 i_j &= \frac{1-s_j}{2} \quad (j=1,2,\dots,n),
\end{align}
into the multinomial
RBM wave function. One of such choices is to set the parameters
in the following manner:
\begin{align}
 a_i &= 0,
\\
 n_h &= 1,
\\
 n_r &= R,
\\
 \exp(\theta_{j=1,k})
 &=
 A^{(1)}_{i_1,k} A^{(2)}_{i_2,k} \cdots A^{(n_v)}_{i_{n_v},k}.
\end{align}
By taking the logarithm in the last equation,
we obtain
\begin{align}
 \theta_{j=1,k}
 =
 \sum_{l=1}^{n_v}
 \left(
 \frac{b_{j=1,k}}{n_v} + W_{l,j=1,k} v_l
 \right)
 =
 \sum_{l=1}^{n_v} \mathrm{Log\,} A^{(l)}_{i_l,k},
\end{align}
where $\mathrm{Log\,} z = \ln |z| + i \arg z$ is the complex logarithm.
When $z=0$, we add a small constant to $z$ to calculate
$\mathrm{Log\,} z$.
Since $v_l=\pm 1$ with $i_l=0,1$ for a spin-$1/2$ system,
formally, it is sufficient to set the parameters as
\begin{align}
\label{eq:rbm_bw_from_mps_a}
 \sum_{l=1}^{n_v}
 \frac{b_{j=1,k}}{n_v}
 &=
 \sum_{l=1}^{n_v}
 \frac{\mathrm{Log\,} A^{(l)}_{i_l=0,k}
     + \mathrm{Log\,} A^{(l)}_{i_l=1,k}}{2},
\\
 W_{l,j=1,k}
 &=
 \frac{\mathrm{Log\,} A^{(l)}_{i_l=0,k}
     - \mathrm{Log\,} A^{(l)}_{i_l=1,k}}{2}
\end{align}
for each $l=1,2,\dots,n_v$ and $k=1,2,\dots,R$.

By introducing a new site-dependent parameter $b_{l,k}$,
corresponding to the parameter $b_{j=1,k}/n_v$
in Eq.~\eqref{eq:rbm_bw_from_mps_a},
we finally obtain the RBM wave function
for $n$ sites with the rank $R$ as
\begin{align}
 \Psi(\{v_l\})
 &=
 \sum_{k=1}^{R}
 \exp\left(
 \sum_{l=1}^{n}
 \theta_{l,k}
 \right),
\\
 \theta_{l,k}
 &=
 b_{l,k} + W_{l,k} v_l,
\end{align}
with the parameters in the CP decomposed MPSs,
\begin{align}
 b_{l,k}
 &=
 \frac{\mathrm{Log\,} A^{(l)}_{i_l=0,k}
     + \mathrm{Log\,} A^{(l)}_{i_l=1,k}}{2},
\\
 W_{l,k}
 &=
 \frac{\mathrm{Log\,} A^{(l)}_{i_l=0,k}
     - \mathrm{Log\,} A^{(l)}_{i_l=1,k}}{2}
\end{align}
for each $l=1,2,\dots,n$ and $k=1,2,\dots,R$.
Therefore,
once we obtain the CP decomposition of MPSs,
we can convert the MPSs into RBM wave functions immediately.
Note that the relation between the CP decomposed tensors
and deep Boltzmann machine wave functions 
with more than one hidden layer is already discussed
before~\cite{clark2018,chen2018}.
Here, we stick to the simplest RBM wave function without
introducing deeper hidden layers.
The expressibility of the RBM wave function is
extended by allowing the hidden unit to have multinomial
values~\cite{salakhutdinov2007,guo2016_arxiv,vlugt2020,pei2021b,pei2024}.

In general, as the number of hidden units increases,
the RBM wave function well approximates the ground state
of the quantum many-body system,
which is known as the universal approximation theorem~\cite{leroux2008}.
By contrast, as for the present RBM wave function
with a single multinomial hidden unit,
this theorem does not necessarily hold.
However, by definition,
the CP decomposition exactly reproduces
the original tensor when $R$ is larger than or equal to
the CP rank of the tensor~\cite{kolda2009};
even for $R$ that is smaller than the CP rank,
the CP decomposition would nearly reconstruct
the original tensor when $R$ is sufficiently
large~\cite{comon2009,erichson2020}.
This fact implies that the RBM wave function with a single hidden unit
has a potential to represent the ground state of the quantum many-body
system when $R$ is sufficiently large,
although $R$ could be exponentially large in the number of sites $n$
in the worst case~\cite{kour2023}.

\subsection{Optimization of RBM wave functions}
\label{subsec:opt_rbm}

In the variational Monte Carlo (VMC) simulation,
we calculate physical quantities using the Markov chain Monte Carlo
sampling over the probability distribution
$p(x) = |\langle x | \Psi \rangle|^2 / \langle \Psi | \Psi \rangle$
with $|x\rangle$ being a real-space spin
configuration~\cite{gubernatis2016,becca2017}.
We optimize the parameters in the RBM wave function
by the gradient-based optimization method.
In particular,
we use the stochastic reconfiguration (SR)
method~\cite{sorella1998,sorella2000},
which is known to be
equivalent~\cite{nomura2017}
to the natural gradient method~\cite{amari1996,amari1998}.
The derivative of the RBM wave function to calculate the gradient
is given by
\begin{align}
 \tilde{\theta}_{k}
 &=
 \sum_{l=1}^{n} \theta_{l,k},
\\
 \frac{\partial \ln\Psi(\{v_i\})}{\partial b_{l,k}}
 &=
 \frac{ \exp \tilde{\theta}_{k} }
 { \sum_{k'=1}^{R} \exp \tilde{\theta}_{k'} },
\\
 \frac{\partial \ln\Psi(\{v_i\})}{\partial W_{l,k}}
 &=
 v_l \frac{ \exp \tilde{\theta}_{k}}
 { \sum_{k'=1}^{R} \exp \tilde{\theta}_{k'} }.
\end{align}
The right-hand side of the above equations
is expressed by the softmax function.
To avoid numerical instabilities such as overflow and underflow,
we typically calculate a maximum of a real part of
$\tilde{\theta}_{k}$
[$m = \max_{k} \mathrm{Re\,}\tilde{\theta}_{k}$]
and estimate the exponentials in the numerator and denominator by
$\exp(\tilde{\theta}_{k} - m)$.
The most time-consuming part of the SR method is
solving the linear equations with a large number of variational parameters.
In general, the computational cost scales quadratically
with the number of variational parameters~\cite{carleo2017}.

\section{Applications to quantum many-body problems:
transverse-field Ising model}
\label{sec:applications}

We examine the applicability of our method
by taking the transverse-field Ising model as an example.
The Hamiltonian of the one-dimensional
transverse-field Ising model is given by
\begin{align}
 H =
 - J \sum_{i} \sigma^z_i \sigma^z_{i+1}
 - h \sum_{i} \sigma^x_i,
\end{align}
where $J$ and $h$ are the strength of the spin-exchange interaction
and the transverse field, respectively.
For open boundary conditions,
the summation over $i$ for the interaction term
is taken over $i = 1, \ldots, n-1$,
where $n$ is the number of sites in the system.
For periodic boundary conditions,
the summation over $i$ for the interaction term
is taken over $i = 1, \ldots, n$,
and the last site is connected to the first site
by the interaction term as
$\sigma^z_{n+1} := \sigma^z_1$.

The ground state of the one-dimensional
transverse-field Ising model can be solved analytically
using the Jordan-Wigner transformation~\cite{pfeuty1970},
which maps spin operators onto fermionic creation and
annihilation operators.
The ground-state energy is given by
\begin{align}
 E
 =
 - J \sum_{k}
 \sqrt{1 + \lambda^2 - 2 \lambda \cos k}
\quad
 (\lambda = h / J).
\end{align}
Hereafter, the energy is expressed in the units of $J$.
For periodic boundary conditions
with even total fermion number,
the summation over $k$ is taken over the set of momenta,
satisfying
\begin{align}
 k = \frac{\pi (2j+1)}{n}
\quad
 (j = 0, 1, 2, \dots, n-1).
\end{align}
For open boundary conditions~\cite{cabrera1987,he2017},
the corresponding $n$ momenta fulfill
\begin{align}
 \lambda \sin[(n+1)k] = \sin nk.
\end{align}

Hereafter, we first consider the model under open boundary conditions
and obtain the ground-state wave function in the MPS representation
using the DMRG method.
We apply the \textsc{TeNPy} library for the DMRG
method~\cite{hauschild2018,hauschild2024}
and choose the bond dimension that is large enough to represent
the true ground-state energy within a desired accuracy,
which is smaller than $10^{-9}$
in the units of $J$ in this study.
We turn off the parity conservation during the DMRG simulation
to purposely keep the wave function primitive.
We then apply the CP decomposition to the MPS representation
of the ground state.
We find that the CP decomposed MPSs are sufficiently close to
the ground-state wave function when the rank of the CP decomposition
is large.
The infidelity between the ground-state MPS and
the approximated CP decomposed tensor decreases at most polynomially
as the rank of the CP decomposition increases.
We then approximately convert the MPS representation into the RBM wave function
consisting of a single hidden unit and apply the VMC
method to optimize the parameters in the RBM wave function.
We demonstrate that the initial state prepared under open boundary
conditions can be efficiently used for the simulation under periodic
boundary conditions.

Note that the ground state of the present transverse-field Ising model
does not have a complex nodal structure.
The amplitude of the ground-state wave function can be taken as non-negative
according to the Perron-Frobenius theorem~\cite{sehayek2019,tasaki2020}.
This fact suggests that one can use positive factors for the CP
decomposition in this study.
However, to analyze the performance of the CP decomposition
more generally, we purposely apply the CP decomposition allowing both
positive and negative factors.
As we see later, the CP decomposition with positive and negative
factors also well approximates the ground-state wave function and
provides a good initial state having complex amplitudes
for the VMC simulation.

\subsection{Accuracy of the CP decomposition of MPSs}
\label{subsec:acc_cpd_mps}

We calculate two types of errors to evaluate
the accuracy of the CP decomposition of MPSs.
One is the infidelity
$\bar{F}$
between the ground-state MPS ($|\mathrm{MPS}\rangle$) and
the approximated CP decomposed tensor ($|\mathrm{CP}\rangle$),
defined by
\begin{align}
 \bar{F}
 &=
 1 - F,
\\
 F
 &=
 \frac{|\langle\mathrm{MPS}|\mathrm{CP}\rangle|^2}
 {\langle\mathrm{MPS}|\mathrm{MPS}\rangle
 \langle\mathrm{CP}|\mathrm{CP}\rangle}.
\end{align}
When the CP decomposition exactly reproduces the ground-state wave function,
the infidelity becomes zero.
For simplicity, all the elements in $|\mathrm{MPS}\rangle$
and $|\mathrm{CP}\rangle$ are assumed to be real
in the transverse-field Ising model.
The other is the energy difference between
the exact ground-state energy $E_{\mathrm{exact}}$ and
the energy calculated from the approximated CP decomposed tensor
($E_{\mathrm{CP}}$), defined by
\begin{align}
 \Delta E
 &=
 \left|
 \frac{E_{\mathrm{CP}} - E_{\mathrm{exact}}}
 {E_{\mathrm{exact}}}
 \right|,
\\
 E_{\mathrm{CP}}
 &=
 \frac{\langle\mathrm{CP}|H|\mathrm{CP}\rangle}
 {\langle\mathrm{CP}|\mathrm{CP}\rangle}.
\end{align}
We see that both of these errors decrease polynomially
as the rank of the CP decomposition increases.

\begin{figure}[!t]
\centering
\includegraphics[width=.95\linewidth]{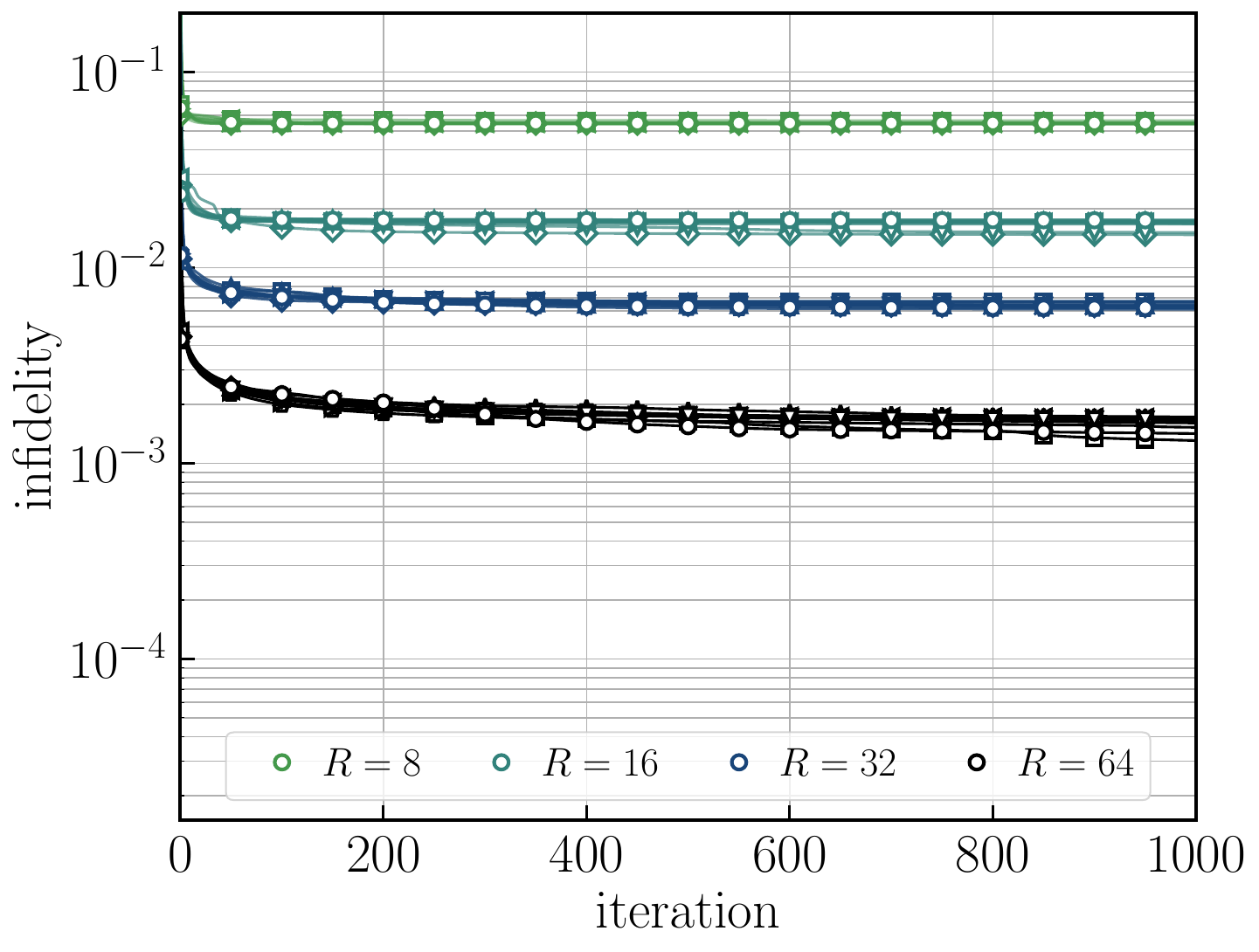}
\caption{%
Infidelity as a function of iterations for $h/J = 2$ and $n=16$.
We show the results for ranks ranging from $R=8$ to $R=64$.
Different colors represent different ranks of the CP decomposition,
whereas different symbols represent different initial factors
for the CP decomposition.
}
\label{fig:cpd_err_it_h2}
\end{figure}

\begin{figure}[!t]
\centering
\includegraphics[width=.9\linewidth]{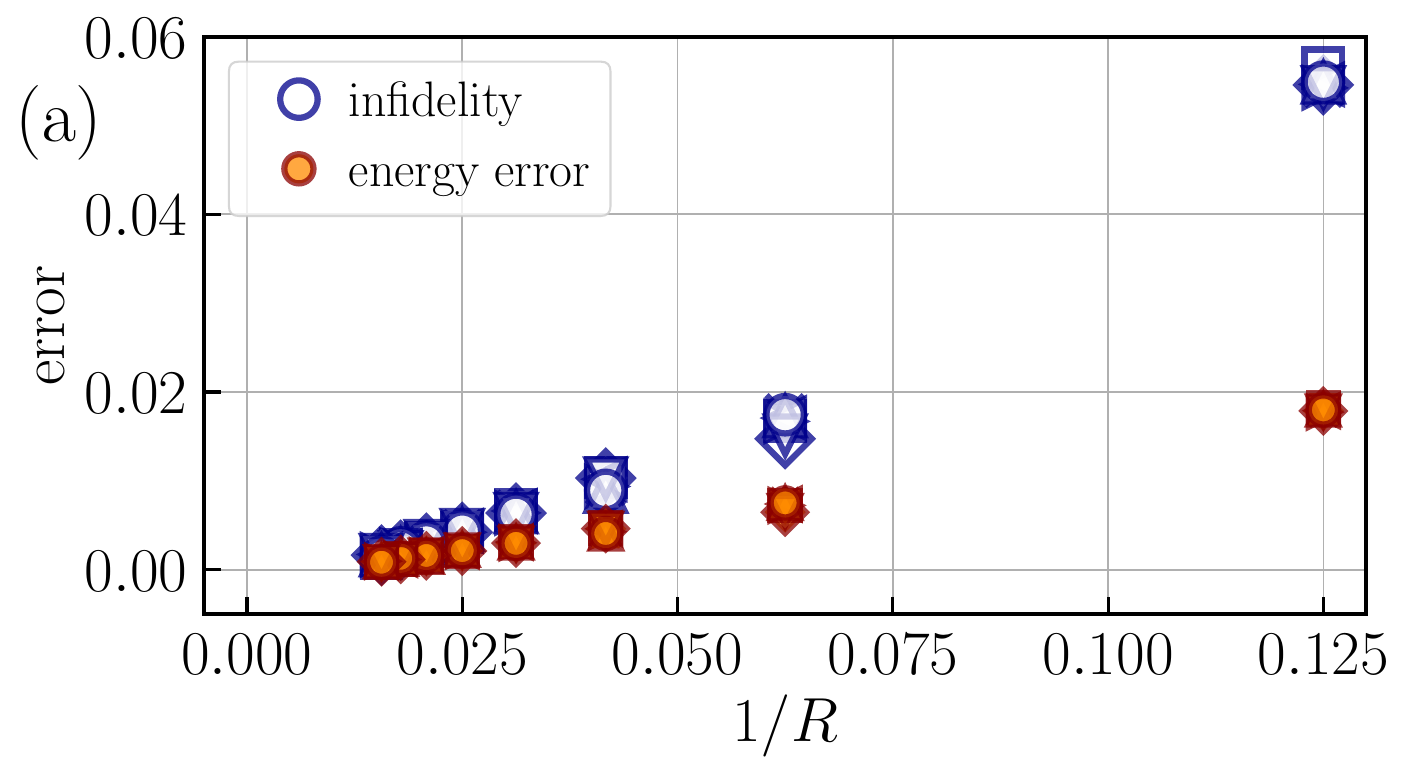}\\
\includegraphics[width=.9\linewidth]{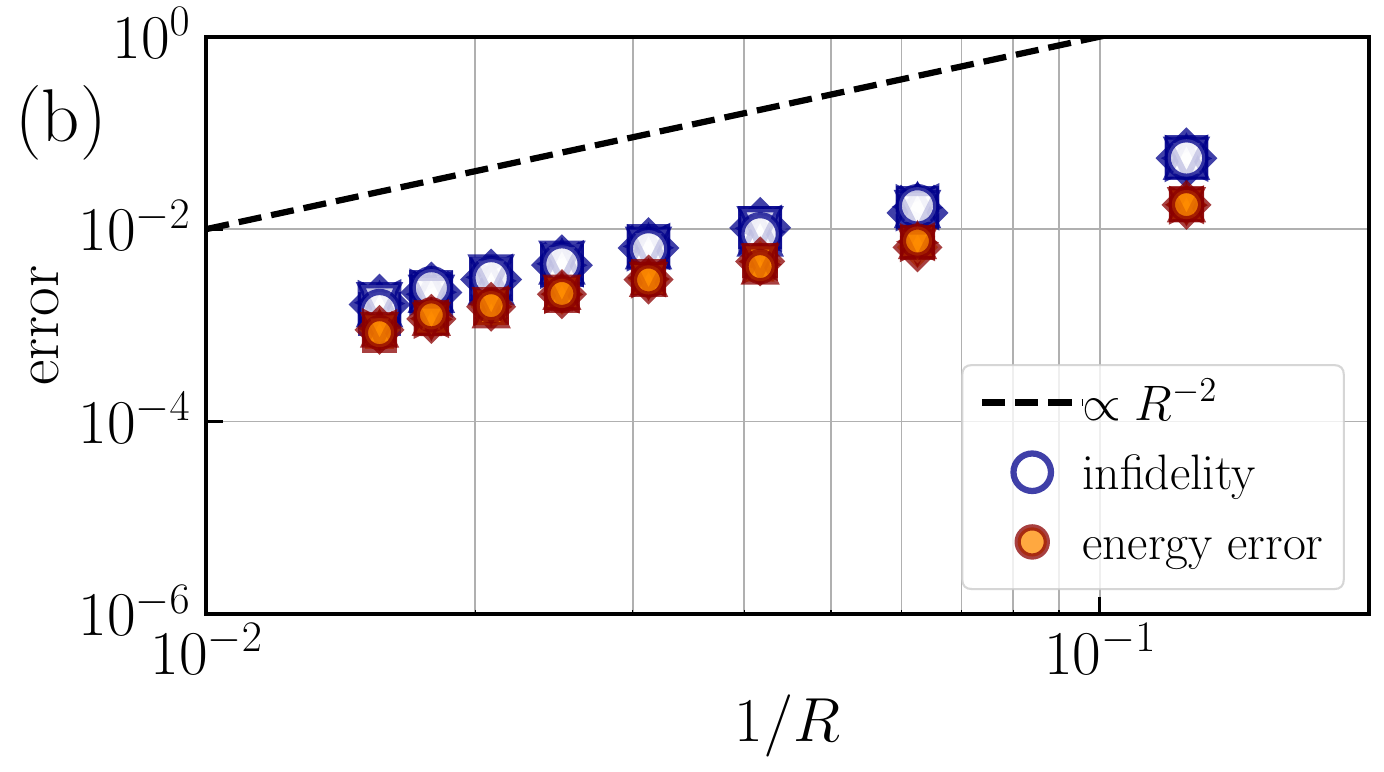}
\caption{%
(a) Infidelity and energy difference
as a function of the rank $R$ of the CP decomposition
for $h/J = 2$ and $n=16$.
Different symbols represent different initial factors
for the CP decomposition.
Open symbols are for the infidelity and
filled symbols are for the energy difference.
(b) Same as panel (a) but shown on a logarithmic scale.
We plot the line proportional to $R^{-2}$ as a reference.
Note that, hereafter, all energy values are expressed in the units of $J$.
}
\label{fig:cpd_err_R_h2}
\end{figure}

\begin{figure}[!t]
\centering
\includegraphics[width=.95\linewidth]{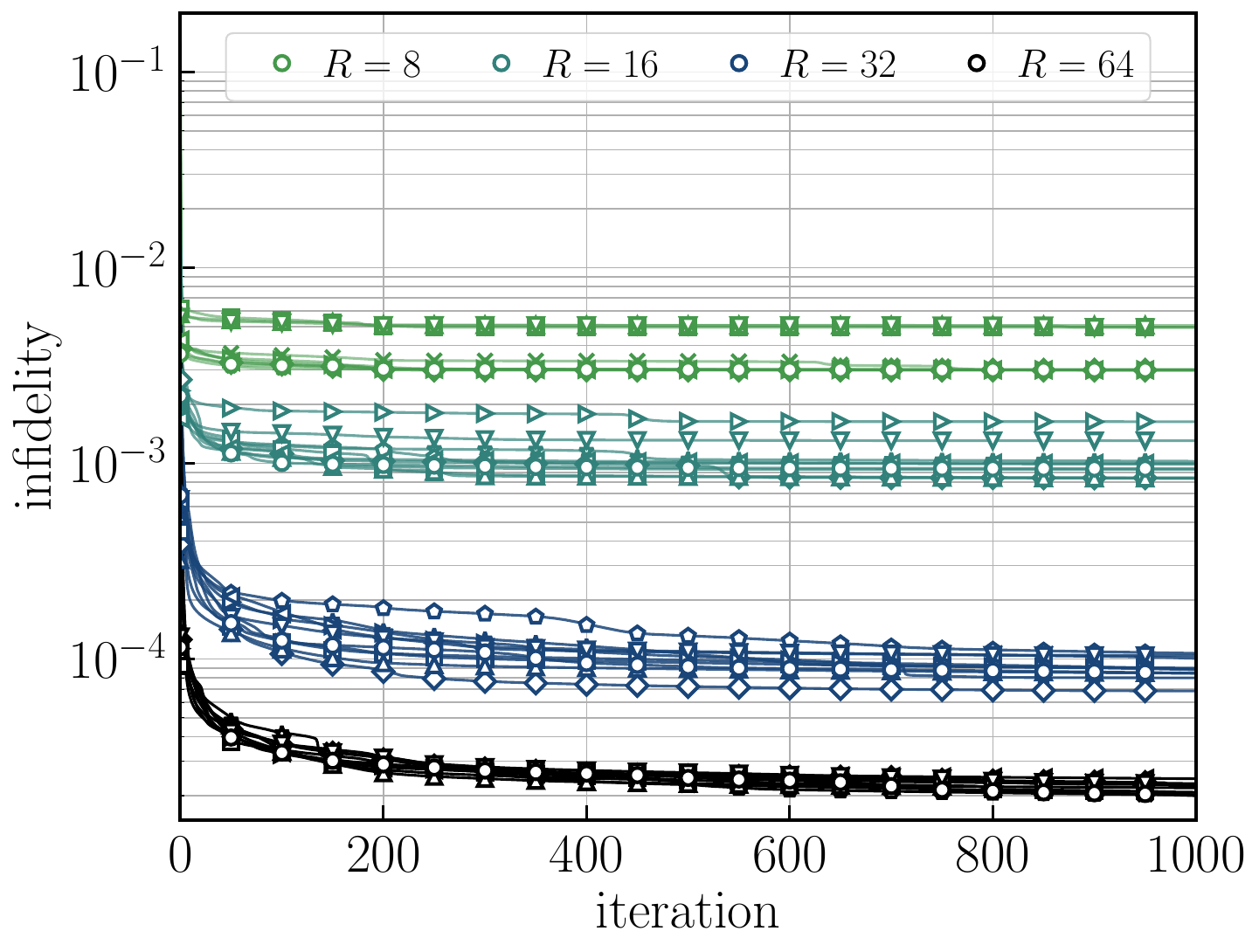}
\caption{%
Infidelity as a function of iterations for $h/J = 0.5$ and $n=16$.
We show the results for ranks ranging from $R=8$ to $R=64$.
Different colors represent different ranks of the CP decomposition,
whereas different symbols represent different initial factors
for the CP decomposition.
}
\label{fig:cpd_err_it_h05}
\end{figure}

\begin{figure}[!t]
\centering
\includegraphics[width=.9\linewidth]{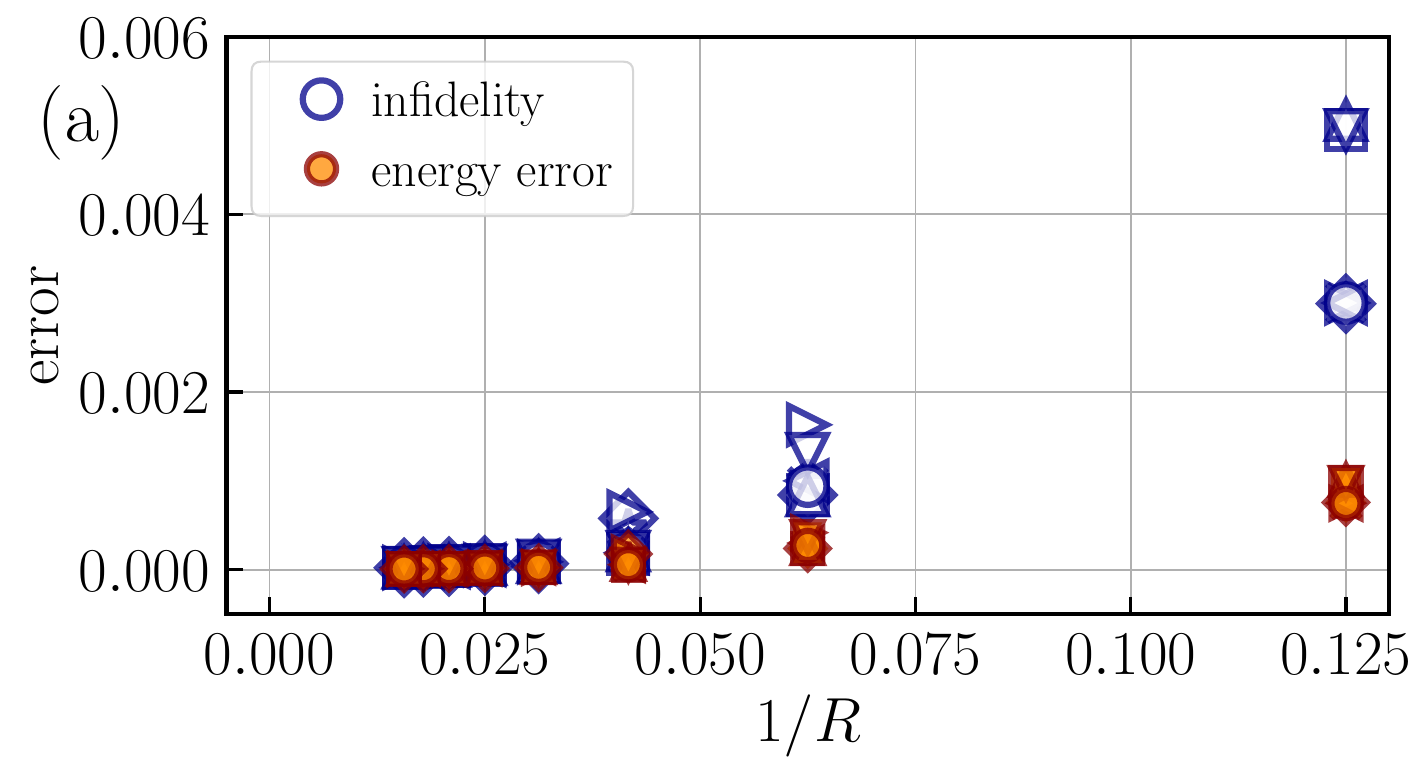}\\
\includegraphics[width=.9\linewidth]{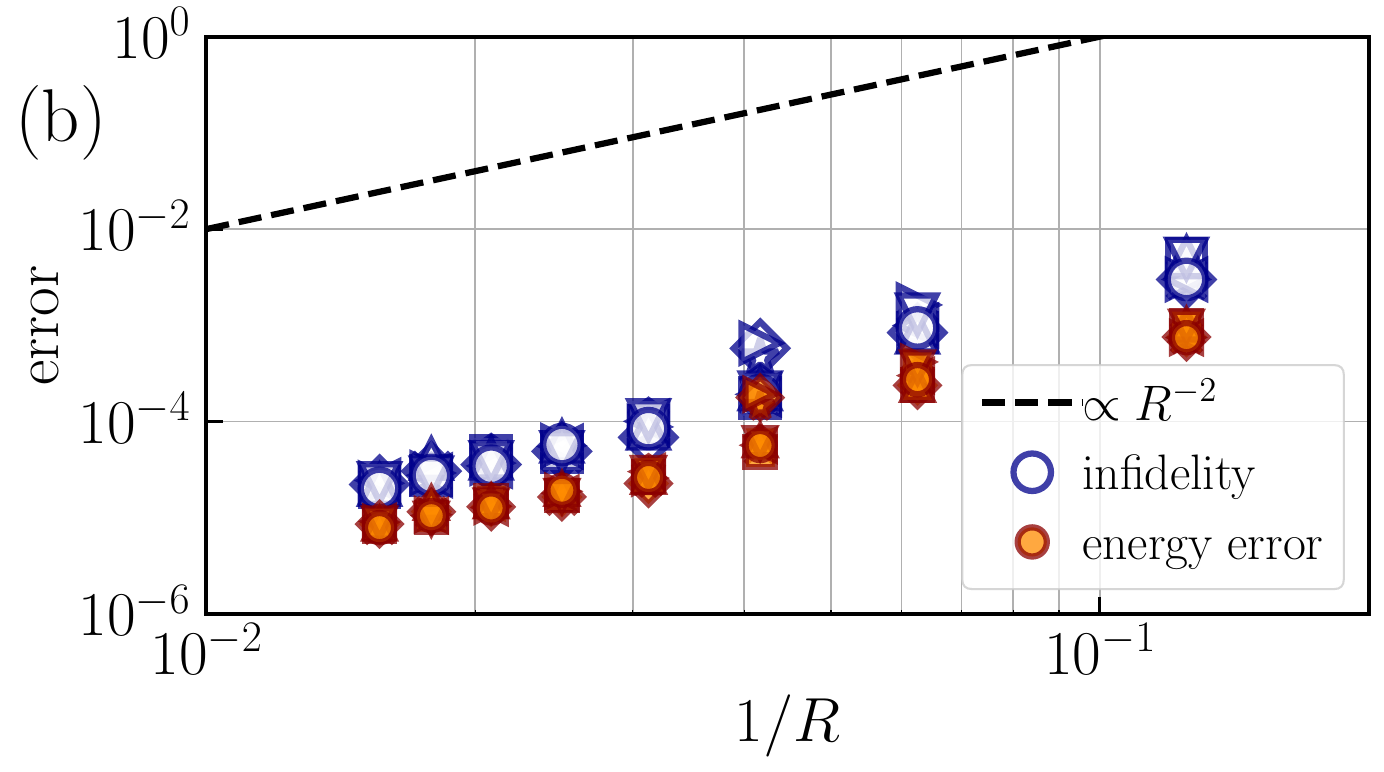}
\caption{%
(a) Infidelity and energy difference
as a function of the rank $R$ of the CP decomposition
for $h/J = 0.5$ and $n=16$.
Different symbols represent different initial factors
for the CP decomposition.
Open symbols are for the infidelity and
filled symbols are for the energy difference.
(b) Same as panel (a) but shown on a logarithmic scale.
We plot the line proportional to $R^{-2}$ as a reference.
}
\label{fig:cpd_err_R_h05}
\end{figure}

\begin{figure}[!t]
\centering
\includegraphics[width=.95\linewidth]{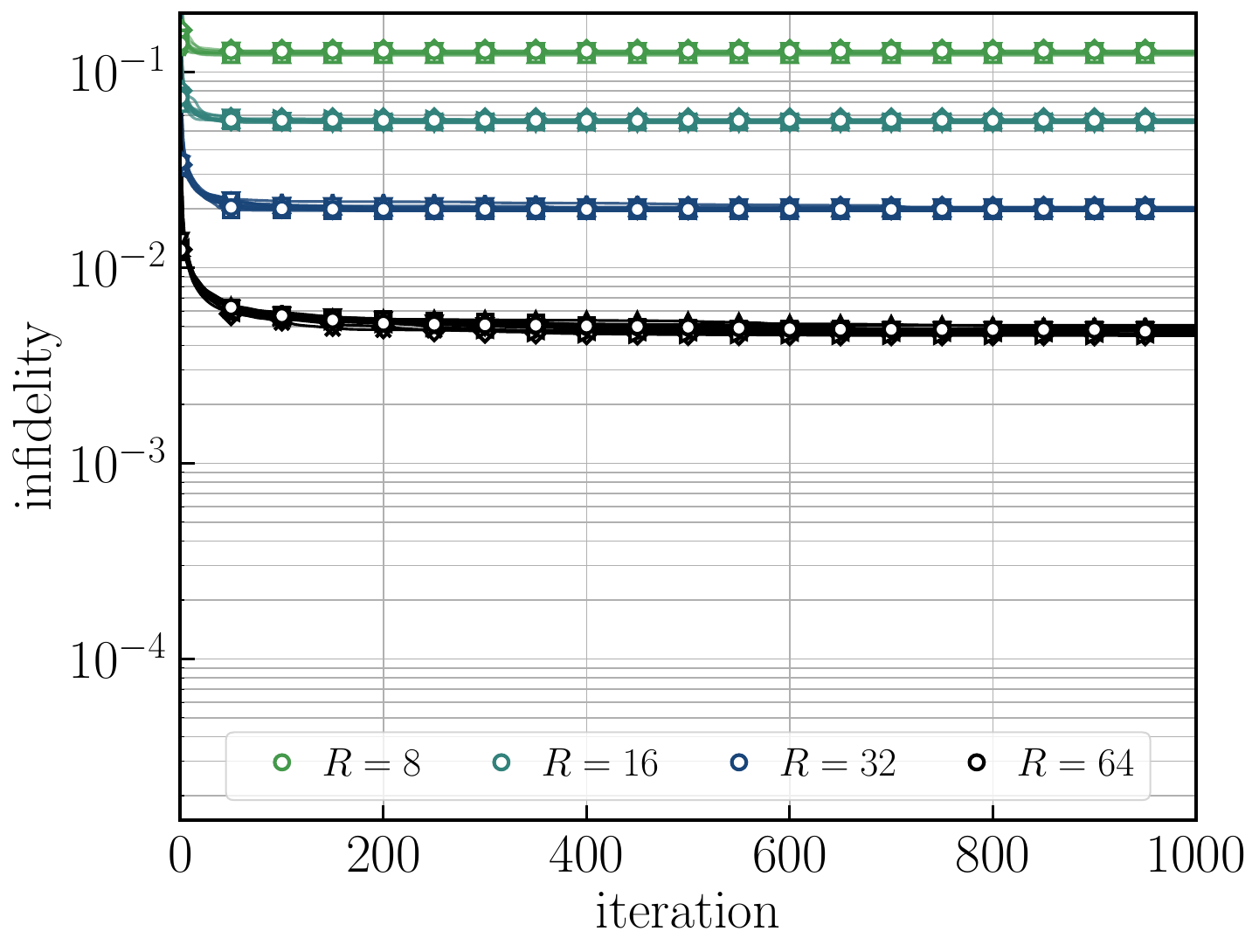}
\caption{%
Infidelity as a function of iterations for $h/J = 1$ and $n=16$.
We show the results for ranks ranging from $R=8$ to $R=64$.
Different colors represent different ranks of the CP decomposition,
whereas different symbols represent different initial factors
for the CP decomposition.
}
\label{fig:cpd_err_it_h1}
\end{figure}

\begin{figure}[!t]
\centering
\includegraphics[width=.9\linewidth]{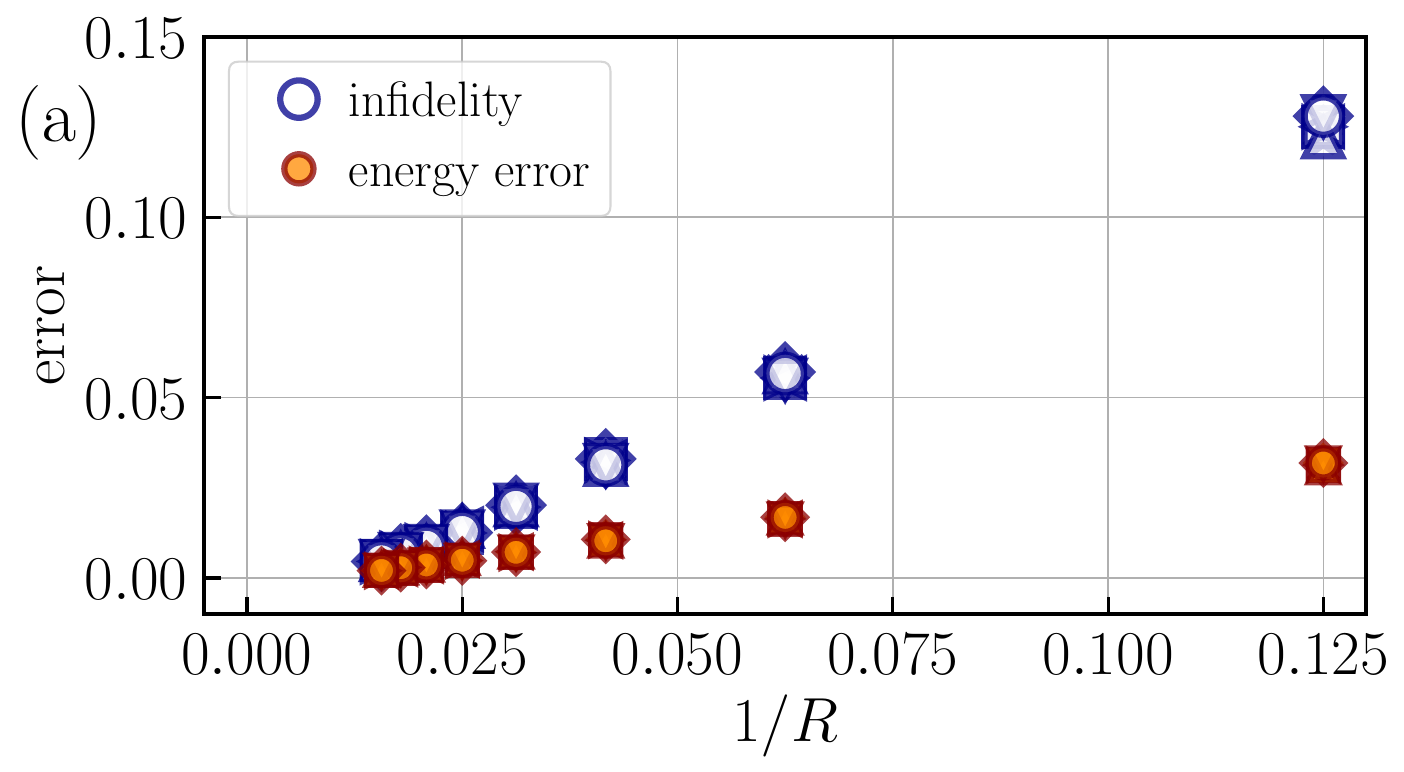}\\
\includegraphics[width=.9\linewidth]{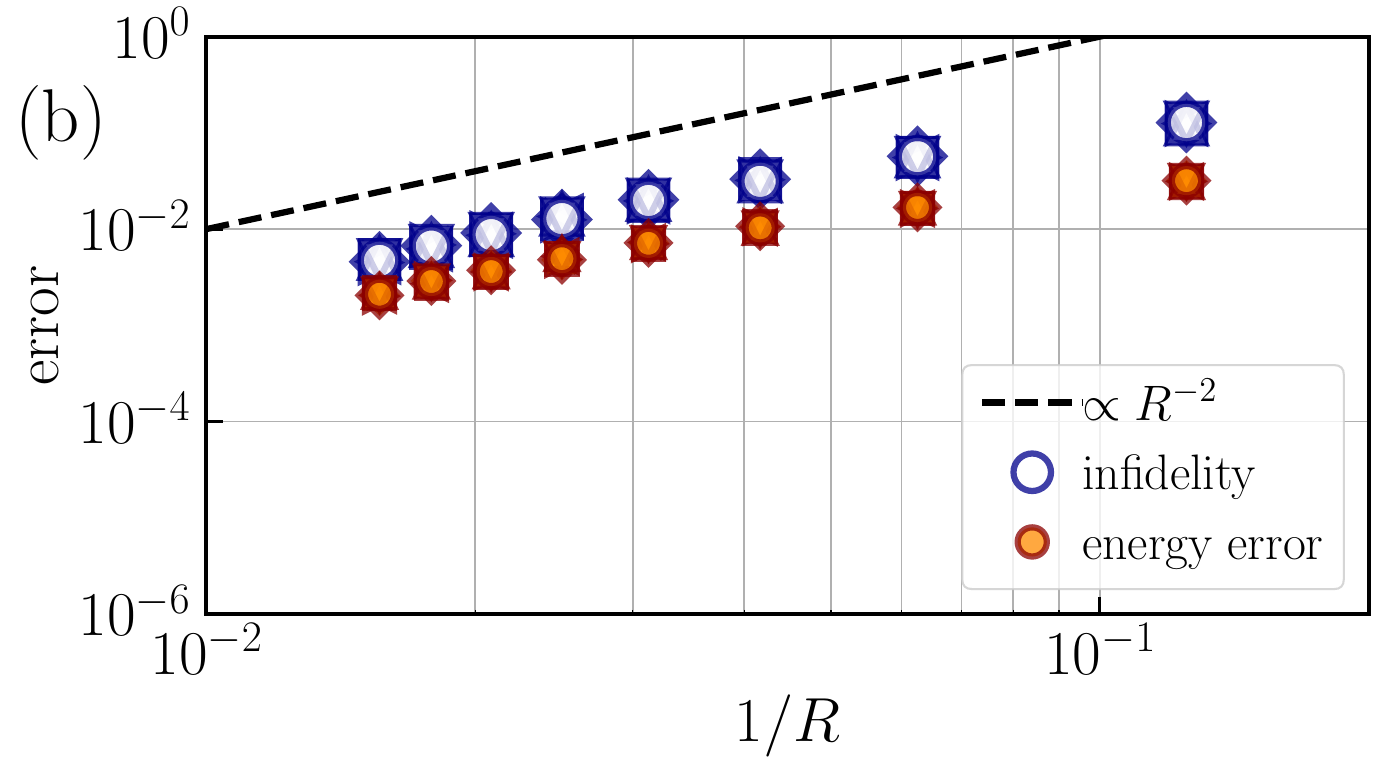}
\caption{%
(a) Infidelity and energy difference
as a function of the rank $R$ of the CP decomposition
for $h/J = 1$ and $n=16$.
Different symbols represent different initial factors
for the CP decomposition.
Open symbols are for the infidelity and
filled symbols are for the energy difference.
(b) Same as panel (a) but shown on a logarithmic scale.
We plot the line proportional to $R^{-2}$ as a reference.
}
\label{fig:cpd_err_R_h1}
\end{figure}

We first examine the model under open boundary conditions
for $h/J = 2$. The system size is chosen as $n = 16$,
and the maximum bond dimension is set as $D_{\mathrm{max}} = 8$
for the DMRG simulation.
For the fixed system size $n$ and
the fixed bond dimension $D_{\mathrm{max}}$,
we increase the rank of the CP decomposition
from $R=8$ to $R=64$ and calculate the errors.
Note that the CP decomposition by the ALS method itself does not require
the knowledge of the Hamiltonian $H$.
The Hamiltonian $H$ is only required to calculate the energy difference
between the ground state and the approximated CP decomposed tensor.

As shown in Fig.~\ref{fig:cpd_err_it_h2},
the infidelity between the ground-state MPS and
the approximated CP decomposed tensor
almost converges to a small constant value
for iterations larger than a few hundred steps.
We consider $10$ different sets of random initial factors
for the CP decomposition.
Regardless of the choice of the initial factors,
the infidelity nearly converges to the same value
for the same rank of the CP decomposition,
suggesting that the CP decomposition is less susceptible
to the random initial state.
As $R$ increases,
the converged value of the infidelity gradually decreases.

We illustrate the $R$ dependence of the errors in
Fig.~\ref{fig:cpd_err_R_h2}.
Both infidelity and energy difference decrease monotonically
as $R$ increases.
The decay appears to be faster than $\sim R^{-1}$.
Indeed, for larger $R$,
$R$ dependencies of both errors are well approximated by the line
proportional to $R^{-2}$.

We then examine the case for $h/J = 0.5$.
As shown in Fig.~\ref{fig:cpd_err_it_h05},
similarly to the case for $h/J = 2$,
the infidelity converges to a small constant value
within a few hundred iterations.
When the rank of the CP decomposition is fixed,
the converged value of the infidelity is typically one digit smaller
than the case for $h/J = 2$.
We also observe that the infidelity is rather sensitive
to the initial factor of the CP decomposition.

We also show the $R$ dependence of the errors in
Fig.~\ref{fig:cpd_err_R_h05}.
Both infidelity and energy difference decrease 
rapidly as $R$ increases.
When $R\ge 24$, the energy error becomes less than $10^{-4}$ and
is almost invisible from the plot.
When plotting the errors in the logarithmic scale,
we confirm that the decay appears to obey
the same $R$ dependence ($\sim R^{-2}$)
as is the case for $h/J = 2$.

Finally, we examine the case for $h/J = 1$,
corresponding to the critical point of the model.
At the critical point, the correlation length diverges,
and the ground-state wave function contains long tails
that do not decay to zero at long distances.
Such a wave function is generally hard to represent
with a small number of parameters.
However, we show that the CP decomposition
is able to represent the ground-state wave function,
just as in the cases for $h/J = 0.5$ and $h/J = 2$.
As shown in Fig.~\ref{fig:cpd_err_it_h1},
the infidelity again converges to a small constant value
within a few hundred iterations.
The value is larger than the cases for $h/J = 0.5$
and $h/J = 2$, but it gradually decreases as $R$ increases.
The overall trend is similar to the case for $h/J = 2$,
and the initial-factor dependence of the CP decomposition
is small even at the critical point.

We show the $R$ dependence of the errors
at the critical point in Fig.~\ref{fig:cpd_err_R_h1}.
Both infidelity and energy difference decrease 
monotonically as $R$ increases.
For larger $R$,
$R$ dependencies of both errors satisfy the power-law decay
$\sim R^{-2}$.
The exponent of the decay is consistent with
what we observed in the cases for $h/J = 0.5$ and $h/J = 2$.
Both infidelity and energy difference behave in the same way
for all parameters $h/J$ that we have considered.
Therefore, both quantities can be viewed as an equivalent indicator of
the errors.

\subsection{VMC simulation under open boundary
conditions}
\label{subsec:vmc_open}

\begin{figure}[!t]
\centering
\includegraphics[width=.95\linewidth]{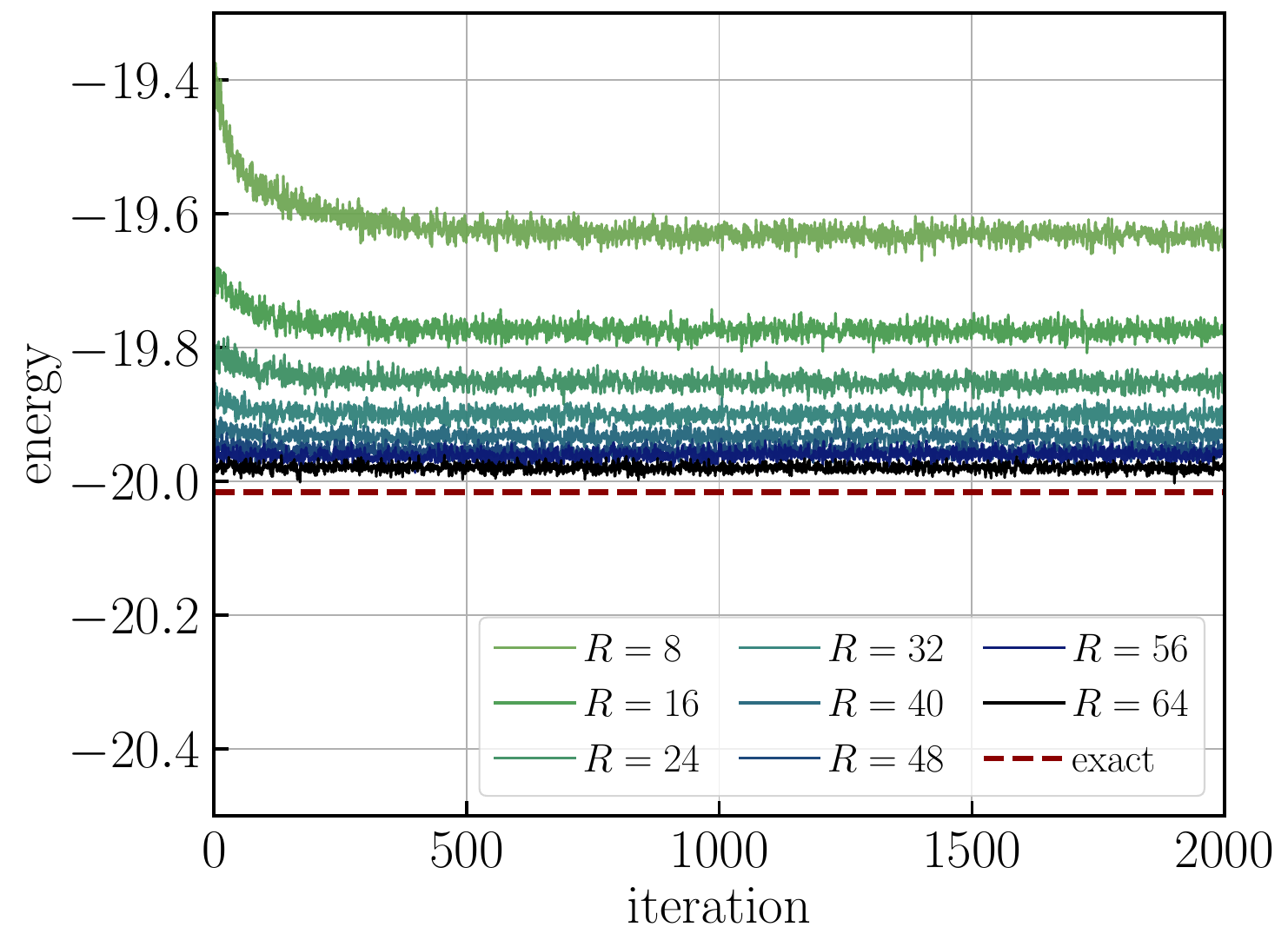}
\caption{%
VMC simulation for $h/J = 1$ and $n = 16$
under open boundary conditions.
The initial state is generated by the CP decomposition of the
ground-state MPS with the maximum bond dimension $D_{\mathrm{max}} = 8$
under open boundary conditions.
With increasing the rank $R$ of the CP decomposition,
the energy approaches the true ground-state energy
for open boundary conditions.
}
\label{fig:vmc_opt_open}
\end{figure}

\begin{figure}[!t]
\centering
\includegraphics[width=.9\linewidth]{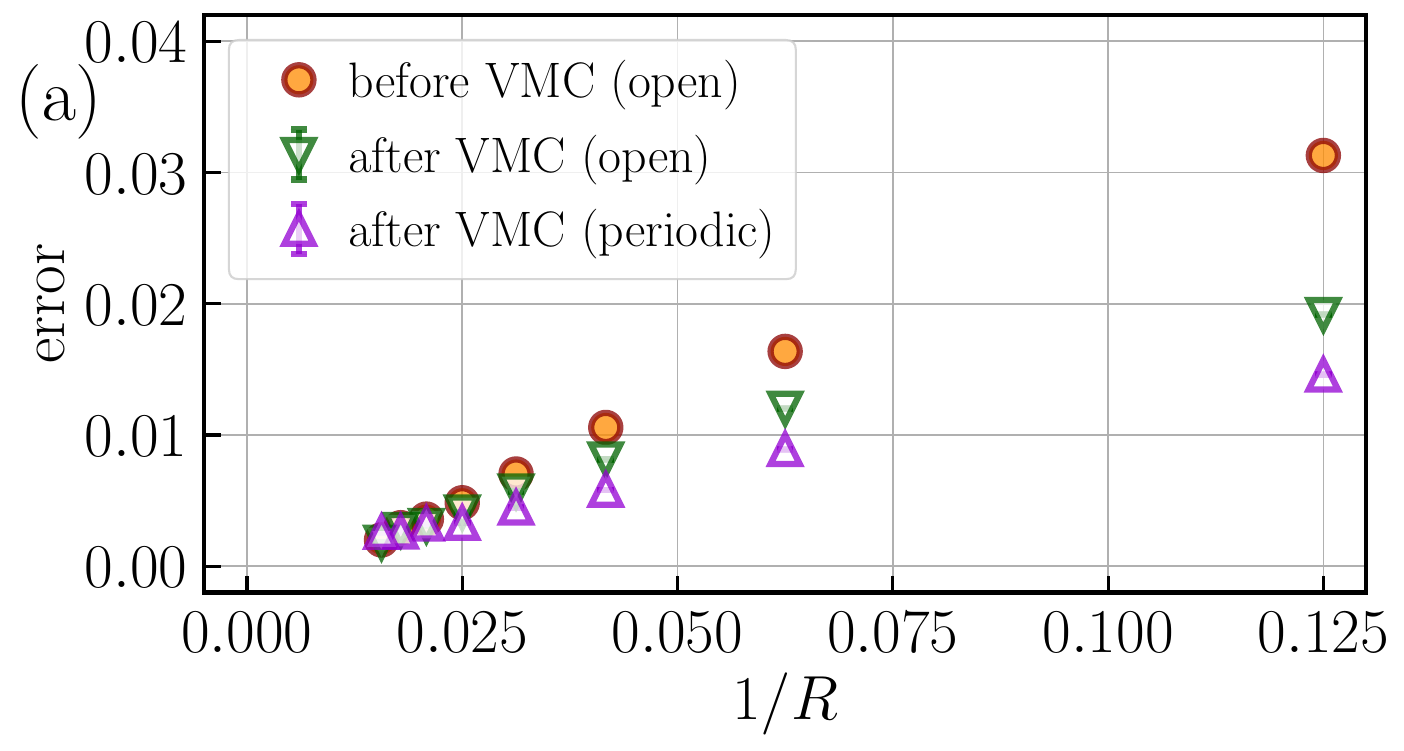}\\
\includegraphics[width=.9\linewidth]{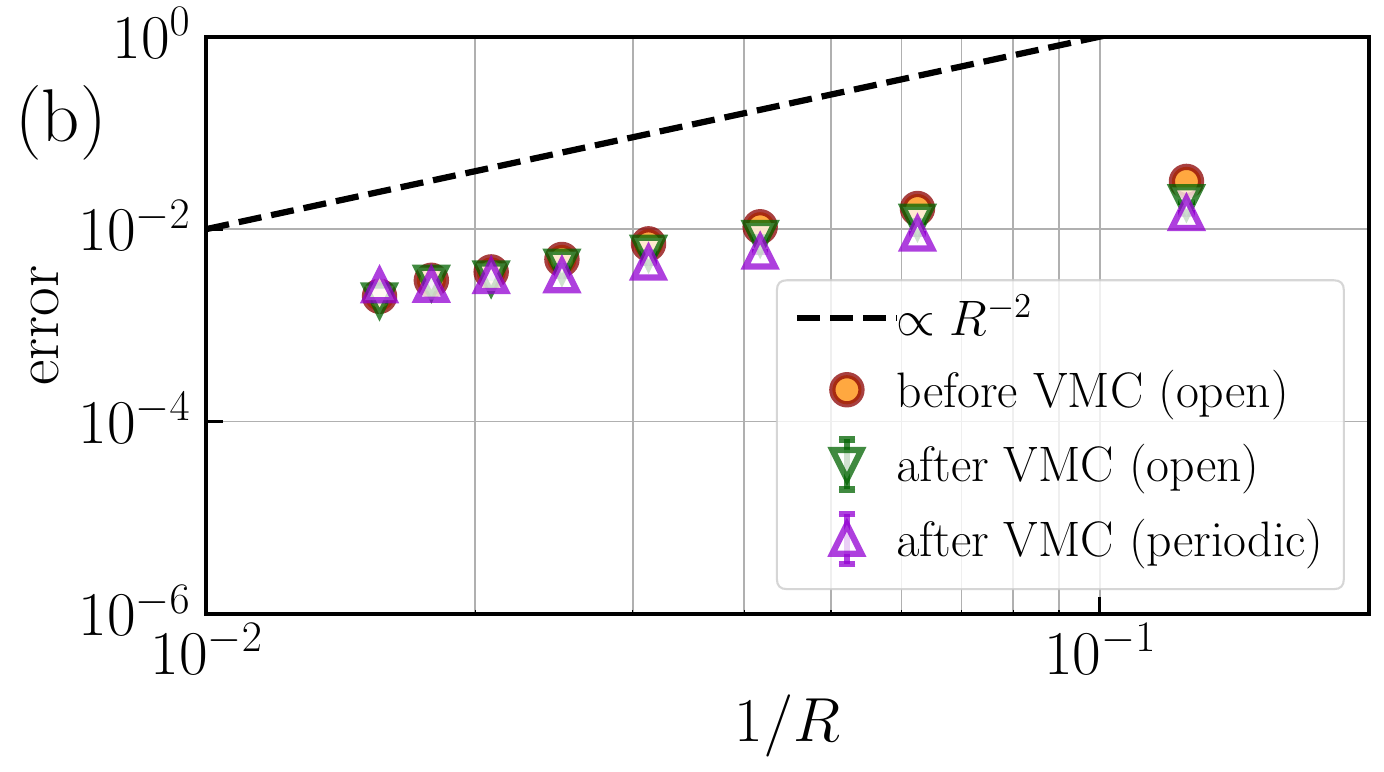}
\caption{%
(a) Energy error
as a function of the rank $R$ of the CP decomposition
for $h/J = 1$ and $n=16$.
Filled circles represent the energy error
($|E_{\mathrm{initial,open}}-E_{\mathrm{exact,open}}|/|E_{\mathrm{exact,open}}|$)
of the initial state for open boundary conditions
used for the VMC method.
Up-pointing and down-pointing triangles represent the energy error
($|E_{\mathrm{VMC,open}}-E_{\mathrm{exact,open}}|/|E_{\mathrm{exact,open}}|$,
$|E_{\mathrm{VMC,periodic}}-E_{\mathrm{exact,periodic}}|/|E_{\mathrm{exact,periodic}}|$)
of the optimized state for open and periodic boundary conditions,
respectively.
(b) Same as panel (a) but shown on a logarithmic scale.
We plot the line proportional to $R^{-2}$ as a reference.
}
\label{fig:vmc_err_R_h1}
\end{figure}

We have observed that the CP decomposition
is able to represent the ground-state wave function
efficiently with increasing the rank of the CP decomposition.
To investigate how close the CP decomposed MPSs
(the initial state) is to
the ground-state wave function
within a given rank of the CP decomposition,
we apply the VMC method to
further optimize the parameters in the RBM wave function.
For each iteration,
we typically thermalize the system for $10^4$ Monte Carlo steps
and evaluate the physical quantities for $10^4$ Monte Carlo steps.
We optimize the parameters in the RBM wave function
for $2000$ iterations and average them for the last $100$ iterations
to obtain the final parameters.
We then calculate the energy and its statistical error
using independent measurements for $32$ bins,
each of which contains $10^5$ Monte Carlo steps for thermalization
and $10^5$ Monte Carlo steps for evaluation of the physical quantities.

Figure~\ref{fig:vmc_opt_open}
shows the optimization by the VMC method
for $h/J = 1$ and $n = 16$.
For smaller $R$, we observe the further decrease of the energy
by optimizing the parameters in the RBM wave function using
the VMC method.
This result suggests that the initial state
obtained by the CP decomposition
can be further refined to improve the accuracy of the wave function.
With increasing $R$, the energy gain by the VMC
method gets smaller.
The initial state prepared by the CP decomposition is already
close to the ground-state wave function,
and the optimization does not change the energy significantly.

We summarize the $R$ dependence of the energy error 
in Fig.~\ref{fig:vmc_err_R_h1}.
After the further optimization by the VMC method,
we can slightly improve the accuracy of the wave function.
The energy error after the optimization still decreases monotonically as
a function of $R$.
Because of the improvement of the accuracy for smaller $R$,
the $R$ dependence of the energy error
becomes less steeper and appears to be
well approximated by the line proportional to $R^{-c}$
with $c\lesssim 2$ being a constant.

\subsection{VMC simulation under periodic boundary
conditions}
\label{subsec:vmc_periodic}

\begin{figure}[!t]
\centering
\includegraphics[width=.95\linewidth]{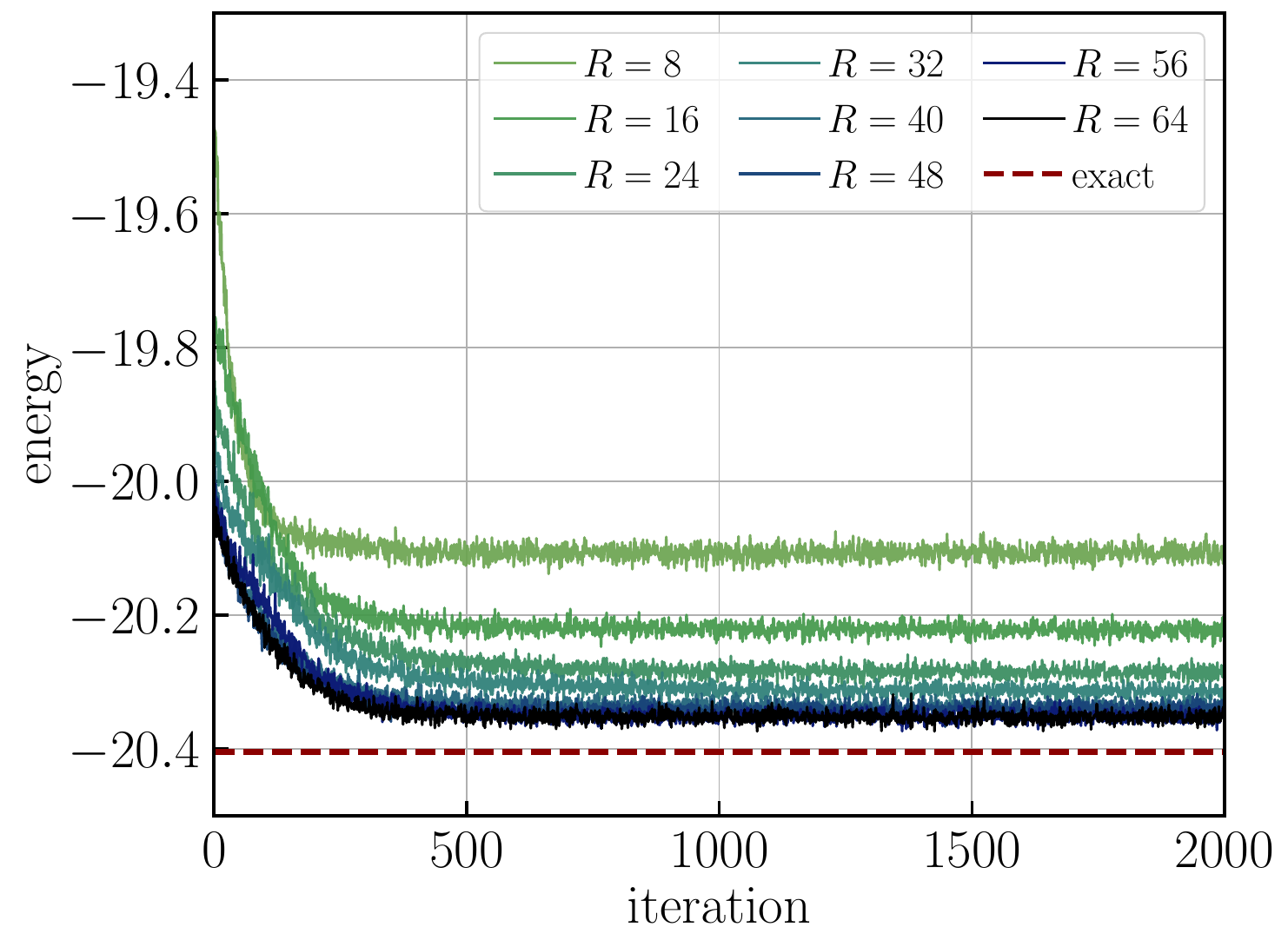}
\caption{%
VMC simulation for $h/J = 1$ and $n = 16$
under periodic boundary conditions.
Similarly to the case in Fig.~\ref{fig:vmc_opt_open},
the initial state is generated by the CP decomposition of the
ground-state MPS with the maximum bond dimension $D_{\mathrm{max}} = 8$
under open boundary conditions.
With increasing the rank $R$ of the CP decomposition,
the energy approaches the true ground-state energy
for periodic boundary conditions.
}
\label{fig:vmc_opt_periodic}
\end{figure}

In general, it is hard to prepare the MPS representation
of the ground-state wave function
under periodic boundary conditions.
However, the ground-state wave function
should be insensitive to boundary conditions
when the system is large enough.
Therefore, one may seed the wave function for periodic boundary
conditions with the initial wave function for open boundary conditions.
We demonstrate that the initial state
obtained by the CP decomposition
for the open-boundary system
can also be efficiently used for the simulation
under periodic boundary conditions.

Figure~\ref{fig:vmc_opt_periodic}
shows the optimization by the VMC method
for $h/J = 1$ and $n = 16$.
We use the same initial state
as in the case in Fig.~\ref{fig:vmc_opt_open}.
We can confirm it from the fact that
the energy at the first iteration
for periodic boundary conditions
is nearly the same as that for open boundary conditions.
For all $R$,
we observe the further decrease of the energy
by optimizing the parameters in the RBM wave function using
the VMC method.
Although the initial state is not necessarily
an accurate representation of the ground-state wave function,
it is close enough to the true ground state
in a sense that the VMC method
can efficiently optimize the parameters.
Optimized energies monotonically decrease with increasing $R$
and approach the true ground-state energy
for periodic boundary conditions.

We also examine the $R$ dependence of the energy error
in Fig.~\ref{fig:vmc_err_R_h1}.
As in the case of open boundary conditions,
the energy error after the optimization decreases monotonically as a
function of $R$.
The improvement under periodic boundary conditions
is comparable to or slightly better than
that under open boundary conditions.
The error is nearly approximated by the line proportional to $R^{-c}$
with $c\lesssim 2$ being a constant.

\subsection{Size dependence of the required rank of the CP decomposition}
\label{subsec:size_dep_rank}

\begin{figure}[!t]
\centering
\includegraphics[width=.95\linewidth]{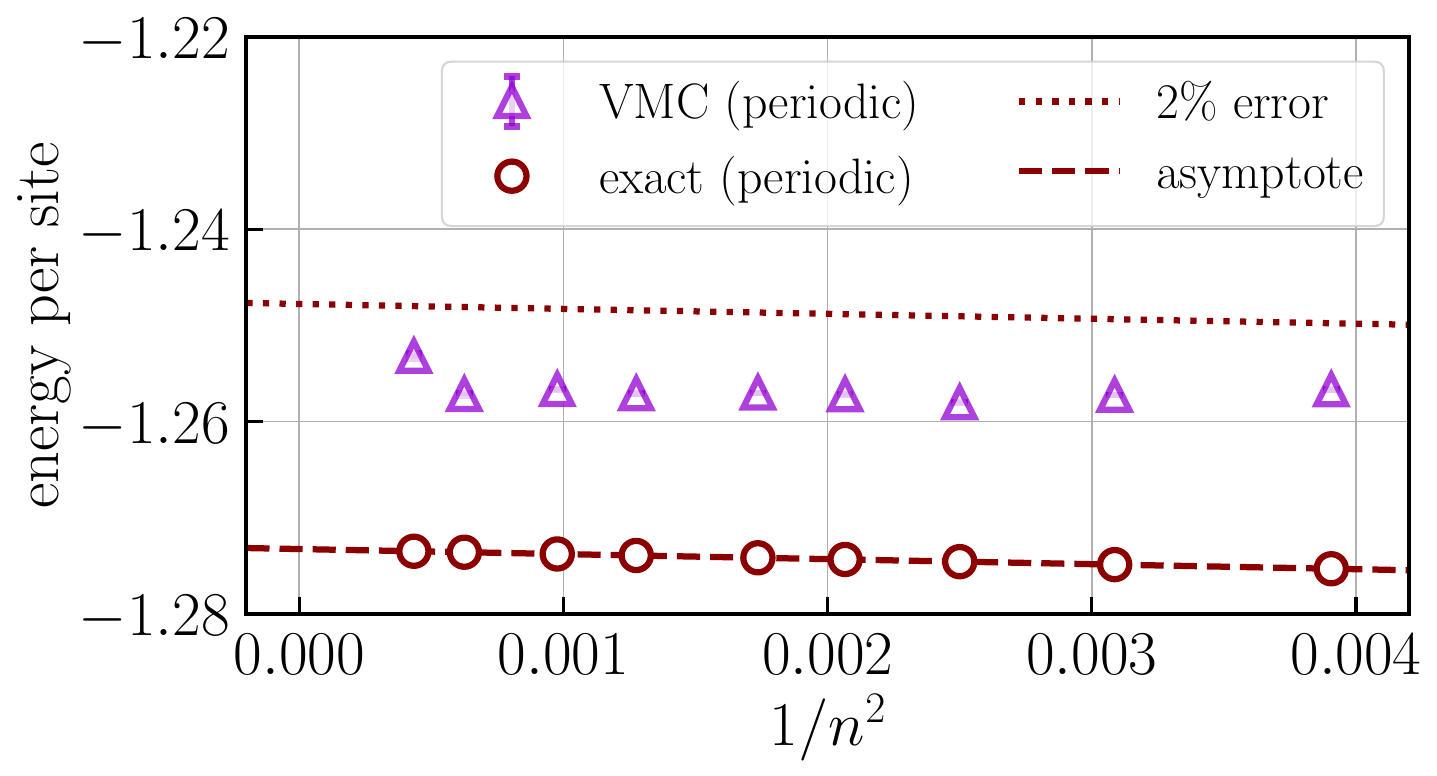}
\caption{%
Size dependence of the ground-state energy per site for $h/J = 1$
under periodic boundary conditions.
The rank $R$ of the CP decomposition,
which gives $2nR$ variational parameters
($b_{l,k}$ and $W_{l,k}$)
in the RBM wave function for the VMC method,
is chosen to be $R = n/2$ for all systems of $n$ sites.
Circles and triangles represent the exact and VMC energy per site,
respectively.
As references, we also plot the asymptotic behavior
of the energy per site by a dashed line and
that multiplied by $0.98$, corresponding to a $2\%$ error,
by a dotted line.
The energy error obtained by the VMC method
is nearly independent of the system size
and is less than $2\%$ for all systems.
}
\label{fig:ene_vs_n}
\end{figure}

\begin{figure}[!t]
\centering
\includegraphics[width=.95\linewidth]{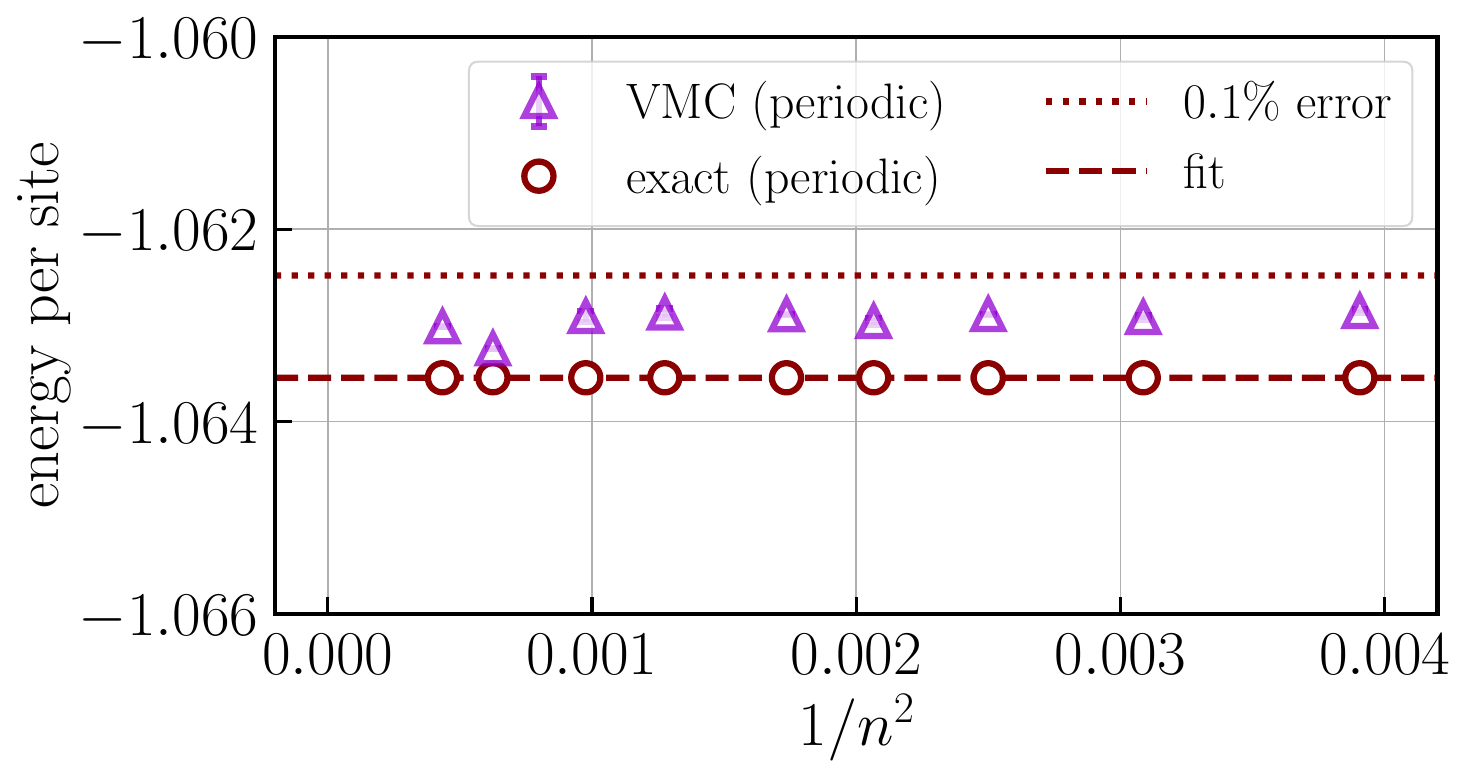}
\caption{%
Size dependence of the ground-state energy per site for $h/J = 0.5$
under periodic boundary conditions.
The rank $R$ of the CP decomposition
is chosen to be $R = n/2$ for all systems of $n$ sites.
Circles and triangles represent the exact and VMC energy per site,
respectively.
As references, we also plot 
a dashed line interpolating the exact data points
and the line multiplied by $0.999$, corresponding to a $0.1\%$ error,
by a dotted line.
}
\label{fig:ene_vs_n_g0.5}
\end{figure}

\begin{figure}[!t]
\centering
\includegraphics[width=.95\linewidth]{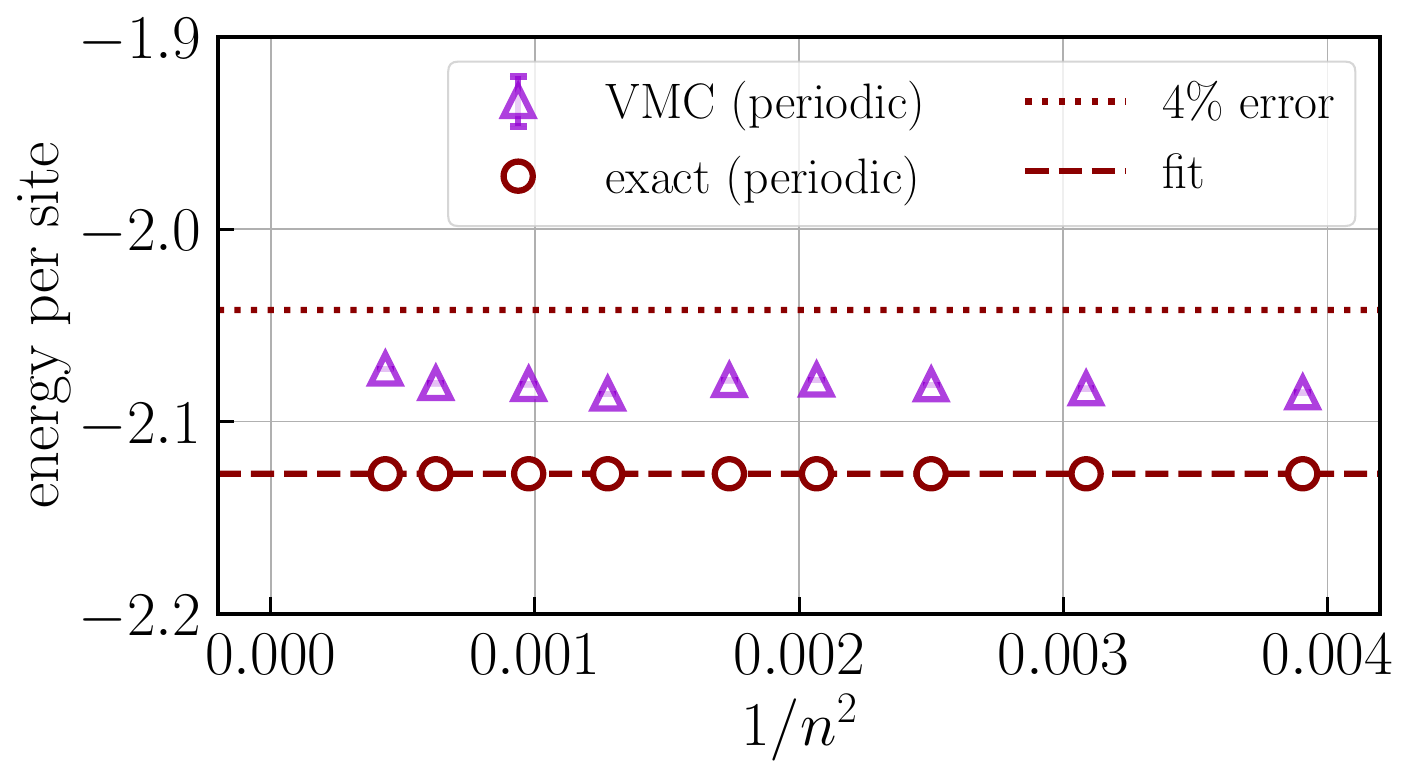}
\caption{%
Size dependence of the ground-state energy per site for $h/J = 2$
under periodic boundary conditions.
The rank $R$ of the CP decomposition
is chosen to be $R = n/2$ for all systems of $n$ sites.
Circles and triangles represent the exact and VMC energy per site,
respectively.
As references, we also plot
a dashed line interpolating the exact data points
and the line multiplied by $0.96$, corresponding to a $4\%$ error,
by a dotted line.
}
\label{fig:ene_vs_n_g2}
\end{figure}

Finally, we numerically investigate how large the rank $R$
of the CP decomposition
should be for arbitrary system sizes.
We mainly focus on a rather difficult case
at the critical point ($h/J = 1$)
of the transverse-field Ising model
under periodic boundary conditions.
The ground state is known to exhibit
a logarithmic correction in the entanglement entropy,
which naturally requires slightly larger bond dimensions
in the MPS representation
compared to bond dimensions required for
conventional gapped ground states.
As we see below,
for all systems of $n$ sites, the condition
$R = \mathcal{O}(n)$ is sufficient to obtain
the ground-state wave function
having a small energy error
that is independent of the system size.

We first prepare the MPS representation
of the ground-state wave function
with the maximum bond dimension $D_{\mathrm{max}} = n/2$
for all systems of $n$ sites
using the DMRG method.
The energy error from the true ground state
is found to be less than $10^{-9}$ in the units of $J$
for any system size that we consider.
We then CP decompose the MPS representation
using the ALS method with $100$ iterations
from the random initial CP factors.
We optimize the parameters in the RBM wave function
generated by the CP decomposed tensor
using the VMC method.
We keep the rank $R$ of the CP decomposition
to $R = D_{\mathrm{max}} (= n/2)$ for all systems of $n$ sites.
The number of variational parameters ($b_{l,k}$ and $W_{l,k}$)
is $2nR=n^2$, which is much smaller than
the dimension of the Hilbert space.
We use $10^6$ ($4\times 10^6$) Monte Carlo steps
for evaluating physical quantities
during the optimization (during the statistical processing).

Figure~\ref{fig:ene_vs_n}
shows the ground-state energy per site
for $h/J = 1$ under periodic boundary conditions.
The exact energy per site ($e_{\mathrm{exact}}$) follows the
asymptotic behavior~\cite{katsura1962,pfeuty1970,ovchinnikov2003,
bloete1986,affleck1986}, given by
\begin{align}
\label{eq:exact_ene_per_site}
 e_{\mathrm{exact}}(n) = - \frac{4}{\pi} - \frac{\pi}{6n^2}
 + \mathcal{O}(n^{-3}),
\end{align}
and the energy per site obtained by the VMC method
nearly shows the same behavior.
The difference between the exact and VMC energy per site
is nearly independent of the system size
and is always less than $2\%$.
Consequently, we numerically confirm that
the condition $R = n/2$ is sufficient to reproduce
the ground-state wave function
having a small energy error
that does not depend on the system size.

Away from the critical point ($h/J = 1$),
the energy error obtained by the VMC method
is also independent of the system size $n$
when the condition $R = \mathcal{O}(n)$ is satisfied.
We show the ground-state energy per site
with $R = n/2$ for $h/J = 0.5$ and $h/J = 2$
under periodic boundary conditions
in Figs.~\ref{fig:ene_vs_n_g0.5} and \ref{fig:ene_vs_n_g2}.
The VMC method combined with the initial wave function
generated by the CP decomposed MPSs
performs effectively regardless of whether
the system is at the critical point.

In general, at the critical point,
the size dependence of the entanglement entropy
contains a logarithmic correction,
which requires more parameters to accurately represent
the corresponding wave function during numerical simulations.
Therefore, one may expect the largest energy error
at the critical point ($h/J = 1$).
However, our numerical results suggest that
the relative energy error is slightly larger
for a parameter region which is away from the critical point.
We do not have a clear explanation for the origin of this behavior;
however, the RBM wave function does not prevent the efficient
representation of quantum states exhibiting entanglement growth that
exceeds area-law behavior.
Because of the flexible structure,
the RBM wave function does not necessarily suffer from the same
limitations as the MPS wave function at corresponding points
in the parameter space.
These results are actually encouraging,
because our approach may enable successful simulations
in parameter regions that are typically difficult to handle
with the MPS wave function.

\section{Summary and outlook}
\label{sec:summary}

We proposed a method for approximately converting
the MPS representation of the ground state
in quantum many-body systems
into the RBM wave function
consisting of a single multinomial hidden unit.
This procedure helped us prepare suitable initial states for the
VMC simulation using NNQSs.
The computational complexity of the method
for converting MPSs into RBM wave functions
using the CP decomposition
scaled polynomially with the number of variational parameters
and sites in the system.
The error of the initial wave function
typically decreases as $\sim R^{-2}$ with increasing the rank $R$
of the CP decomposition.
We examined the applicability of the method
by taking the transverse-field Ising model as an example
and found that the RBM wave function with a single hidden unit
well approximated the ground state
of the quantum many-body system.
Even when the energy of the initial state was slightly away from
that of the true ground state,
the VMC method was able to efficiently optimize the parameters
in the RBM wave function
and provided a better energy estimate after the optimization,
as we confirmed in the case of periodic boundary conditions.
This observation suggests that
there remains a high likelihood of obtaining an accurate ground state
through the VMC method
even if the DMRG simulation is inaccurate,
causing the corresponding initial state to be slightly away
from the ideal state.
We also numerically investigated the required rank $R$
of the CP decomposition for arbitrary system sizes.
We empirically found that the condition $R = \mathcal{O}(n)$
is sufficient for any $n$ sites
to obtain the ground-state wave function
having a small energy error (less than $2\%$
for $R = n/2$ at the critical point)
that does not depend on the system size.

Our approach can be extended to more general NNQSs,
such as those combining the multinomial RBM
[$\Psi_{\mathrm{mRBM}}(\{v_i\})$]
and conventional binomial RBM
[$\Psi_{\mathrm{bRBM}}(\{v_i\})$]
wave functions:
\begin{align}
 \Psi(\{v_i\})
 &=
 \Psi_{\mathrm{mRBM}}(\{v_i\})
 \Psi_{\mathrm{bRBM}}(\{v_i\}),
\\
 \Psi_{\mathrm{mRBM}}(\{v_i\})
 &=
 \sum_{j=1}^{R}
 \exp\left[
 \sum_{i=1}^{n}
 \left(
 b_{i,j} + W_{i,j} v_i
 \right)
 \right],
\\
 \Psi_{\mathrm{bRBM}}(\{v_i\})
 &=
 \exp\left(
 \sum_{i=1}^{n} a'_i v_i
 \right)
 \prod_{j=1}^{n_h}
 2\cosh\left(
 b'_j + \sum_{i=1}^{n} W'_{i,j} v_i
 \right),
\end{align}
where $b_{i,j}$, $W_{i,j}$, $a'_i$, $b'_j$, and $W'_{i,j}$
are variational parameters.
The binomial RBM wave function further improves the approximation
of the ground state in quantum many-body systems.
The conventional binomial RBM wave function
would efficiently represent quantum states that
exhibit volume-law entanglement entropy~\cite{sharir2022,deng2017a}.
While the multinomial RBM wave function
may struggle to capture sufficient
entanglement entropy needed for accurately modeling quantum states
in some unfortunate cases,
combining the binomial RBM wave function
can offer a more effective alternative.
The integrated RBMs are unlikely to significantly hinder
the variational wave function's convergence
toward the ground state when the initial state is
already close to the target state.

In strongly correlated electron systems,
the low-energy eigenstates often exhibit complex nodal
structures~\cite{ceperley1991}.
In most of the VMC simulations,
we imitate such nodal structures
by using one-body wave functions constructed from the Slater determinant
or the Pfaffian,
derived from ground states of noninteracting fermion systems
or mean-field solutions of Fermi-Hubbard
systems~\cite{leblanc2015,wu2024}.
The CP decomposition has recently been employed to enhance fermionic
trial wave functions within the VMC method~\cite{bortone2025}.
Our method would provide a complementary way
to prepare wave functions with complex nodal structures,
which will be more accurate starting points
than the one-body wave functions consisting of
the Slater determinant or the Pfaffian,
once we have accurate tensor networks representing
the low-energy eigenstates in strongly correlated electron systems.

We have thoroughly examined the performance of the proposed method
in the simplest quantum one-dimensional systems
to deepen the understanding of how one can efficiently
transform tensor network states into shallow RBM wave functions.
It is an interesting future direction
to extend our method to higher-dimensional systems.
In higher dimensions, optimized tensor network states
could be inaccurate, causing the corresponding initial RBM wave
functions to be far from the ground state.
Nevertheless, this initial energy is expected to be
much closer to the ground-state energy
than those obtained from random initializations
and those prepared by the conventional mean-field approximation.
Then, the optimization from the initial RBM wave function
prepared by our method is more likely to be successful.
Investigating the performance of our method
in general higher-dimensional systems requires
much more careful analyses and will be left for future work.

In the present study, we mainly focused on wave functions
that explicitly break the lattice translational symmetry.
In translational invariant systems,
we may use the translational invariant RBM wave function
that has a much smaller number of variational parameters.
Such wave functions should be generated
using the symmetric CP decomposition~\cite{sherman2020}
of translational invariant tensor networks,
including the infinite MPS (iMPS)~\cite{vidal2007}
in one-dimensional systems
and the infinite PEPS (iPEPS)~\cite{nishino2001,verstraete2004a_arxiv,
verstraete2008,corboz2011,orus2014,orus2019}
in two-dimensional systems.
One may apply the ALS method to iMPS or iPEPS to obtain
translationally invariant factors for the CP decomposed tensor networks.
It will be interesting to see to what extent
the CP decomposition of iMPS or iPEPS helps us prepare suitable initial states
for the translational invariant RBM wave function.
However, this is beyond the scope of this paper and is left for future work.

Note that,
even for systems that break the translational symmetry,
tensor network states in higher dimensions possess
structural and computational characteristics
that may differ significantly from the present shallow RBM wave function
and the 1D tensor network state.
Therefore,
generalizing our method to higher dimensions is not straightforward,
and one may develop a more specific method
for converting PEPSs into RBM wave functions.
Moreover, the optimization of PEPS itself remains a challenging and active
research field.
Addressing these challenges is an interesting future direction.

Finally, we would like to point out that
the CP decomposition of MPSs
and tensor trains~\cite{baxter1968,white1992,
vidal2007,
schollwock2011,oseledets2011,
hauschild2018,hauschild2024,fishman2022,
verstraete2008,orus2014}
is a general technique that can be applied to
other research fields, such as information science and
engineering~\cite{wang2023_arxiv,hamreras2025_arxiv}.
One may consider the Monte Carlo approach
to obtain the CP decomposition of MPSs or tensor trains
using the RBM-type neural network.
Indeed, extended Boltzmann machine representations
of the CP decomposition have been proposed for
non-negative tensors, formulated as a convex optimization problem
based on information geometry
recently~\cite{sugiyama2019,pang2020_arxiv}.
Such approach is also an interesting future direction.

\begin{acknowledgments}
This work was financially supported by
MEXT KAKENHI, Grant-in-Aid for Transformative Research Area
(Grants No.\ JP22H05111 and No.\ JP22H05114).
R.K.\ was supported by
JSPS KAKENHI
(Grants No.\ JP24H00973 and No.\ JP25K07157).
S.G.\ was supported by
JSPS KAKENHI
(Grant No.\ JP23K13042).
The numerical computations were performed on computers at
the Yukawa Institute Computer Facility,
Kyoto University
and on computers at
the Supercomputer Center, the Institute for Solid State Physics,
the University of Tokyo.
\end{acknowledgments}

\section*{Data availability}
The data that support the findings of this article are openly
available~\cite{data_availability}.

\onecolumngrid


\begin{thebibliography}{117}%
\makeatletter
\providecommand \@ifxundefined [1]{%
 \@ifx{#1\undefined}
}%
\providecommand \@ifnum [1]{%
 \ifnum #1\expandafter \@firstoftwo
 \else \expandafter \@secondoftwo
 \fi
}%
\providecommand \@ifx [1]{%
 \ifx #1\expandafter \@firstoftwo
 \else \expandafter \@secondoftwo
 \fi
}%
\providecommand \natexlab [1]{#1}%
\providecommand \enquote  [1]{``#1''}%
\providecommand \bibnamefont  [1]{#1}%
\providecommand \bibfnamefont [1]{#1}%
\providecommand \citenamefont [1]{#1}%
\providecommand \href@noop [0]{\@secondoftwo}%
\providecommand \href [0]{\begingroup \@sanitize@url \@href}%
\providecommand \@href[1]{\@@startlink{#1}\@@href}%
\providecommand \@@href[1]{\endgroup#1\@@endlink}%
\providecommand \@sanitize@url [0]{\catcode `\\12\catcode `\$12\catcode
  `\&12\catcode `\#12\catcode `\^12\catcode `\_12\catcode `\%12\relax}%
\providecommand \@@startlink[1]{}%
\providecommand \@@endlink[0]{}%
\providecommand \url  [0]{\begingroup\@sanitize@url \@url }%
\providecommand \@url [1]{\endgroup\@href {#1}{\urlprefix }}%
\providecommand \urlprefix  [0]{URL }%
\providecommand \Eprint [0]{\href }%
\providecommand \doibase [0]{https://doi.org/}%
\providecommand \selectlanguage [0]{\@gobble}%
\providecommand \bibinfo  [0]{\@secondoftwo}%
\providecommand \bibfield  [0]{\@secondoftwo}%
\providecommand \translation [1]{[#1]}%
\providecommand \BibitemOpen [0]{}%
\providecommand \bibitemStop [0]{}%
\providecommand \bibitemNoStop [0]{.\EOS\space}%
\providecommand \EOS [0]{\spacefactor3000\relax}%
\providecommand \BibitemShut  [1]{\csname bibitem#1\endcsname}%
\let\auto@bib@innerbib\@empty
\bibitem [{\citenamefont {Baxter}(1968)}]{baxter1968}%
  \BibitemOpen
  \bibfield  {author} {\bibinfo {author} {\bibfnamefont {R.~J.}\ \bibnamefont
  {Baxter}},\ }\bibfield  {title} {\bibinfo {title} {\emph{Dimers on a
  Rectangular Lattice}},\ }\href {https://doi.org/10.1063/1.1664623} {\bibfield
   {journal} {\bibinfo  {journal} {J. Math. Phys.}\ }\textbf {\bibinfo {volume}
  {9}},\ \bibinfo {pages} {650} (\bibinfo {year} {1968})}\BibitemShut {NoStop}%
\bibitem [{\citenamefont {White}(1992)}]{white1992}%
  \BibitemOpen
  \bibfield  {author} {\bibinfo {author} {\bibfnamefont {S.~R.}\ \bibnamefont
  {White}},\ }\bibfield  {title} {\bibinfo {title} {\emph{Density matrix
  formulation for quantum renormalization groups}},\ }\href
  {https://doi.org/10.1103/PhysRevLett.69.2863} {\bibfield  {journal} {\bibinfo
   {journal} {Phys. Rev. Lett.}\ }\textbf {\bibinfo {volume} {69}},\ \bibinfo
  {pages} {2863} (\bibinfo {year} {1992})}\BibitemShut {NoStop}%
\bibitem [{\citenamefont {Vidal}(2007)}]{vidal2007}%
  \BibitemOpen
  \bibfield  {author} {\bibinfo {author} {\bibfnamefont {G.}~\bibnamefont
  {Vidal}},\ }\bibfield  {title} {\bibinfo {title} {\emph{Classical Simulation
  of Infinite-Size Quantum Lattice Systems in One Spatial Dimension}},\ }\href
  {https://doi.org/10.1103/PhysRevLett.98.070201} {\bibfield  {journal}
  {\bibinfo  {journal} {Phys. Rev. Lett.}\ }\textbf {\bibinfo {volume} {98}},\
  \bibinfo {pages} {070201} (\bibinfo {year} {2007})}\BibitemShut {NoStop}%
\bibitem [{\citenamefont {Schollw{\"o}ck}(2011)}]{schollwock2011}%
  \BibitemOpen
  \bibfield  {author} {\bibinfo {author} {\bibfnamefont {U.}~\bibnamefont
  {Schollw{\"o}ck}},\ }\bibfield  {title} {\bibinfo {title} {\emph{The
  density-matrix renormalization group in the age of matrix product states}},\
  }\href {https://doi.org/10.1016/j.aop.2010.09.012} {\bibfield  {journal}
  {\bibinfo  {journal} {Ann. Phys.}\ }\textbf {\bibinfo {volume} {326}},\
  \bibinfo {pages} {96} (\bibinfo {year} {2011})}\BibitemShut {NoStop}%
\bibitem [{\citenamefont {Oseledets}(2011)}]{oseledets2011}%
  \BibitemOpen
  \bibfield  {author} {\bibinfo {author} {\bibfnamefont {I.~V.}\ \bibnamefont
  {Oseledets}},\ }\bibfield  {title} {\bibinfo {title} {\emph{Tensor-Train
  Decomposition}},\ }\href {https://doi.org/10.1137/090752286} {\bibfield
  {journal} {\bibinfo  {journal} {SIAM J. Sci. Comput.}\ }\textbf {\bibinfo
  {volume} {33}},\ \bibinfo {pages} {2295} (\bibinfo {year}
  {2011})}\BibitemShut {NoStop}%
\bibitem [{\citenamefont {Hauschild}\ and\ \citenamefont
  {Pollmann}(2018)}]{hauschild2018}%
  \BibitemOpen
  \bibfield  {author} {\bibinfo {author} {\bibfnamefont {J.}~\bibnamefont
  {Hauschild}}\ and\ \bibinfo {author} {\bibfnamefont {F.}~\bibnamefont
  {Pollmann}},\ }\bibfield  {title} {\bibinfo {title} {\emph{{Efficient
  numerical simulations with Tensor Networks: Tensor Network Python
  (TeNPy)}}},\ }\href {https://doi.org/10.21468/SciPostPhysLectNotes.5}
  {\bibfield  {journal} {\bibinfo  {journal} {SciPost Phys. Lect. Notes}\ ,\
  \bibinfo {pages} {5}} (\bibinfo {year} {2018})}\BibitemShut {NoStop}%
\bibitem [{\citenamefont {Hauschild}\ \emph {et~al.}(2024)\citenamefont
  {Hauschild}, \citenamefont {Unfried}, \citenamefont {Anand}, \citenamefont
  {Andrews}, \citenamefont {Bintz}, \citenamefont {Borla}, \citenamefont
  {Divic}, \citenamefont {Drescher}, \citenamefont {Geiger}, \citenamefont
  {Hefel}, \citenamefont {H\'{e}mery}, \citenamefont {Kadow}, \citenamefont
  {Kemp}, \citenamefont {Kirchner}, \citenamefont {Liu}, \citenamefont
  {M\"{o}ller}, \citenamefont {Parker}, \citenamefont {Rader}, \citenamefont
  {Romen}, \citenamefont {Scalet}, \citenamefont {Schoonderwoerd},
  \citenamefont {Schulz}, \citenamefont {Soejima}, \citenamefont {Thoma},
  \citenamefont {Wu}, \citenamefont {Zechmann}, \citenamefont {Zweng},
  \citenamefont {Mong}, \citenamefont {Zaletel},\ and\ \citenamefont
  {Pollmann}}]{hauschild2024}%
  \BibitemOpen
  \bibfield  {author} {\bibinfo {author} {\bibfnamefont {J.}~\bibnamefont
  {Hauschild}}, \bibinfo {author} {\bibfnamefont {J.}~\bibnamefont {Unfried}},
  \bibinfo {author} {\bibfnamefont {S.}~\bibnamefont {Anand}}, \bibinfo
  {author} {\bibfnamefont {B.}~\bibnamefont {Andrews}}, \bibinfo {author}
  {\bibfnamefont {M.}~\bibnamefont {Bintz}}, \bibinfo {author} {\bibfnamefont
  {U.}~\bibnamefont {Borla}}, \bibinfo {author} {\bibfnamefont
  {S.}~\bibnamefont {Divic}}, \bibinfo {author} {\bibfnamefont
  {M.}~\bibnamefont {Drescher}}, \bibinfo {author} {\bibfnamefont
  {J.}~\bibnamefont {Geiger}}, \bibinfo {author} {\bibfnamefont
  {M.}~\bibnamefont {Hefel}}, \bibinfo {author} {\bibfnamefont
  {K.}~\bibnamefont {H\'{e}mery}}, \bibinfo {author} {\bibfnamefont
  {W.}~\bibnamefont {Kadow}}, \bibinfo {author} {\bibfnamefont
  {J.}~\bibnamefont {Kemp}}, \bibinfo {author} {\bibfnamefont {N.}~\bibnamefont
  {Kirchner}}, \bibinfo {author} {\bibfnamefont {V.~S.}\ \bibnamefont {Liu}},
  \bibinfo {author} {\bibfnamefont {G.}~\bibnamefont {M\"{o}ller}}, \bibinfo
  {author} {\bibfnamefont {D.}~\bibnamefont {Parker}}, \bibinfo {author}
  {\bibfnamefont {M.}~\bibnamefont {Rader}}, \bibinfo {author} {\bibfnamefont
  {A.}~\bibnamefont {Romen}}, \bibinfo {author} {\bibfnamefont
  {S.}~\bibnamefont {Scalet}}, \bibinfo {author} {\bibfnamefont
  {L.}~\bibnamefont {Schoonderwoerd}}, \bibinfo {author} {\bibfnamefont
  {M.}~\bibnamefont {Schulz}}, \bibinfo {author} {\bibfnamefont
  {T.}~\bibnamefont {Soejima}}, \bibinfo {author} {\bibfnamefont
  {P.}~\bibnamefont {Thoma}}, \bibinfo {author} {\bibfnamefont
  {Y.}~\bibnamefont {Wu}}, \bibinfo {author} {\bibfnamefont {P.}~\bibnamefont
  {Zechmann}}, \bibinfo {author} {\bibfnamefont {L.}~\bibnamefont {Zweng}},
  \bibinfo {author} {\bibfnamefont {R.~S.~K.}\ \bibnamefont {Mong}}, \bibinfo
  {author} {\bibfnamefont {M.~P.}\ \bibnamefont {Zaletel}},\ and\ \bibinfo
  {author} {\bibfnamefont {F.}~\bibnamefont {Pollmann}},\ }\bibfield  {title}
  {\bibinfo {title} {\emph{{Tensor network Python (TeNPy) version 1}}},\ }\href
  {https://doi.org/10.21468/SciPostPhysCodeb.41} {\bibfield  {journal}
  {\bibinfo  {journal} {SciPost Phys. Codebases}\ ,\ \bibinfo {pages} {41}}
  (\bibinfo {year} {2024})}\BibitemShut {NoStop}%
\bibitem [{\citenamefont {Fishman}\ \emph {et~al.}(2022)\citenamefont
  {Fishman}, \citenamefont {White},\ and\ \citenamefont
  {Stoudenmire}}]{fishman2022}%
  \BibitemOpen
  \bibfield  {author} {\bibinfo {author} {\bibfnamefont {M.}~\bibnamefont
  {Fishman}}, \bibinfo {author} {\bibfnamefont {S.~R.}\ \bibnamefont {White}},\
  and\ \bibinfo {author} {\bibfnamefont {E.~M.}\ \bibnamefont {Stoudenmire}},\
  }\bibfield  {title} {\bibinfo {title} {\emph{{The ITensor Software Library
  for Tensor Network Calculations}}},\ }\href
  {https://doi.org/10.21468/SciPostPhysCodeb.4} {\bibfield  {journal} {\bibinfo
   {journal} {SciPost Phys. Codebases}\ ,\ \bibinfo {pages} {4}} (\bibinfo
  {year} {2022})}\BibitemShut {NoStop}%
\bibitem [{\citenamefont {Verstraete}\ \emph {et~al.}(2008)\citenamefont
  {Verstraete}, \citenamefont {Murg},\ and\ \citenamefont
  {Cirac}}]{verstraete2008}%
  \BibitemOpen
  \bibfield  {author} {\bibinfo {author} {\bibfnamefont {F.}~\bibnamefont
  {Verstraete}}, \bibinfo {author} {\bibfnamefont {V.}~\bibnamefont {Murg}},\
  and\ \bibinfo {author} {\bibfnamefont {J.~I.}\ \bibnamefont {Cirac}},\
  }\bibfield  {title} {\bibinfo {title} {\emph{{Matrix product states,
  projected entangled pair states, and variational renormalization group
  methods for quantum spin systems}}},\ }\href
  {https://doi.org/10.1080/14789940801912366} {\bibfield  {journal} {\bibinfo
  {journal} {Adv. Phys.}\ }\textbf {\bibinfo {volume} {57}},\ \bibinfo {pages}
  {143} (\bibinfo {year} {2008})}\BibitemShut {NoStop}%
\bibitem [{\citenamefont {Or{\'u}s}(2014)}]{orus2014}%
  \BibitemOpen
  \bibfield  {author} {\bibinfo {author} {\bibfnamefont {R.}~\bibnamefont
  {Or{\'u}s}},\ }\bibfield  {title} {\bibinfo {title} {\emph{{A practical
  introduction to tensor networks: Matrix product states and projected
  entangled pair states}}},\ }\href {https://doi.org/10.1016/j.aop.2014.06.013}
  {\bibfield  {journal} {\bibinfo  {journal} {Ann. Phys.}\ }\textbf {\bibinfo
  {volume} {349}},\ \bibinfo {pages} {117} (\bibinfo {year}
  {2014})}\BibitemShut {NoStop}%
\bibitem [{\citenamefont {Nishino}\ \emph {et~al.}(2001)\citenamefont
  {Nishino}, \citenamefont {Hieida}, \citenamefont {Okunishi}, \citenamefont
  {Maeshima}, \citenamefont {Akutsu},\ and\ \citenamefont
  {Gendiar}}]{nishino2001}%
  \BibitemOpen
  \bibfield  {author} {\bibinfo {author} {\bibfnamefont {T.}~\bibnamefont
  {Nishino}}, \bibinfo {author} {\bibfnamefont {Y.}~\bibnamefont {Hieida}},
  \bibinfo {author} {\bibfnamefont {K.}~\bibnamefont {Okunishi}}, \bibinfo
  {author} {\bibfnamefont {N.}~\bibnamefont {Maeshima}}, \bibinfo {author}
  {\bibfnamefont {Y.}~\bibnamefont {Akutsu}},\ and\ \bibinfo {author}
  {\bibfnamefont {A.}~\bibnamefont {Gendiar}},\ }\bibfield  {title} {\bibinfo
  {title} {\emph{{Two-dimensional tensor product variational formulation}}},\
  }\href {https://doi.org/10.1143/PTP.105.409} {\bibfield  {journal} {\bibinfo
  {journal} {Prog. Theor. Phys.}\ }\textbf {\bibinfo {volume} {105}},\ \bibinfo
  {pages} {409} (\bibinfo {year} {2001})}\BibitemShut {NoStop}%
\bibitem [{\citenamefont {Verstraete}\ and\ \citenamefont
  {Cirac}()}]{verstraete2004a_arxiv}%
  \BibitemOpen
  \bibfield  {author} {\bibinfo {author} {\bibfnamefont {F.}~\bibnamefont
  {Verstraete}}\ and\ \bibinfo {author} {\bibfnamefont {J.~I.}\ \bibnamefont
  {Cirac}},\ }\href@noop {} {\bibinfo {title} {\emph{{Renormalization
  algorithms for quantum-many body systems in two and higher dimensions}}}},\
  \Eprint {https://arxiv.org/abs/cond-mat/0407066} {arXiv:cond-mat/0407066}
  \BibitemShut {NoStop}%
\bibitem [{\citenamefont {Corboz}\ \emph {et~al.}(2011)\citenamefont {Corboz},
  \citenamefont {White}, \citenamefont {Vidal},\ and\ \citenamefont
  {Troyer}}]{corboz2011}%
  \BibitemOpen
  \bibfield  {author} {\bibinfo {author} {\bibfnamefont {P.}~\bibnamefont
  {Corboz}}, \bibinfo {author} {\bibfnamefont {S.~R.}\ \bibnamefont {White}},
  \bibinfo {author} {\bibfnamefont {G.}~\bibnamefont {Vidal}},\ and\ \bibinfo
  {author} {\bibfnamefont {M.}~\bibnamefont {Troyer}},\ }\bibfield  {title}
  {\bibinfo {title} {\emph{Stripes in the two-dimensional $t$-$J$ model with
  infinite projected entangled-pair states}},\ }\href
  {https://doi.org/10.1103/PhysRevB.84.041108} {\bibfield  {journal} {\bibinfo
  {journal} {Phys. Rev. B}\ }\textbf {\bibinfo {volume} {84}},\ \bibinfo
  {pages} {041108} (\bibinfo {year} {2011})}\BibitemShut {NoStop}%
\bibitem [{\citenamefont {Or{\'u}s}(2019)}]{orus2019}%
  \BibitemOpen
  \bibfield  {author} {\bibinfo {author} {\bibfnamefont {R.}~\bibnamefont
  {Or{\'u}s}},\ }\bibfield  {title} {\bibinfo {title} {\emph{{Tensor networks
  for complex quantum systems}}},\ }\href
  {https://doi.org/10.1038/s42254-019-0086-7} {\bibfield  {journal} {\bibinfo
  {journal} {Nat. Rev. Phys.}\ }\textbf {\bibinfo {volume} {1}},\ \bibinfo
  {pages} {538} (\bibinfo {year} {2019})}\BibitemShut {NoStop}%
\bibitem [{\citenamefont {Fannes}\ \emph {et~al.}(1992)\citenamefont {Fannes},
  \citenamefont {Nachtergaele},\ and\ \citenamefont {Werner}}]{fannes1992}%
  \BibitemOpen
  \bibfield  {author} {\bibinfo {author} {\bibfnamefont {M.}~\bibnamefont
  {Fannes}}, \bibinfo {author} {\bibfnamefont {B.}~\bibnamefont
  {Nachtergaele}},\ and\ \bibinfo {author} {\bibfnamefont {R.~F.}\ \bibnamefont
  {Werner}},\ }\bibfield  {title} {\bibinfo {title} {\emph{Ground states of VBS
  models on Cayley trees}},\ }\href
  {https://doi.org/https://doi.org/10.1007/BF01055710} {\bibfield  {journal}
  {\bibinfo  {journal} {J. Stat. Phys}\ }\textbf {\bibinfo {volume} {66}},\
  \bibinfo {pages} {939} (\bibinfo {year} {1992})}\BibitemShut {NoStop}%
\bibitem [{\citenamefont {Shi}\ \emph {et~al.}(2006)\citenamefont {Shi},
  \citenamefont {Duan},\ and\ \citenamefont {Vidal}}]{shi2006}%
  \BibitemOpen
  \bibfield  {author} {\bibinfo {author} {\bibfnamefont {Y.-Y.}\ \bibnamefont
  {Shi}}, \bibinfo {author} {\bibfnamefont {L.-M.}\ \bibnamefont {Duan}},\ and\
  \bibinfo {author} {\bibfnamefont {G.}~\bibnamefont {Vidal}},\ }\bibfield
  {title} {\bibinfo {title} {\emph{Classical simulation of quantum many-body
  systems with a tree tensor network}},\ }\href
  {https://doi.org/10.1103/PhysRevA.74.022320} {\bibfield  {journal} {\bibinfo
  {journal} {Phys. Rev. A}\ }\textbf {\bibinfo {volume} {74}},\ \bibinfo
  {pages} {022320} (\bibinfo {year} {2006})}\BibitemShut {NoStop}%
\bibitem [{\citenamefont {Murg}\ \emph {et~al.}(2010)\citenamefont {Murg},
  \citenamefont {Verstraete}, \citenamefont {Legeza},\ and\ \citenamefont
  {Noack}}]{murg2010}%
  \BibitemOpen
  \bibfield  {author} {\bibinfo {author} {\bibfnamefont {V.}~\bibnamefont
  {Murg}}, \bibinfo {author} {\bibfnamefont {F.}~\bibnamefont {Verstraete}},
  \bibinfo {author} {\bibfnamefont {O.}~\bibnamefont {Legeza}},\ and\ \bibinfo
  {author} {\bibfnamefont {R.~M.}\ \bibnamefont {Noack}},\ }\bibfield  {title}
  {\bibinfo {title} {\emph{Simulating strongly correlated quantum systems with
  tree tensor networks}},\ }\href {https://doi.org/10.1103/PhysRevB.82.205105}
  {\bibfield  {journal} {\bibinfo  {journal} {Phys. Rev. B}\ }\textbf {\bibinfo
  {volume} {82}},\ \bibinfo {pages} {205105} (\bibinfo {year}
  {2010})}\BibitemShut {NoStop}%
\bibitem [{\citenamefont {Nakatani}\ and\ \citenamefont
  {Chan}(2013)}]{nakatani2013}%
  \BibitemOpen
  \bibfield  {author} {\bibinfo {author} {\bibfnamefont {N.}~\bibnamefont
  {Nakatani}}\ and\ \bibinfo {author} {\bibfnamefont {G.~K.-L.}\ \bibnamefont
  {Chan}},\ }\bibfield  {title} {\bibinfo {title} {\emph{Efficient tree tensor
  network states (TTNS) for quantum chemistry: Generalizations of the density
  matrix renormalization group algorithm}},\ }\href
  {https://doi.org/10.1063/1.4798639} {\bibfield  {journal} {\bibinfo
  {journal} {J. Chem. Phys.}\ }\textbf {\bibinfo {volume} {138}},\ \bibinfo
  {pages} {134113} (\bibinfo {year} {2013})}\BibitemShut {NoStop}%
\bibitem [{\citenamefont {Carleo}\ and\ \citenamefont
  {Troyer}(2017)}]{carleo2017}%
  \BibitemOpen
  \bibfield  {author} {\bibinfo {author} {\bibfnamefont {G.}~\bibnamefont
  {Carleo}}\ and\ \bibinfo {author} {\bibfnamefont {M.}~\bibnamefont
  {Troyer}},\ }\bibfield  {title} {\bibinfo {title} {\emph{Solving the quantum
  many-body problem with artificial neural networks}},\ }\href
  {https://doi.org/10.1126/science.aag2302} {\bibfield  {journal} {\bibinfo
  {journal} {Science}\ }\textbf {\bibinfo {volume} {355}},\ \bibinfo {pages}
  {602} (\bibinfo {year} {2017})}\BibitemShut {NoStop}%
\bibitem [{\citenamefont {Jeswal}\ and\ \citenamefont
  {Chakraverty}(2018)}]{jeswal2018}%
  \BibitemOpen
  \bibfield  {author} {\bibinfo {author} {\bibfnamefont {S.~K.}\ \bibnamefont
  {Jeswal}}\ and\ \bibinfo {author} {\bibfnamefont {S.}~\bibnamefont
  {Chakraverty}},\ }\bibfield  {title} {\bibinfo {title} {\emph{Recent
  Developments and Applications in Quantum Neural Network: A Review}},\ }\href
  {https://doi.org/10.1007/s11831-018-9269-0} {\bibfield  {journal} {\bibinfo
  {journal} {Arch. Comput. Method. Eng.}\ }\textbf {\bibinfo {volume} {26}},\
  \bibinfo {pages} {793} (\bibinfo {year} {2018})}\BibitemShut {NoStop}%
\bibitem [{\citenamefont {Jia}\ \emph {et~al.}(2019)\citenamefont {Jia},
  \citenamefont {Yi}, \citenamefont {Zhai}, \citenamefont {Wu}, \citenamefont
  {Guo},\ and\ \citenamefont {Guo}}]{jia2019}%
  \BibitemOpen
  \bibfield  {author} {\bibinfo {author} {\bibfnamefont {Z.-A.}\ \bibnamefont
  {Jia}}, \bibinfo {author} {\bibfnamefont {B.}~\bibnamefont {Yi}}, \bibinfo
  {author} {\bibfnamefont {R.}~\bibnamefont {Zhai}}, \bibinfo {author}
  {\bibfnamefont {Y.-C.}\ \bibnamefont {Wu}}, \bibinfo {author} {\bibfnamefont
  {G.-C.}\ \bibnamefont {Guo}},\ and\ \bibinfo {author} {\bibfnamefont {G.-P.}\
  \bibnamefont {Guo}},\ }\bibfield  {title} {\bibinfo {title} {\emph{Quantum
  Neural Network States: A Brief Review of Methods and Applications}},\ }\href
  {https://doi.org/10.1002/qute.201800077} {\bibfield  {journal} {\bibinfo
  {journal} {Adv. Quantum Technol.}\ }\textbf {\bibinfo {volume} {2}},\
  \bibinfo {pages} {1800077} (\bibinfo {year} {2019})}\BibitemShut {NoStop}%
\bibitem [{\citenamefont {Carleo}\ \emph {et~al.}(2019)\citenamefont {Carleo},
  \citenamefont {Cirac}, \citenamefont {Cranmer}, \citenamefont {Daudet},
  \citenamefont {Schuld}, \citenamefont {Tishby}, \citenamefont
  {Vogt-Maranto},\ and\ \citenamefont {Zdeborov\'a}}]{carleo2019}%
  \BibitemOpen
  \bibfield  {author} {\bibinfo {author} {\bibfnamefont {G.}~\bibnamefont
  {Carleo}}, \bibinfo {author} {\bibfnamefont {I.}~\bibnamefont {Cirac}},
  \bibinfo {author} {\bibfnamefont {K.}~\bibnamefont {Cranmer}}, \bibinfo
  {author} {\bibfnamefont {L.}~\bibnamefont {Daudet}}, \bibinfo {author}
  {\bibfnamefont {M.}~\bibnamefont {Schuld}}, \bibinfo {author} {\bibfnamefont
  {N.}~\bibnamefont {Tishby}}, \bibinfo {author} {\bibfnamefont
  {L.}~\bibnamefont {Vogt-Maranto}},\ and\ \bibinfo {author} {\bibfnamefont
  {L.}~\bibnamefont {Zdeborov\'a}},\ }\bibfield  {title} {\bibinfo {title}
  {\emph{Machine learning and the physical sciences}},\ }\href
  {https://doi.org/10.1103/RevModPhys.91.045002} {\bibfield  {journal}
  {\bibinfo  {journal} {Rev. Mod. Phys.}\ }\textbf {\bibinfo {volume} {91}},\
  \bibinfo {pages} {045002} (\bibinfo {year} {2019})}\BibitemShut {NoStop}%
\bibitem [{\citenamefont {Carrasquilla}\ and\ \citenamefont
  {Torlai}(2021)}]{carrasquilla2021}%
  \BibitemOpen
  \bibfield  {author} {\bibinfo {author} {\bibfnamefont {J.}~\bibnamefont
  {Carrasquilla}}\ and\ \bibinfo {author} {\bibfnamefont {G.}~\bibnamefont
  {Torlai}},\ }\bibfield  {title} {\bibinfo {title} {\emph{How To Use Neural
  Networks To Investigate Quantum Many-Body Physics}},\ }\href
  {https://doi.org/10.1103/PRXQuantum.2.040201} {\bibfield  {journal} {\bibinfo
   {journal} {PRX Quantum}\ }\textbf {\bibinfo {volume} {2}},\ \bibinfo {pages}
  {040201} (\bibinfo {year} {2021})}\BibitemShut {NoStop}%
\bibitem [{\citenamefont {Medvidovi\'{c}}\ and\ \citenamefont
  {Moreno}(2024)}]{medvidovi2024}%
  \BibitemOpen
  \bibfield  {author} {\bibinfo {author} {\bibfnamefont {M.}~\bibnamefont
  {Medvidovi\'{c}}}\ and\ \bibinfo {author} {\bibfnamefont {J.~R.}\
  \bibnamefont {Moreno}},\ }\bibfield  {title} {\bibinfo {title}
  {\emph{Neural-network quantum states for many-body physics}},\ }\href
  {https://doi.org/10.1140/epjp/s13360-024-05311-y} {\bibfield  {journal}
  {\bibinfo  {journal} {Eur. Phys. J. Plus}\ }\textbf {\bibinfo {volume}
  {139}},\ \bibinfo {pages} {631} (\bibinfo {year} {2024})}\BibitemShut
  {NoStop}%
\bibitem [{\citenamefont {Lange}\ \emph {et~al.}(2024)\citenamefont {Lange},
  \citenamefont {Van~de Walle}, \citenamefont {Abedinnia},\ and\ \citenamefont
  {Bohrdt}}]{lange2024}%
  \BibitemOpen
  \bibfield  {author} {\bibinfo {author} {\bibfnamefont {H.}~\bibnamefont
  {Lange}}, \bibinfo {author} {\bibfnamefont {A.}~\bibnamefont {Van~de Walle}},
  \bibinfo {author} {\bibfnamefont {A.}~\bibnamefont {Abedinnia}},\ and\
  \bibinfo {author} {\bibfnamefont {A.}~\bibnamefont {Bohrdt}},\ }\bibfield
  {title} {\bibinfo {title} {\emph{From architectures to applications: a review
  of neural quantum states}},\ }\href
  {https://doi.org/10.1088/2058-9565/ad7168} {\bibfield  {journal} {\bibinfo
  {journal} {Quantum Sci. Technol.}\ }\textbf {\bibinfo {volume} {9}},\
  \bibinfo {pages} {040501} (\bibinfo {year} {2024})}\BibitemShut {NoStop}%
\bibitem [{\citenamefont {Nomura}\ and\ \citenamefont
  {Imada}(2025)}]{nomura2025}%
  \BibitemOpen
  \bibfield  {author} {\bibinfo {author} {\bibfnamefont {Y.}~\bibnamefont
  {Nomura}}\ and\ \bibinfo {author} {\bibfnamefont {M.}~\bibnamefont {Imada}},\
  }\bibfield  {title} {\bibinfo {title} {\emph{Quantum Many-Body Solver Using
  Artificial Neural Networks and its Applications to Strongly Correlated
  Electron Systems}},\ }\href {https://doi.org/10.7566/jpsj.94.031001}
  {\bibfield  {journal} {\bibinfo  {journal} {J. Phys. Soc. Jpn.}\ }\textbf
  {\bibinfo {volume} {94}},\ \bibinfo {pages} {031001} (\bibinfo {year}
  {2025})}\BibitemShut {NoStop}%
\bibitem [{\citenamefont {Nomura}\ \emph {et~al.}(2017)\citenamefont {Nomura},
  \citenamefont {Darmawan}, \citenamefont {Yamaji},\ and\ \citenamefont
  {Imada}}]{nomura2017}%
  \BibitemOpen
  \bibfield  {author} {\bibinfo {author} {\bibfnamefont {Y.}~\bibnamefont
  {Nomura}}, \bibinfo {author} {\bibfnamefont {A.~S.}\ \bibnamefont
  {Darmawan}}, \bibinfo {author} {\bibfnamefont {Y.}~\bibnamefont {Yamaji}},\
  and\ \bibinfo {author} {\bibfnamefont {M.}~\bibnamefont {Imada}},\ }\bibfield
   {title} {\bibinfo {title} {\emph{Restricted Boltzmann machine learning for
  solving strongly correlated quantum systems}},\ }\href
  {https://doi.org/10.1103/PhysRevB.96.205152} {\bibfield  {journal} {\bibinfo
  {journal} {Phys. Rev. B}\ }\textbf {\bibinfo {volume} {96}},\ \bibinfo
  {pages} {205152} (\bibinfo {year} {2017})}\BibitemShut {NoStop}%
\bibitem [{\citenamefont {Torlai}\ \emph {et~al.}(2018)\citenamefont {Torlai},
  \citenamefont {Mazzola}, \citenamefont {Carrasquilla}, \citenamefont
  {Troyer}, \citenamefont {Melko},\ and\ \citenamefont {Carleo}}]{torlai2018}%
  \BibitemOpen
  \bibfield  {author} {\bibinfo {author} {\bibfnamefont {G.}~\bibnamefont
  {Torlai}}, \bibinfo {author} {\bibfnamefont {G.}~\bibnamefont {Mazzola}},
  \bibinfo {author} {\bibfnamefont {J.}~\bibnamefont {Carrasquilla}}, \bibinfo
  {author} {\bibfnamefont {M.}~\bibnamefont {Troyer}}, \bibinfo {author}
  {\bibfnamefont {R.}~\bibnamefont {Melko}},\ and\ \bibinfo {author}
  {\bibfnamefont {G.}~\bibnamefont {Carleo}},\ }\bibfield  {title} {\bibinfo
  {title} {\emph{Neural-network quantum state tomography}},\ }\href
  {https://doi.org/10.1038/s41567-018-0048-5} {\bibfield  {journal} {\bibinfo
  {journal} {Nat. Phys.}\ }\textbf {\bibinfo {volume} {14}},\ \bibinfo {pages}
  {447} (\bibinfo {year} {2018})}\BibitemShut {NoStop}%
\bibitem [{\citenamefont {Sehayek}\ \emph {et~al.}(2019)\citenamefont
  {Sehayek}, \citenamefont {Golubeva}, \citenamefont {Albergo}, \citenamefont
  {Kulchytskyy}, \citenamefont {Torlai},\ and\ \citenamefont
  {Melko}}]{sehayek2019}%
  \BibitemOpen
  \bibfield  {author} {\bibinfo {author} {\bibfnamefont {D.}~\bibnamefont
  {Sehayek}}, \bibinfo {author} {\bibfnamefont {A.}~\bibnamefont {Golubeva}},
  \bibinfo {author} {\bibfnamefont {M.~S.}\ \bibnamefont {Albergo}}, \bibinfo
  {author} {\bibfnamefont {B.}~\bibnamefont {Kulchytskyy}}, \bibinfo {author}
  {\bibfnamefont {G.}~\bibnamefont {Torlai}},\ and\ \bibinfo {author}
  {\bibfnamefont {R.~G.}\ \bibnamefont {Melko}},\ }\bibfield  {title} {\bibinfo
  {title} {\emph{Learnability scaling of quantum states: Restricted Boltzmann
  machines}},\ }\href {https://doi.org/10.1103/PhysRevB.100.195125} {\bibfield
  {journal} {\bibinfo  {journal} {Phys. Rev. B}\ }\textbf {\bibinfo {volume}
  {100}},\ \bibinfo {pages} {195125} (\bibinfo {year} {2019})}\BibitemShut
  {NoStop}%
\bibitem [{\citenamefont {Huang}\ and\ \citenamefont
  {Moore}(2021)}]{huang2021}%
  \BibitemOpen
  \bibfield  {author} {\bibinfo {author} {\bibfnamefont {Y.}~\bibnamefont
  {Huang}}\ and\ \bibinfo {author} {\bibfnamefont {J.~E.}\ \bibnamefont
  {Moore}},\ }\bibfield  {title} {\bibinfo {title} {\emph{Neural Network
  Representation of Tensor Network and Chiral States}},\ }\href
  {https://doi.org/10.1103/PhysRevLett.127.170601} {\bibfield  {journal}
  {\bibinfo  {journal} {Phys. Rev. Lett.}\ }\textbf {\bibinfo {volume} {127}},\
  \bibinfo {pages} {170601} (\bibinfo {year} {2021})}\BibitemShut {NoStop}%
\bibitem [{\citenamefont {Nomura}(2021)}]{nomura2021}%
  \BibitemOpen
  \bibfield  {author} {\bibinfo {author} {\bibfnamefont {Y.}~\bibnamefont
  {Nomura}},\ }\bibfield  {title} {\bibinfo {title} {\emph{Helping restricted
  Boltzmann machines with quantum-state representation by restoring
  symmetry}},\ }\href {https://doi.org/10.1088/1361-648x/abe268} {\bibfield
  {journal} {\bibinfo  {journal} {J. Phys.: Condens. Matter.}\ }\textbf
  {\bibinfo {volume} {33}},\ \bibinfo {pages} {174003} (\bibinfo {year}
  {2021})}\BibitemShut {NoStop}%
\bibitem [{\citenamefont {Rrapaj}\ and\ \citenamefont
  {Roggero}(2021)}]{rrapaj2021}%
  \BibitemOpen
  \bibfield  {author} {\bibinfo {author} {\bibfnamefont {E.}~\bibnamefont
  {Rrapaj}}\ and\ \bibinfo {author} {\bibfnamefont {A.}~\bibnamefont
  {Roggero}},\ }\bibfield  {title} {\bibinfo {title} {\emph{Exact
  representations of many-body interactions with restricted-Boltzmann-machine
  neural networks}},\ }\href {https://doi.org/10.1103/PhysRevE.103.013302}
  {\bibfield  {journal} {\bibinfo  {journal} {Phys. Rev. E}\ }\textbf {\bibinfo
  {volume} {103}},\ \bibinfo {pages} {013302} (\bibinfo {year}
  {2021})}\BibitemShut {NoStop}%
\bibitem [{\citenamefont {Nomura}(2022)}]{nomura2022}%
  \BibitemOpen
  \bibfield  {author} {\bibinfo {author} {\bibfnamefont {Y.}~\bibnamefont
  {Nomura}},\ }\bibfield  {title} {\bibinfo {title} {\emph{Investigating
  Network Parameters in Neural-Network Quantum States}},\ }\href
  {https://doi.org/10.7566/jpsj.91.054709} {\bibfield  {journal} {\bibinfo
  {journal} {J. Phys. Soc. Jpn.}\ }\textbf {\bibinfo {volume} {91}},\ \bibinfo
  {pages} {054709} (\bibinfo {year} {2022})}\BibitemShut {NoStop}%
\bibitem [{\citenamefont {Golubeva}\ and\ \citenamefont
  {Melko}(2022)}]{golubeva2022}%
  \BibitemOpen
  \bibfield  {author} {\bibinfo {author} {\bibfnamefont {A.}~\bibnamefont
  {Golubeva}}\ and\ \bibinfo {author} {\bibfnamefont {R.~G.}\ \bibnamefont
  {Melko}},\ }\bibfield  {title} {\bibinfo {title} {\emph{Pruning a restricted
  Boltzmann machine for quantum state reconstruction}},\ }\href
  {https://doi.org/10.1103/PhysRevB.105.125124} {\bibfield  {journal} {\bibinfo
   {journal} {Phys. Rev. B}\ }\textbf {\bibinfo {volume} {105}},\ \bibinfo
  {pages} {125124} (\bibinfo {year} {2022})}\BibitemShut {NoStop}%
\bibitem [{\citenamefont {Kaubruegger}\ \emph {et~al.}(2018)\citenamefont
  {Kaubruegger}, \citenamefont {Pastori},\ and\ \citenamefont
  {Budich}}]{kaubruegger2018}%
  \BibitemOpen
  \bibfield  {author} {\bibinfo {author} {\bibfnamefont {R.}~\bibnamefont
  {Kaubruegger}}, \bibinfo {author} {\bibfnamefont {L.}~\bibnamefont
  {Pastori}},\ and\ \bibinfo {author} {\bibfnamefont {J.~C.}\ \bibnamefont
  {Budich}},\ }\bibfield  {title} {\bibinfo {title} {\emph{Chiral topological
  phases from artificial neural networks}},\ }\href
  {https://doi.org/10.1103/PhysRevB.97.195136} {\bibfield  {journal} {\bibinfo
  {journal} {Phys. Rev. B}\ }\textbf {\bibinfo {volume} {97}},\ \bibinfo
  {pages} {195136} (\bibinfo {year} {2018})}\BibitemShut {NoStop}%
\bibitem [{\citenamefont {Pastori}\ \emph {et~al.}(2019)\citenamefont
  {Pastori}, \citenamefont {Kaubruegger},\ and\ \citenamefont
  {Budich}}]{pastori2019}%
  \BibitemOpen
  \bibfield  {author} {\bibinfo {author} {\bibfnamefont {L.}~\bibnamefont
  {Pastori}}, \bibinfo {author} {\bibfnamefont {R.}~\bibnamefont
  {Kaubruegger}},\ and\ \bibinfo {author} {\bibfnamefont {J.~C.}\ \bibnamefont
  {Budich}},\ }\bibfield  {title} {\bibinfo {title} {\emph{Generalized transfer
  matrix states from artificial neural networks}},\ }\href
  {https://doi.org/10.1103/PhysRevB.99.165123} {\bibfield  {journal} {\bibinfo
  {journal} {Phys. Rev. B}\ }\textbf {\bibinfo {volume} {99}},\ \bibinfo
  {pages} {165123} (\bibinfo {year} {2019})}\BibitemShut {NoStop}%
\bibitem [{\citenamefont {Lu}\ \emph {et~al.}(2019)\citenamefont {Lu},
  \citenamefont {Gao},\ and\ \citenamefont {Duan}}]{lu2019}%
  \BibitemOpen
  \bibfield  {author} {\bibinfo {author} {\bibfnamefont {S.}~\bibnamefont
  {Lu}}, \bibinfo {author} {\bibfnamefont {X.}~\bibnamefont {Gao}},\ and\
  \bibinfo {author} {\bibfnamefont {L.-M.}\ \bibnamefont {Duan}},\ }\bibfield
  {title} {\bibinfo {title} {\emph{Efficient representation of topologically
  ordered states with restricted Boltzmann machines}},\ }\href
  {https://doi.org/10.1103/PhysRevB.99.155136} {\bibfield  {journal} {\bibinfo
  {journal} {Phys. Rev. B}\ }\textbf {\bibinfo {volume} {99}},\ \bibinfo
  {pages} {155136} (\bibinfo {year} {2019})}\BibitemShut {NoStop}%
\bibitem [{\citenamefont {Wu}\ \emph {et~al.}(2023)\citenamefont {Wu},
  \citenamefont {Rossi}, \citenamefont {Vicentini},\ and\ \citenamefont
  {Carleo}}]{wu2023}%
  \BibitemOpen
  \bibfield  {author} {\bibinfo {author} {\bibfnamefont {D.}~\bibnamefont
  {Wu}}, \bibinfo {author} {\bibfnamefont {R.}~\bibnamefont {Rossi}}, \bibinfo
  {author} {\bibfnamefont {F.}~\bibnamefont {Vicentini}},\ and\ \bibinfo
  {author} {\bibfnamefont {G.}~\bibnamefont {Carleo}},\ }\bibfield  {title}
  {\bibinfo {title} {\emph{From tensor-network quantum states to tensorial
  recurrent neural networks}},\ }\href
  {https://doi.org/10.1103/PhysRevResearch.5.L032001} {\bibfield  {journal}
  {\bibinfo  {journal} {Phys. Rev. Res.}\ }\textbf {\bibinfo {volume} {5}},\
  \bibinfo {pages} {L032001} (\bibinfo {year} {2023})}\BibitemShut {NoStop}%
\bibitem [{\citenamefont {Carleo}\ \emph {et~al.}(2018)\citenamefont {Carleo},
  \citenamefont {Nomura},\ and\ \citenamefont {Imada}}]{carleo2018}%
  \BibitemOpen
  \bibfield  {author} {\bibinfo {author} {\bibfnamefont {G.}~\bibnamefont
  {Carleo}}, \bibinfo {author} {\bibfnamefont {Y.}~\bibnamefont {Nomura}},\
  and\ \bibinfo {author} {\bibfnamefont {M.}~\bibnamefont {Imada}},\ }\bibfield
   {title} {\bibinfo {title} {\emph{Constructing exact representations of
  quantum many-body systems with deep neural networks}},\ }\href
  {https://doi.org/10.1038/s41467-018-07520-3} {\bibfield  {journal} {\bibinfo
  {journal} {Nature Commun.}\ }\textbf {\bibinfo {volume} {9}},\ \bibinfo
  {pages} {5322} (\bibinfo {year} {2018})}\BibitemShut {NoStop}%
\bibitem [{\citenamefont {Levine}\ \emph {et~al.}(2019)\citenamefont {Levine},
  \citenamefont {Sharir}, \citenamefont {Cohen},\ and\ \citenamefont
  {Shashua}}]{levine2019}%
  \BibitemOpen
  \bibfield  {author} {\bibinfo {author} {\bibfnamefont {Y.}~\bibnamefont
  {Levine}}, \bibinfo {author} {\bibfnamefont {O.}~\bibnamefont {Sharir}},
  \bibinfo {author} {\bibfnamefont {N.}~\bibnamefont {Cohen}},\ and\ \bibinfo
  {author} {\bibfnamefont {A.}~\bibnamefont {Shashua}},\ }\bibfield  {title}
  {\bibinfo {title} {\emph{Quantum Entanglement in Deep Learning
  Architectures}},\ }\href {https://doi.org/10.1103/PhysRevLett.122.065301}
  {\bibfield  {journal} {\bibinfo  {journal} {Phys. Rev. Lett.}\ }\textbf
  {\bibinfo {volume} {122}},\ \bibinfo {pages} {065301} (\bibinfo {year}
  {2019})}\BibitemShut {NoStop}%
\bibitem [{\citenamefont {Schmitt}\ and\ \citenamefont
  {Heyl}(2020)}]{schmitt2020}%
  \BibitemOpen
  \bibfield  {author} {\bibinfo {author} {\bibfnamefont {M.}~\bibnamefont
  {Schmitt}}\ and\ \bibinfo {author} {\bibfnamefont {M.}~\bibnamefont {Heyl}},\
  }\bibfield  {title} {\bibinfo {title} {\emph{Quantum Many-Body Dynamics in
  Two Dimensions with Artificial Neural Networks}},\ }\href
  {https://doi.org/10.1103/PhysRevLett.125.100503} {\bibfield  {journal}
  {\bibinfo  {journal} {Phys. Rev. Lett.}\ }\textbf {\bibinfo {volume} {125}},\
  \bibinfo {pages} {100503} (\bibinfo {year} {2020})}\BibitemShut {NoStop}%
\bibitem [{\citenamefont {Medina}\ \emph {et~al.}(2021)\citenamefont {Medina},
  \citenamefont {Vasseur},\ and\ \citenamefont {Serbyn}}]{medina2021}%
  \BibitemOpen
  \bibfield  {author} {\bibinfo {author} {\bibfnamefont {R.}~\bibnamefont
  {Medina}}, \bibinfo {author} {\bibfnamefont {R.}~\bibnamefont {Vasseur}},\
  and\ \bibinfo {author} {\bibfnamefont {M.}~\bibnamefont {Serbyn}},\
  }\bibfield  {title} {\bibinfo {title} {\emph{Entanglement transitions from
  restricted Boltzmann machines}},\ }\href
  {https://doi.org/10.1103/PhysRevB.104.104205} {\bibfield  {journal} {\bibinfo
   {journal} {Phys. Rev. B}\ }\textbf {\bibinfo {volume} {104}},\ \bibinfo
  {pages} {104205} (\bibinfo {year} {2021})}\BibitemShut {NoStop}%
\bibitem [{\citenamefont {Sharir}\ \emph {et~al.}(2022)\citenamefont {Sharir},
  \citenamefont {Shashua},\ and\ \citenamefont {Carleo}}]{sharir2022}%
  \BibitemOpen
  \bibfield  {author} {\bibinfo {author} {\bibfnamefont {O.}~\bibnamefont
  {Sharir}}, \bibinfo {author} {\bibfnamefont {A.}~\bibnamefont {Shashua}},\
  and\ \bibinfo {author} {\bibfnamefont {G.}~\bibnamefont {Carleo}},\
  }\bibfield  {title} {\bibinfo {title} {\emph{Neural tensor contractions and
  the expressive power of deep neural quantum states}},\ }\href
  {https://doi.org/10.1103/PhysRevB.106.205136} {\bibfield  {journal} {\bibinfo
   {journal} {Phys. Rev. B}\ }\textbf {\bibinfo {volume} {106}},\ \bibinfo
  {pages} {205136} (\bibinfo {year} {2022})}\BibitemShut {NoStop}%
\bibitem [{\citenamefont {Passetti}\ \emph {et~al.}(2023)\citenamefont
  {Passetti}, \citenamefont {Hofmann}, \citenamefont {Neitemeier},
  \citenamefont {Grunwald}, \citenamefont {Sentef},\ and\ \citenamefont
  {Kennes}}]{passetti2023}%
  \BibitemOpen
  \bibfield  {author} {\bibinfo {author} {\bibfnamefont {G.}~\bibnamefont
  {Passetti}}, \bibinfo {author} {\bibfnamefont {D.}~\bibnamefont {Hofmann}},
  \bibinfo {author} {\bibfnamefont {P.}~\bibnamefont {Neitemeier}}, \bibinfo
  {author} {\bibfnamefont {L.}~\bibnamefont {Grunwald}}, \bibinfo {author}
  {\bibfnamefont {M.~A.}\ \bibnamefont {Sentef}},\ and\ \bibinfo {author}
  {\bibfnamefont {D.~M.}\ \bibnamefont {Kennes}},\ }\bibfield  {title}
  {\bibinfo {title} {\emph{Can Neural Quantum States Learn Volume-Law Ground
  States?}},\ }\href {https://doi.org/10.1103/PhysRevLett.131.036502}
  {\bibfield  {journal} {\bibinfo  {journal} {Phys. Rev. Lett.}\ }\textbf
  {\bibinfo {volume} {131}},\ \bibinfo {pages} {036502} (\bibinfo {year}
  {2023})}\BibitemShut {NoStop}%
\bibitem [{\citenamefont {Wurst}\ \emph {et~al.}(2025)\citenamefont {Wurst},
  \citenamefont {Kennes},\ and\ \citenamefont {Profe}}]{wurst2025}%
  \BibitemOpen
  \bibfield  {author} {\bibinfo {author} {\bibfnamefont {B.~J.}\ \bibnamefont
  {Wurst}}, \bibinfo {author} {\bibfnamefont {D.~M.}\ \bibnamefont {Kennes}},\
  and\ \bibinfo {author} {\bibfnamefont {J.~B.}\ \bibnamefont {Profe}},\
  }\bibfield  {title} {\bibinfo {title} {\emph{Efficiency of the hidden fermion
  determinant states Ansatz in the light of different complexity measures}},\
  }\href {https://doi.org/10.1103/PhysRevResearch.7.023316} {\bibfield
  {journal} {\bibinfo  {journal} {Phys. Rev. Res.}\ }\textbf {\bibinfo {volume}
  {7}},\ \bibinfo {pages} {023316} (\bibinfo {year} {2025})}\BibitemShut
  {NoStop}%
\bibitem [{\citenamefont {Denis}\ \emph {et~al.}(2025)\citenamefont {Denis},
  \citenamefont {Sinibaldi},\ and\ \citenamefont {Carleo}}]{denis2025}%
  \BibitemOpen
  \bibfield  {author} {\bibinfo {author} {\bibfnamefont {Z.}~\bibnamefont
  {Denis}}, \bibinfo {author} {\bibfnamefont {A.}~\bibnamefont {Sinibaldi}},\
  and\ \bibinfo {author} {\bibfnamefont {G.}~\bibnamefont {Carleo}},\
  }\bibfield  {title} {\bibinfo {title} {\emph{Comment on ``Can Neural Quantum
  States Learn Volume-Law Ground States?''}},\ }\href
  {https://doi.org/10.1103/PhysRevLett.134.079701} {\bibfield  {journal}
  {\bibinfo  {journal} {Phys. Rev. Lett.}\ }\textbf {\bibinfo {volume} {134}},\
  \bibinfo {pages} {079701} (\bibinfo {year} {2025})}\BibitemShut {NoStop}%
\bibitem [{\citenamefont {Deng}\ \emph
  {et~al.}(2017{\natexlab{a}})\citenamefont {Deng}, \citenamefont {Li},\ and\
  \citenamefont {Das~Sarma}}]{deng2017a}%
  \BibitemOpen
  \bibfield  {author} {\bibinfo {author} {\bibfnamefont {D.-L.}\ \bibnamefont
  {Deng}}, \bibinfo {author} {\bibfnamefont {X.}~\bibnamefont {Li}},\ and\
  \bibinfo {author} {\bibfnamefont {S.}~\bibnamefont {Das~Sarma}},\ }\bibfield
  {title} {\bibinfo {title} {\emph{Quantum Entanglement in Neural Network
  States}},\ }\href {https://doi.org/10.1103/PhysRevX.7.021021} {\bibfield
  {journal} {\bibinfo  {journal} {Phys. Rev. X}\ }\textbf {\bibinfo {volume}
  {7}},\ \bibinfo {pages} {021021} (\bibinfo {year}
  {2017}{\natexlab{a}})}\BibitemShut {NoStop}%
\bibitem [{\citenamefont {Pei}\ and\ \citenamefont
  {Clark}(2021{\natexlab{a}})}]{pei2021b}%
  \BibitemOpen
  \bibfield  {author} {\bibinfo {author} {\bibfnamefont {M.~Y.}\ \bibnamefont
  {Pei}}\ and\ \bibinfo {author} {\bibfnamefont {S.~R.}\ \bibnamefont
  {Clark}},\ }\bibfield  {title} {\bibinfo {title} {\emph{Compact
  neural-network quantum state representations of Jastrow and stabilizer
  states}},\ }\href {https://doi.org/10.1088/1751-8121/ac1f3d} {\bibfield
  {journal} {\bibinfo  {journal} {J. Phys. A: Math. Theor.}\ }\textbf {\bibinfo
  {volume} {54}},\ \bibinfo {pages} {405304} (\bibinfo {year}
  {2021}{\natexlab{a}})}\BibitemShut {NoStop}%
\bibitem [{\citenamefont {Chen}\ \emph {et~al.}(2018)\citenamefont {Chen},
  \citenamefont {Cheng}, \citenamefont {Xie}, \citenamefont {Wang},\ and\
  \citenamefont {Xiang}}]{chen2018}%
  \BibitemOpen
  \bibfield  {author} {\bibinfo {author} {\bibfnamefont {J.}~\bibnamefont
  {Chen}}, \bibinfo {author} {\bibfnamefont {S.}~\bibnamefont {Cheng}},
  \bibinfo {author} {\bibfnamefont {H.}~\bibnamefont {Xie}}, \bibinfo {author}
  {\bibfnamefont {L.}~\bibnamefont {Wang}},\ and\ \bibinfo {author}
  {\bibfnamefont {T.}~\bibnamefont {Xiang}},\ }\bibfield  {title} {\bibinfo
  {title} {\emph{Equivalence of restricted Boltzmann machines and tensor
  network states}},\ }\href {https://doi.org/10.1103/PhysRevB.97.085104}
  {\bibfield  {journal} {\bibinfo  {journal} {Phys. Rev. B}\ }\textbf {\bibinfo
  {volume} {97}},\ \bibinfo {pages} {085104} (\bibinfo {year}
  {2018})}\BibitemShut {NoStop}%
\bibitem [{\citenamefont {Gao}\ and\ \citenamefont {Duan}(2017)}]{gao2017}%
  \BibitemOpen
  \bibfield  {author} {\bibinfo {author} {\bibfnamefont {X.}~\bibnamefont
  {Gao}}\ and\ \bibinfo {author} {\bibfnamefont {L.-M.}\ \bibnamefont {Duan}},\
  }\bibfield  {title} {\bibinfo {title} {\emph{Efficient representation of
  quantum many-body states with deep neural networks}},\ }\href
  {https://doi.org/10.1038/s41467-017-00705-2} {\bibfield  {journal} {\bibinfo
  {journal} {Nature Commun.}\ }\textbf {\bibinfo {volume} {8}},\ \bibinfo
  {pages} {662} (\bibinfo {year} {2017})}\BibitemShut {NoStop}%
\bibitem [{\citenamefont {Astrakhantsev}\ \emph {et~al.}(2021)\citenamefont
  {Astrakhantsev}, \citenamefont {Westerhout}, \citenamefont {Tiwari},
  \citenamefont {Choo}, \citenamefont {Chen}, \citenamefont {Fischer},
  \citenamefont {Carleo},\ and\ \citenamefont {Neupert}}]{astrakhantsev2021}%
  \BibitemOpen
  \bibfield  {author} {\bibinfo {author} {\bibfnamefont {N.}~\bibnamefont
  {Astrakhantsev}}, \bibinfo {author} {\bibfnamefont {T.}~\bibnamefont
  {Westerhout}}, \bibinfo {author} {\bibfnamefont {A.}~\bibnamefont {Tiwari}},
  \bibinfo {author} {\bibfnamefont {K.}~\bibnamefont {Choo}}, \bibinfo {author}
  {\bibfnamefont {A.}~\bibnamefont {Chen}}, \bibinfo {author} {\bibfnamefont
  {M.~H.}\ \bibnamefont {Fischer}}, \bibinfo {author} {\bibfnamefont
  {G.}~\bibnamefont {Carleo}},\ and\ \bibinfo {author} {\bibfnamefont
  {T.}~\bibnamefont {Neupert}},\ }\bibfield  {title} {\bibinfo {title}
  {\emph{Broken-Symmetry Ground States of the Heisenberg Model on the
  Pyrochlore Lattice}},\ }\href {https://doi.org/10.1103/PhysRevX.11.041021}
  {\bibfield  {journal} {\bibinfo  {journal} {Phys. Rev. X}\ }\textbf {\bibinfo
  {volume} {11}},\ \bibinfo {pages} {041021} (\bibinfo {year}
  {2021})}\BibitemShut {NoStop}%
\bibitem [{\citenamefont {Pohle}\ \emph {et~al.}()\citenamefont {Pohle},
  \citenamefont {Yamaji},\ and\ \citenamefont {Imada}}]{pohle2023_arxiv}%
  \BibitemOpen
  \bibfield  {author} {\bibinfo {author} {\bibfnamefont {R.}~\bibnamefont
  {Pohle}}, \bibinfo {author} {\bibfnamefont {Y.}~\bibnamefont {Yamaji}},\ and\
  \bibinfo {author} {\bibfnamefont {M.}~\bibnamefont {Imada}},\ }\href@noop {}
  {\bibinfo {title} {\emph{Ground state of the $S$=1/2 pyrochlore Heisenberg
  antiferromagnet: A quantum spin liquid emergent from dimensional
  reduction}}},\ \Eprint {https://arxiv.org/abs/2311.11561} {arXiv:2311.11561}
  \BibitemShut {NoStop}%
\bibitem [{\citenamefont {Machaczek}\ \emph {et~al.}(2025)\citenamefont
  {Machaczek}, \citenamefont {Pollet},\ and\ \citenamefont
  {Liu}}]{machaczek2025}%
  \BibitemOpen
  \bibfield  {author} {\bibinfo {author} {\bibfnamefont {M.}~\bibnamefont
  {Machaczek}}, \bibinfo {author} {\bibfnamefont {L.}~\bibnamefont {Pollet}},\
  and\ \bibinfo {author} {\bibfnamefont {K.}~\bibnamefont {Liu}},\ }\bibfield
  {title} {\bibinfo {title} {\emph{{Neural quantum state study of fracton
  models}}},\ }\href {https://doi.org/10.21468/SciPostPhys.18.3.112} {\bibfield
   {journal} {\bibinfo  {journal} {SciPost Phys.}\ }\textbf {\bibinfo {volume}
  {18}},\ \bibinfo {pages} {112} (\bibinfo {year} {2025})}\BibitemShut
  {NoStop}%
\bibitem [{\citenamefont {Deng}\ \emph
  {et~al.}(2017{\natexlab{b}})\citenamefont {Deng}, \citenamefont {Li},\ and\
  \citenamefont {Das~Sarma}}]{deng2017b}%
  \BibitemOpen
  \bibfield  {author} {\bibinfo {author} {\bibfnamefont {D.-L.}\ \bibnamefont
  {Deng}}, \bibinfo {author} {\bibfnamefont {X.}~\bibnamefont {Li}},\ and\
  \bibinfo {author} {\bibfnamefont {S.}~\bibnamefont {Das~Sarma}},\ }\bibfield
  {title} {\bibinfo {title} {\emph{Machine learning topological states}},\
  }\href {https://doi.org/10.1103/PhysRevB.96.195145} {\bibfield  {journal}
  {\bibinfo  {journal} {Phys. Rev. B}\ }\textbf {\bibinfo {volume} {96}},\
  \bibinfo {pages} {195145} (\bibinfo {year} {2017}{\natexlab{b}})}\BibitemShut
  {NoStop}%
\bibitem [{\citenamefont {Zheng}\ \emph {et~al.}(2019)\citenamefont {Zheng},
  \citenamefont {He}, \citenamefont {Regnault},\ and\ \citenamefont
  {Bernevig}}]{zheng2019}%
  \BibitemOpen
  \bibfield  {author} {\bibinfo {author} {\bibfnamefont {Y.}~\bibnamefont
  {Zheng}}, \bibinfo {author} {\bibfnamefont {H.}~\bibnamefont {He}}, \bibinfo
  {author} {\bibfnamefont {N.}~\bibnamefont {Regnault}},\ and\ \bibinfo
  {author} {\bibfnamefont {B.~A.}\ \bibnamefont {Bernevig}},\ }\bibfield
  {title} {\bibinfo {title} {\emph{Restricted Boltzmann machines and matrix
  product states of one-dimensional translationally invariant stabilizer
  codes}},\ }\href {https://doi.org/10.1103/PhysRevB.99.155129} {\bibfield
  {journal} {\bibinfo  {journal} {Phys. Rev. B}\ }\textbf {\bibinfo {volume}
  {99}},\ \bibinfo {pages} {155129} (\bibinfo {year} {2019})}\BibitemShut
  {NoStop}%
\bibitem [{\citenamefont {Pei}\ and\ \citenamefont
  {Clark}(2021{\natexlab{b}})}]{pei2021a}%
  \BibitemOpen
  \bibfield  {author} {\bibinfo {author} {\bibfnamefont {M.~Y.}\ \bibnamefont
  {Pei}}\ and\ \bibinfo {author} {\bibfnamefont {S.~R.}\ \bibnamefont
  {Clark}},\ }\bibfield  {title} {\bibinfo {title} {\emph{Neural-Network
  Quantum States for Spin-1 Systems: Spin-Basis and Parameterization Effects on
  Compactness of Representations}},\ }\href {https://doi.org/10.3390/e23070879}
  {\bibfield  {journal} {\bibinfo  {journal} {Entropy}\ }\textbf {\bibinfo
  {volume} {23}},\ \bibinfo {pages} {879} (\bibinfo {year}
  {2021}{\natexlab{b}})}\BibitemShut {NoStop}%
\bibitem [{\citenamefont {Harney}\ \emph {et~al.}(2021)\citenamefont {Harney},
  \citenamefont {Paternostro},\ and\ \citenamefont {Pirandola}}]{harney2021}%
  \BibitemOpen
  \bibfield  {author} {\bibinfo {author} {\bibfnamefont {C.}~\bibnamefont
  {Harney}}, \bibinfo {author} {\bibfnamefont {M.}~\bibnamefont
  {Paternostro}},\ and\ \bibinfo {author} {\bibfnamefont {S.}~\bibnamefont
  {Pirandola}},\ }\bibfield  {title} {\bibinfo {title} {\emph{Mixed state
  entanglement classification using artificial neural networks}},\ }\href
  {https://doi.org/10.1088/1367-2630/ac0388} {\bibfield  {journal} {\bibinfo
  {journal} {New J. Phys.}\ }\textbf {\bibinfo {volume} {23}},\ \bibinfo
  {pages} {063033} (\bibinfo {year} {2021})}\BibitemShut {NoStop}%
\bibitem [{\citenamefont {Zhang}\ and\ \citenamefont
  {Di~Ventra}(2022)}]{zhang2022}%
  \BibitemOpen
  \bibfield  {author} {\bibinfo {author} {\bibfnamefont {Y.-H.}\ \bibnamefont
  {Zhang}}\ and\ \bibinfo {author} {\bibfnamefont {M.}~\bibnamefont
  {Di~Ventra}},\ }\bibfield  {title} {\bibinfo {title} {\emph{Efficient quantum
  state tomography with mode-assisted training}},\ }\href
  {https://doi.org/10.1103/PhysRevA.106.042420} {\bibfield  {journal} {\bibinfo
   {journal} {Phys. Rev. A}\ }\textbf {\bibinfo {volume} {106}},\ \bibinfo
  {pages} {042420} (\bibinfo {year} {2022})}\BibitemShut {NoStop}%
\bibitem [{\citenamefont {Glasser}\ \emph {et~al.}(2018)\citenamefont
  {Glasser}, \citenamefont {Pancotti}, \citenamefont {August}, \citenamefont
  {Rodriguez},\ and\ \citenamefont {Cirac}}]{glasser2018}%
  \BibitemOpen
  \bibfield  {author} {\bibinfo {author} {\bibfnamefont {I.}~\bibnamefont
  {Glasser}}, \bibinfo {author} {\bibfnamefont {N.}~\bibnamefont {Pancotti}},
  \bibinfo {author} {\bibfnamefont {M.}~\bibnamefont {August}}, \bibinfo
  {author} {\bibfnamefont {I.~D.}\ \bibnamefont {Rodriguez}},\ and\ \bibinfo
  {author} {\bibfnamefont {J.~I.}\ \bibnamefont {Cirac}},\ }\bibfield  {title}
  {\bibinfo {title} {\emph{Neural-Network Quantum States, String-Bond States,
  and Chiral Topological States}},\ }\href
  {https://doi.org/10.1103/PhysRevX.8.011006} {\bibfield  {journal} {\bibinfo
  {journal} {Phys. Rev. X}\ }\textbf {\bibinfo {volume} {8}},\ \bibinfo {pages}
  {011006} (\bibinfo {year} {2018})}\BibitemShut {NoStop}%
\bibitem [{\citenamefont {He}\ \emph {et~al.}()\citenamefont {He},
  \citenamefont {Zheng}, \citenamefont {Bernevig},\ and\ \citenamefont
  {Sierra}}]{he2019_arxiv}%
  \BibitemOpen
  \bibfield  {author} {\bibinfo {author} {\bibfnamefont {H.}~\bibnamefont
  {He}}, \bibinfo {author} {\bibfnamefont {Y.}~\bibnamefont {Zheng}}, \bibinfo
  {author} {\bibfnamefont {B.~A.}\ \bibnamefont {Bernevig}},\ and\ \bibinfo
  {author} {\bibfnamefont {G.}~\bibnamefont {Sierra}},\ }\href@noop {}
  {\bibinfo {title} {\emph{Multi-Layer Restricted Boltzmann Machine
  Representation of 1D Quantum Many-Body Wave Functions}}},\ \Eprint
  {https://arxiv.org/abs/1910.13454} {arXiv:1910.13454} \BibitemShut {NoStop}%
\bibitem [{\citenamefont {Clark}(2018)}]{clark2018}%
  \BibitemOpen
  \bibfield  {author} {\bibinfo {author} {\bibfnamefont {S.~R.}\ \bibnamefont
  {Clark}},\ }\bibfield  {title} {\bibinfo {title} {\emph{Unifying
  neural-network quantum states and correlator product states via tensor
  networks}},\ }\href {https://doi.org/10.1088/1751-8121/aaaaf2} {\bibfield
  {journal} {\bibinfo  {journal} {J. Phys. A: Math. Theor.}\ }\textbf {\bibinfo
  {volume} {51}},\ \bibinfo {pages} {135301} (\bibinfo {year}
  {2018})}\BibitemShut {NoStop}%
\bibitem [{\citenamefont {Glasser}\ \emph {et~al.}(2020)\citenamefont
  {Glasser}, \citenamefont {Pancotti},\ and\ \citenamefont
  {Cirac}}]{glasser2020}%
  \BibitemOpen
  \bibfield  {author} {\bibinfo {author} {\bibfnamefont {I.}~\bibnamefont
  {Glasser}}, \bibinfo {author} {\bibfnamefont {N.}~\bibnamefont {Pancotti}},\
  and\ \bibinfo {author} {\bibfnamefont {J.~I.}\ \bibnamefont {Cirac}},\
  }\bibfield  {title} {\bibinfo {title} {\emph{From Probabilistic Graphical
  Models to Generalized Tensor Networks for Supervised Learning}},\ }\href
  {https://doi.org/10.1109/access.2020.2986279} {\bibfield  {journal} {\bibinfo
   {journal} {IEEE Access}\ }\textbf {\bibinfo {volume} {8}},\ \bibinfo {pages}
  {68169} (\bibinfo {year} {2020})}\BibitemShut {NoStop}%
\bibitem [{\citenamefont {Sharir}\ \emph {et~al.}(2020)\citenamefont {Sharir},
  \citenamefont {Levine}, \citenamefont {Wies}, \citenamefont {Carleo},\ and\
  \citenamefont {Shashua}}]{sharir2020}%
  \BibitemOpen
  \bibfield  {author} {\bibinfo {author} {\bibfnamefont {O.}~\bibnamefont
  {Sharir}}, \bibinfo {author} {\bibfnamefont {Y.}~\bibnamefont {Levine}},
  \bibinfo {author} {\bibfnamefont {N.}~\bibnamefont {Wies}}, \bibinfo {author}
  {\bibfnamefont {G.}~\bibnamefont {Carleo}},\ and\ \bibinfo {author}
  {\bibfnamefont {A.}~\bibnamefont {Shashua}},\ }\bibfield  {title} {\bibinfo
  {title} {\emph{Deep Autoregressive Models for the Efficient Variational
  Simulation of Many-Body Quantum Systems}},\ }\href
  {https://doi.org/10.1103/PhysRevLett.124.020503} {\bibfield  {journal}
  {\bibinfo  {journal} {Phys. Rev. Lett.}\ }\textbf {\bibinfo {volume} {124}},\
  \bibinfo {pages} {020503} (\bibinfo {year} {2020})}\BibitemShut {NoStop}%
\bibitem [{\citenamefont {Li}\ \emph {et~al.}(2021)\citenamefont {Li},
  \citenamefont {Pan}, \citenamefont {Zhou},\ and\ \citenamefont
  {Zhang}}]{li2021}%
  \BibitemOpen
  \bibfield  {author} {\bibinfo {author} {\bibfnamefont {S.}~\bibnamefont
  {Li}}, \bibinfo {author} {\bibfnamefont {F.}~\bibnamefont {Pan}}, \bibinfo
  {author} {\bibfnamefont {P.}~\bibnamefont {Zhou}},\ and\ \bibinfo {author}
  {\bibfnamefont {P.}~\bibnamefont {Zhang}},\ }\bibfield  {title} {\bibinfo
  {title} {\emph{Boltzmann machines as two-dimensional tensor networks}},\
  }\href {https://doi.org/10.1103/PhysRevB.104.075154} {\bibfield  {journal}
  {\bibinfo  {journal} {Phys. Rev. B}\ }\textbf {\bibinfo {volume} {104}},\
  \bibinfo {pages} {075154} (\bibinfo {year} {2021})}\BibitemShut {NoStop}%
\bibitem [{\citenamefont {Aaronson}\ and\ \citenamefont
  {Gottesman}(2004)}]{aaronson2004}%
  \BibitemOpen
  \bibfield  {author} {\bibinfo {author} {\bibfnamefont {S.}~\bibnamefont
  {Aaronson}}\ and\ \bibinfo {author} {\bibfnamefont {D.}~\bibnamefont
  {Gottesman}},\ }\bibfield  {title} {\bibinfo {title} {\emph{Improved
  simulation of stabilizer circuits}},\ }\href
  {https://doi.org/10.1103/PhysRevA.70.052328} {\bibfield  {journal} {\bibinfo
  {journal} {Phys. Rev. A}\ }\textbf {\bibinfo {volume} {70}},\ \bibinfo
  {pages} {052328} (\bibinfo {year} {2004})}\BibitemShut {NoStop}%
\bibitem [{\citenamefont {Greenberger}\ \emph {et~al.}(1989)\citenamefont
  {Greenberger}, \citenamefont {Horne},\ and\ \citenamefont
  {Zeilinger}}]{greenberger1989}%
  \BibitemOpen
  \bibfield  {author} {\bibinfo {author} {\bibfnamefont {D.~M.}\ \bibnamefont
  {Greenberger}}, \bibinfo {author} {\bibfnamefont {M.~A.}\ \bibnamefont
  {Horne}},\ and\ \bibinfo {author} {\bibfnamefont {A.}~\bibnamefont
  {Zeilinger}},\ }\bibinfo {title} {\emph{Going Beyond Bell's Theorem}},\ in\
  \href {https://doi.org/10.1007/978-94-017-0849-4_10} {\emph {\bibinfo
  {booktitle} {Bell's Theorem, Quantum Theory and Conceptions of the
  Universe}}}\ (\bibinfo  {publisher} {Springer Netherlands},\ \bibinfo {year}
  {1989})\ pp.\ \bibinfo {pages} {69--72}\BibitemShut {NoStop}%
\bibitem [{\citenamefont {Greenberger}\ \emph {et~al.}(1990)\citenamefont
  {Greenberger}, \citenamefont {Horne}, \citenamefont {Shimony},\ and\
  \citenamefont {Zeilinger}}]{greenberger1990}%
  \BibitemOpen
  \bibfield  {author} {\bibinfo {author} {\bibfnamefont {D.~M.}\ \bibnamefont
  {Greenberger}}, \bibinfo {author} {\bibfnamefont {M.~A.}\ \bibnamefont
  {Horne}}, \bibinfo {author} {\bibfnamefont {A.}~\bibnamefont {Shimony}},\
  and\ \bibinfo {author} {\bibfnamefont {A.}~\bibnamefont {Zeilinger}},\
  }\bibfield  {title} {\bibinfo {title} {\emph{Bell's theorem without
  inequalities}},\ }\href {https://doi.org/10.1119/1.16243} {\bibfield
  {journal} {\bibinfo  {journal} {Am. J. Phys.}\ }\textbf {\bibinfo {volume}
  {58}},\ \bibinfo {pages} {1131} (\bibinfo {year} {1990})}\BibitemShut
  {NoStop}%
\bibitem [{\citenamefont {D\"ur}\ \emph {et~al.}(2000)\citenamefont {D\"ur},
  \citenamefont {Vidal},\ and\ \citenamefont {Cirac}}]{duer2000}%
  \BibitemOpen
  \bibfield  {author} {\bibinfo {author} {\bibfnamefont {W.}~\bibnamefont
  {D\"ur}}, \bibinfo {author} {\bibfnamefont {G.}~\bibnamefont {Vidal}},\ and\
  \bibinfo {author} {\bibfnamefont {J.~I.}\ \bibnamefont {Cirac}},\ }\bibfield
  {title} {\bibinfo {title} {\emph{Three qubits can be entangled in two
  inequivalent ways}},\ }\href {https://doi.org/10.1103/PhysRevA.62.062314}
  {\bibfield  {journal} {\bibinfo  {journal} {Phys. Rev. A}\ }\textbf {\bibinfo
  {volume} {62}},\ \bibinfo {pages} {062314} (\bibinfo {year}
  {2000})}\BibitemShut {NoStop}%
\bibitem [{\citenamefont {Hitchcock}(1927)}]{hitchcock1927}%
  \BibitemOpen
  \bibfield  {author} {\bibinfo {author} {\bibfnamefont {F.~L.}\ \bibnamefont
  {Hitchcock}},\ }\bibfield  {title} {\bibinfo {title} {\emph{The Expression of
  a Tensor or a Polyadic as a Sum of Products}},\ }\href
  {https://doi.org/10.1002/sapm192761164} {\bibfield  {journal} {\bibinfo
  {journal} {J. Math. Phys.}\ }\textbf {\bibinfo {volume} {6}},\ \bibinfo
  {pages} {164} (\bibinfo {year} {1927})}\BibitemShut {NoStop}%
\bibitem [{\citenamefont {Hitchcock}(1928)}]{hitchcock1928}%
  \BibitemOpen
  \bibfield  {author} {\bibinfo {author} {\bibfnamefont {F.~L.}\ \bibnamefont
  {Hitchcock}},\ }\bibfield  {title} {\bibinfo {title} {\emph{Multiple
  Invariants and Generalized Rank of a P-Way Matrix or Tensor}},\ }\href
  {https://doi.org/10.1002/sapm19287139} {\bibfield  {journal} {\bibinfo
  {journal} {J. Math. Phys.}\ }\textbf {\bibinfo {volume} {7}},\ \bibinfo
  {pages} {39} (\bibinfo {year} {1928})}\BibitemShut {NoStop}%
\bibitem [{\citenamefont {Cattell}(1944)}]{cattell1944}%
  \BibitemOpen
  \bibfield  {author} {\bibinfo {author} {\bibfnamefont {R.~B.}\ \bibnamefont
  {Cattell}},\ }\bibfield  {title} {\bibinfo {title} {\emph{``Parallel
  Proportional Profiles'' and other Principles for Determining the Choice of
  Factors by Rotation}},\ }\href {https://doi.org/10.1007/bf02288739}
  {\bibfield  {journal} {\bibinfo  {journal} {Psychometrika}\ }\textbf
  {\bibinfo {volume} {9}},\ \bibinfo {pages} {267} (\bibinfo {year}
  {1944})}\BibitemShut {NoStop}%
\bibitem [{\citenamefont {Cattell}(1952)}]{cattell1952}%
  \BibitemOpen
  \bibfield  {author} {\bibinfo {author} {\bibfnamefont {R.~B.}\ \bibnamefont
  {Cattell}},\ }\bibfield  {title} {\bibinfo {title} {\emph{The three basic
  factor-analytic research designs--their interrelations and derivatives.}},\
  }\href {https://doi.org/10.1037/h0054245} {\bibfield  {journal} {\bibinfo
  {journal} {Psychol. Bull.}\ }\textbf {\bibinfo {volume} {49}},\ \bibinfo
  {pages} {499} (\bibinfo {year} {1952})}\BibitemShut {NoStop}%
\bibitem [{\citenamefont {Carroll}\ and\ \citenamefont
  {Chang}(1970)}]{carroll1970}%
  \BibitemOpen
  \bibfield  {author} {\bibinfo {author} {\bibfnamefont {J.~D.}\ \bibnamefont
  {Carroll}}\ and\ \bibinfo {author} {\bibfnamefont {J.-J.}\ \bibnamefont
  {Chang}},\ }\bibfield  {title} {\bibinfo {title} {\emph{Analysis of
  Individual Differences in Multidimensional Scaling Via an N-way
  Generalization of ``Eckart-Young'' Decomposition}},\ }\href
  {https://doi.org/10.1007/bf02310791} {\bibfield  {journal} {\bibinfo
  {journal} {Psychometrika}\ }\textbf {\bibinfo {volume} {35}},\ \bibinfo
  {pages} {283} (\bibinfo {year} {1970})}\BibitemShut {NoStop}%
\bibitem [{\citenamefont {Harshman}(1970)}]{harshman1970}%
  \BibitemOpen
  \bibfield  {author} {\bibinfo {author} {\bibfnamefont {R.~A.}\ \bibnamefont
  {Harshman}},\ }\bibfield  {title} {\bibinfo {title} {\emph{Foundations of the
  PARAFAC procedure: Models and conditions for an ``explanatory'' multi-modal
  factor analysis}},\ }\href
  {https://www.psychology.uwo.ca/faculty/harshman/wpppfac0.pdf} {\bibfield
  {journal} {\bibinfo  {journal} {UCLA working papers in phonetics}\ }\textbf
  {\bibinfo {volume} {16}},\ \bibinfo {pages} {84} (\bibinfo {year}
  {1970})}\BibitemShut {NoStop}%
\bibitem [{\citenamefont {Mocks}(1988)}]{mocks1988}%
  \BibitemOpen
  \bibfield  {author} {\bibinfo {author} {\bibfnamefont {J.}~\bibnamefont
  {Mocks}},\ }\bibfield  {title} {\bibinfo {title} {\emph{Topographic
  components model for event-related potentials and some biophysical
  considerations}},\ }\href {https://doi.org/10.1109/10.2119} {\bibfield
  {journal} {\bibinfo  {journal} {IEEE Trans. Biomed. Eng.}\ }\textbf {\bibinfo
  {volume} {35}},\ \bibinfo {pages} {482} (\bibinfo {year} {1988})}\BibitemShut
  {NoStop}%
\bibitem [{\citenamefont {Kiers}(2000)}]{kiers2000}%
  \BibitemOpen
  \bibfield  {author} {\bibinfo {author} {\bibfnamefont {H.~A.~L.}\
  \bibnamefont {Kiers}},\ }\bibfield  {title} {\bibinfo {title} {\emph{Towards
  a standardized notation and terminology in multiway analysis}},\ }\href
  {https://doi.org/10.1002/1099-128x(200005/06)14:3<105::aid-cem582>3.0.co;2-i}
  {\bibfield  {journal} {\bibinfo  {journal} {J. Chemom.}\ }\textbf {\bibinfo
  {volume} {14}},\ \bibinfo {pages} {105} (\bibinfo {year} {2000})}\BibitemShut
  {NoStop}%
\bibitem [{\citenamefont {Kolda}\ and\ \citenamefont
  {Bader}(2009)}]{kolda2009}%
  \BibitemOpen
  \bibfield  {author} {\bibinfo {author} {\bibfnamefont {T.~G.}\ \bibnamefont
  {Kolda}}\ and\ \bibinfo {author} {\bibfnamefont {B.~W.}\ \bibnamefont
  {Bader}},\ }\bibfield  {title} {\bibinfo {title} {\emph{Tensor Decompositions
  and Applications}},\ }\href {https://doi.org/10.1137/07070111x} {\bibfield
  {journal} {\bibinfo  {journal} {SIAM Review}\ }\textbf {\bibinfo {volume}
  {51}},\ \bibinfo {pages} {455} (\bibinfo {year} {2009})}\BibitemShut
  {NoStop}%
\bibitem [{\citenamefont {Comon}\ \emph {et~al.}(2009)\citenamefont {Comon},
  \citenamefont {Luciani},\ and\ \citenamefont {de~Almeida}}]{comon2009}%
  \BibitemOpen
  \bibfield  {author} {\bibinfo {author} {\bibfnamefont {P.}~\bibnamefont
  {Comon}}, \bibinfo {author} {\bibfnamefont {X.}~\bibnamefont {Luciani}},\
  and\ \bibinfo {author} {\bibfnamefont {A.~L.~F.}\ \bibnamefont
  {de~Almeida}},\ }\bibfield  {title} {\bibinfo {title} {\emph{Tensor
  decompositions, alternating least squares and other tales}},\ }\href
  {https://doi.org/10.1002/cem.1236} {\bibfield  {journal} {\bibinfo  {journal}
  {J. Chemom.}\ }\textbf {\bibinfo {volume} {23}},\ \bibinfo {pages} {393}
  (\bibinfo {year} {2009})}\BibitemShut {NoStop}%
\bibitem [{\citenamefont {Erichson}\ \emph {et~al.}(2020)\citenamefont
  {Erichson}, \citenamefont {Manohar}, \citenamefont {Brunton},\ and\
  \citenamefont {Kutz}}]{erichson2020}%
  \BibitemOpen
  \bibfield  {author} {\bibinfo {author} {\bibfnamefont {N.~B.}\ \bibnamefont
  {Erichson}}, \bibinfo {author} {\bibfnamefont {K.}~\bibnamefont {Manohar}},
  \bibinfo {author} {\bibfnamefont {S.~L.}\ \bibnamefont {Brunton}},\ and\
  \bibinfo {author} {\bibfnamefont {J.~N.}\ \bibnamefont {Kutz}},\ }\bibfield
  {title} {\bibinfo {title} {\emph{Randomized CP tensor decomposition}},\
  }\href {https://doi.org/10.1088/2632-2153/ab8240} {\bibfield  {journal}
  {\bibinfo  {journal} {Mach. Learn.: Sci. Technol.}\ }\textbf {\bibinfo
  {volume} {1}},\ \bibinfo {pages} {025012} (\bibinfo {year}
  {2020})}\BibitemShut {NoStop}%
\bibitem [{\citenamefont {H\r{a}stad}(1990)}]{hastad1990}%
  \BibitemOpen
  \bibfield  {author} {\bibinfo {author} {\bibfnamefont {J.}~\bibnamefont
  {H\r{a}stad}},\ }\bibfield  {title} {\bibinfo {title} {\emph{Tensor rank is
  NP-complete}},\ }\href {https://doi.org/10.1016/0196-6774(90)90014-6}
  {\bibfield  {journal} {\bibinfo  {journal} {J. Algor.}\ }\textbf {\bibinfo
  {volume} {11}},\ \bibinfo {pages} {644} (\bibinfo {year} {1990})}\BibitemShut
  {NoStop}%
\bibitem [{\citenamefont {Auddy}\ \emph {et~al.}(2025)\citenamefont {Auddy},
  \citenamefont {Xia},\ and\ \citenamefont {Yuan}}]{auddy2025}%
  \BibitemOpen
  \bibfield  {author} {\bibinfo {author} {\bibfnamefont {A.}~\bibnamefont
  {Auddy}}, \bibinfo {author} {\bibfnamefont {D.}~\bibnamefont {Xia}},\ and\
  \bibinfo {author} {\bibfnamefont {M.}~\bibnamefont {Yuan}},\ }\bibfield
  {title} {\bibinfo {title} {\emph{Tensors in High-Dimensional Data Analysis:
  Methodological Opportunities and Theoretical Challenges}},\ }\href
  {https://doi.org/10.1146/annurev-statistics-112723-034548} {\bibfield
  {journal} {\bibinfo  {journal} {Annu. Rev. Stat. Appl.}\ }\textbf {\bibinfo
  {volume} {12}},\ \bibinfo {pages} {527} (\bibinfo {year} {2025})}\BibitemShut
  {NoStop}%
\bibitem [{\citenamefont {Diniz}(2019)}]{diniz2019}%
  \BibitemOpen
  \bibfield  {author} {\bibinfo {author} {\bibfnamefont {F.}~\bibnamefont
  {Diniz}},\ }\href {https://doi.org/10.13140/RG.2.2.14982.91200} {\bibinfo
  {title} {\emph{PhD Thesis: Tensor decompositions and algorithms, with
  applications to tensor learning}}} (\bibinfo {year} {2019})\BibitemShut
  {NoStop}%
\bibitem [{\citenamefont {Zniyed}\ \emph {et~al.}(2020)\citenamefont {Zniyed},
  \citenamefont {Boyer}, \citenamefont {de~Almeida},\ and\ \citenamefont
  {Favier}}]{zniyed2020}%
  \BibitemOpen
  \bibfield  {author} {\bibinfo {author} {\bibfnamefont {Y.}~\bibnamefont
  {Zniyed}}, \bibinfo {author} {\bibfnamefont {R.}~\bibnamefont {Boyer}},
  \bibinfo {author} {\bibfnamefont {A.~L.}\ \bibnamefont {de~Almeida}},\ and\
  \bibinfo {author} {\bibfnamefont {G.}~\bibnamefont {Favier}},\ }\bibfield
  {title} {\bibinfo {title} {\emph{High-order tensor estimation via trains of
  coupled third-order CP and Tucker decompositions}},\ }\href
  {https://doi.org/10.1016/j.laa.2019.11.005} {\bibfield  {journal} {\bibinfo
  {journal} {Linear Algebra Appl.}\ }\textbf {\bibinfo {volume} {588}},\
  \bibinfo {pages} {304} (\bibinfo {year} {2020})}\BibitemShut {NoStop}%
\bibitem [{\citenamefont {Giraud}\ \emph {et~al.}(2023)\citenamefont {Giraud},
  \citenamefont {Itier}, \citenamefont {Boyer}, \citenamefont {Zniyed},\ and\
  \citenamefont {Almeida}}]{giraud2023}%
  \BibitemOpen
  \bibfield  {author} {\bibinfo {author} {\bibfnamefont {M.}~\bibnamefont
  {Giraud}}, \bibinfo {author} {\bibfnamefont {V.}~\bibnamefont {Itier}},
  \bibinfo {author} {\bibfnamefont {R.}~\bibnamefont {Boyer}}, \bibinfo
  {author} {\bibfnamefont {Y.}~\bibnamefont {Zniyed}},\ and\ \bibinfo {author}
  {\bibfnamefont {A.~L.~d.}\ \bibnamefont {Almeida}},\ }\bibfield  {title}
  {\bibinfo {title} {\emph{Tucker Decomposition Based on a Tensor Train of
  Coupled and Constrained CP Cores}},\ }\href
  {https://doi.org/10.1109/lsp.2023.3287144} {\bibfield  {journal} {\bibinfo
  {journal} {IEEE Signal Process. Lett.}\ }\textbf {\bibinfo {volume} {30}},\
  \bibinfo {pages} {758} (\bibinfo {year} {2023})}\BibitemShut {NoStop}%
\bibitem [{\citenamefont {Pr\'{e}vost}\ and\ \citenamefont
  {Chainais}(2025)}]{prvost2025}%
  \BibitemOpen
  \bibfield  {author} {\bibinfo {author} {\bibfnamefont {C.}~\bibnamefont
  {Pr\'{e}vost}}\ and\ \bibinfo {author} {\bibfnamefont {P.}~\bibnamefont
  {Chainais}},\ }\bibfield  {title} {\bibinfo {title} {\emph{Optimal estimation
  of the canonical polyadic decomposition from low-rank tensor trains}},\
  }\href {https://doi.org/10.1016/j.sigpro.2025.110001} {\bibfield  {journal}
  {\bibinfo  {journal} {Signal Process.}\ }\textbf {\bibinfo {volume} {234}},\
  \bibinfo {pages} {110001} (\bibinfo {year} {2025})}\BibitemShut {NoStop}%
\bibitem [{\citenamefont {Biamonte}\ \emph {et~al.}(2011)\citenamefont
  {Biamonte}, \citenamefont {Clark},\ and\ \citenamefont
  {Jaksch}}]{biamonte2011}%
  \BibitemOpen
  \bibfield  {author} {\bibinfo {author} {\bibfnamefont {J.~D.}\ \bibnamefont
  {Biamonte}}, \bibinfo {author} {\bibfnamefont {S.~R.}\ \bibnamefont
  {Clark}},\ and\ \bibinfo {author} {\bibfnamefont {D.}~\bibnamefont
  {Jaksch}},\ }\bibfield  {title} {\bibinfo {title} {\emph{Categorical Tensor
  Network States}},\ }\href {https://doi.org/10.1063/1.3672009} {\bibfield
  {journal} {\bibinfo  {journal} {AIP Advances}\ }\textbf {\bibinfo {volume}
  {1}},\ \bibinfo {pages} {042172} (\bibinfo {year} {2011})}\BibitemShut
  {NoStop}%
\bibitem [{\citenamefont {Denny}\ \emph {et~al.}(2011)\citenamefont {Denny},
  \citenamefont {Biamonte}, \citenamefont {Jaksch},\ and\ \citenamefont
  {Clark}}]{denny2011}%
  \BibitemOpen
  \bibfield  {author} {\bibinfo {author} {\bibfnamefont {S.~J.}\ \bibnamefont
  {Denny}}, \bibinfo {author} {\bibfnamefont {J.~D.}\ \bibnamefont {Biamonte}},
  \bibinfo {author} {\bibfnamefont {D.}~\bibnamefont {Jaksch}},\ and\ \bibinfo
  {author} {\bibfnamefont {S.~R.}\ \bibnamefont {Clark}},\ }\bibfield  {title}
  {\bibinfo {title} {\emph{Algebraically contractible topological tensor
  network states}},\ }\href {https://doi.org/10.1088/1751-8113/45/1/015309}
  {\bibfield  {journal} {\bibinfo  {journal} {J. Phys. A: Math. Theor.}\
  }\textbf {\bibinfo {volume} {45}},\ \bibinfo {pages} {015309} (\bibinfo
  {year} {2011})}\BibitemShut {NoStop}%
\bibitem [{\citenamefont {Salakhutdinov}\ \emph {et~al.}(2007)\citenamefont
  {Salakhutdinov}, \citenamefont {Mnih},\ and\ \citenamefont
  {Hinton}}]{salakhutdinov2007}%
  \BibitemOpen
  \bibfield  {author} {\bibinfo {author} {\bibfnamefont {R.}~\bibnamefont
  {Salakhutdinov}}, \bibinfo {author} {\bibfnamefont {A.}~\bibnamefont
  {Mnih}},\ and\ \bibinfo {author} {\bibfnamefont {G.}~\bibnamefont {Hinton}},\
  }\bibfield  {title} {\bibinfo {title} {\emph{Restricted Boltzmann machines
  for collaborative filtering}},\ }in\ \href
  {https://doi.org/10.1145/1273496.1273596} {\emph {\bibinfo {booktitle}
  {Proceedings of the 24th international conference on Machine learning}}},\
  \bibinfo {series and number} {ICML '07 \& ILP '07}\ (\bibinfo  {publisher}
  {ACM},\ \bibinfo {year} {2007})\ pp.\ \bibinfo {pages} {791--798}\BibitemShut
  {NoStop}%
\bibitem [{\citenamefont {Guo}\ and\ \citenamefont
  {Berkhahn}()}]{guo2016_arxiv}%
  \BibitemOpen
  \bibfield  {author} {\bibinfo {author} {\bibfnamefont {C.}~\bibnamefont
  {Guo}}\ and\ \bibinfo {author} {\bibfnamefont {F.}~\bibnamefont {Berkhahn}},\
  }\href@noop {} {\bibinfo {title} {\emph{Entity Embeddings of Categorical
  Variables}}},\ \Eprint {https://arxiv.org/abs/1604.06737} {arXiv:1604.06737}
  \BibitemShut {NoStop}%
\bibitem [{\citenamefont {De~Vlugt}\ \emph {et~al.}(2020)\citenamefont
  {De~Vlugt}, \citenamefont {Iouchtchenko}, \citenamefont {Merali},
  \citenamefont {Roy},\ and\ \citenamefont {Melko}}]{vlugt2020}%
  \BibitemOpen
  \bibfield  {author} {\bibinfo {author} {\bibfnamefont {I.~J.~S.}\
  \bibnamefont {De~Vlugt}}, \bibinfo {author} {\bibfnamefont {D.}~\bibnamefont
  {Iouchtchenko}}, \bibinfo {author} {\bibfnamefont {E.}~\bibnamefont
  {Merali}}, \bibinfo {author} {\bibfnamefont {P.-N.}\ \bibnamefont {Roy}},\
  and\ \bibinfo {author} {\bibfnamefont {R.~G.}\ \bibnamefont {Melko}},\
  }\bibfield  {title} {\bibinfo {title} {\emph{Reconstructing quantum molecular
  rotor ground states}},\ }\href {https://doi.org/10.1103/PhysRevB.102.035108}
  {\bibfield  {journal} {\bibinfo  {journal} {Phys. Rev. B}\ }\textbf {\bibinfo
  {volume} {102}},\ \bibinfo {pages} {035108} (\bibinfo {year}
  {2020})}\BibitemShut {NoStop}%
\bibitem [{\citenamefont {Y~Pei}\ and\ \citenamefont
  {R~Clark}(2024)}]{pei2024}%
  \BibitemOpen
  \bibfield  {author} {\bibinfo {author} {\bibfnamefont {M.}~\bibnamefont
  {Y~Pei}}\ and\ \bibinfo {author} {\bibfnamefont {S.}~\bibnamefont
  {R~Clark}},\ }\bibfield  {title} {\bibinfo {title} {\emph{Specialising
  neural-network quantum states for the Bose Hubbard model}},\ }\href
  {https://doi.org/10.1088/1361-6455/ad7e87} {\bibfield  {journal} {\bibinfo
  {journal} {J. Phys. B: Atom. Molec. Opt. Phys.}\ }\textbf {\bibinfo {volume}
  {57}},\ \bibinfo {pages} {215301} (\bibinfo {year} {2024})}\BibitemShut
  {NoStop}%
\bibitem [{\citenamefont {Le~Roux}\ and\ \citenamefont
  {Bengio}(2008)}]{leroux2008}%
  \BibitemOpen
  \bibfield  {author} {\bibinfo {author} {\bibfnamefont {N.}~\bibnamefont
  {Le~Roux}}\ and\ \bibinfo {author} {\bibfnamefont {Y.}~\bibnamefont
  {Bengio}},\ }\bibfield  {title} {\bibinfo {title} {\emph{Representational
  Power of Restricted Boltzmann Machines and Deep Belief Networks}},\ }\href
  {https://doi.org/10.1162/neco.2008.04-07-510} {\bibfield  {journal} {\bibinfo
   {journal} {Neural Comput.}\ }\textbf {\bibinfo {volume} {20}},\ \bibinfo
  {pages} {1631} (\bibinfo {year} {2008})}\BibitemShut {NoStop}%
\bibitem [{\citenamefont {Kour}\ \emph {et~al.}(2023)\citenamefont {Kour},
  \citenamefont {Dolgov}, \citenamefont {Stoll},\ and\ \citenamefont
  {Benner}}]{kour2023}%
  \BibitemOpen
  \bibfield  {author} {\bibinfo {author} {\bibfnamefont {K.}~\bibnamefont
  {Kour}}, \bibinfo {author} {\bibfnamefont {S.}~\bibnamefont {Dolgov}},
  \bibinfo {author} {\bibfnamefont {M.}~\bibnamefont {Stoll}},\ and\ \bibinfo
  {author} {\bibfnamefont {P.}~\bibnamefont {Benner}},\ }\bibfield  {title}
  {\bibinfo {title} {\emph{Efficient Structure-preserving Support Tensor Train
  Machine}},\ }\href {https://doi.org/10.48550/arXiv.2002.05079} {\bibfield
  {journal} {\bibinfo  {journal} {J. Mach. Learn. Res.}\ }\textbf {\bibinfo
  {volume} {24}},\ \bibinfo {pages} {1} (\bibinfo {year} {2023})}\BibitemShut
  {NoStop}%
\bibitem [{\citenamefont {Gubernatis}\ \emph {et~al.}(2016)\citenamefont
  {Gubernatis}, \citenamefont {Kawashima},\ and\ \citenamefont
  {Werner}}]{gubernatis2016}%
  \BibitemOpen
  \bibfield  {author} {\bibinfo {author} {\bibfnamefont {J.}~\bibnamefont
  {Gubernatis}}, \bibinfo {author} {\bibfnamefont {N.}~\bibnamefont
  {Kawashima}},\ and\ \bibinfo {author} {\bibfnamefont {P.}~\bibnamefont
  {Werner}},\ }\href {https://doi.org/10.1017/cbo9780511902581} {\emph
  {\bibinfo {title} {Quantum Monte Carlo Methods: Algorithms for Lattice
  Models}}}\ (\bibinfo  {publisher} {Cambridge University Press},\ \bibinfo
  {year} {2016})\BibitemShut {NoStop}%
\bibitem [{\citenamefont {Becca}\ and\ \citenamefont
  {Sorella}(2017)}]{becca2017}%
  \BibitemOpen
  \bibfield  {author} {\bibinfo {author} {\bibfnamefont {F.}~\bibnamefont
  {Becca}}\ and\ \bibinfo {author} {\bibfnamefont {S.}~\bibnamefont
  {Sorella}},\ }\href {https://doi.org/10.1017/9781316417041} {\emph {\bibinfo
  {title} {Quantum Monte Carlo Approaches for Correlated Systems}}}\ (\bibinfo
  {publisher} {Cambridge University Press},\ \bibinfo {year}
  {2017})\BibitemShut {NoStop}%
\bibitem [{\citenamefont {Sorella}(1998)}]{sorella1998}%
  \BibitemOpen
  \bibfield  {author} {\bibinfo {author} {\bibfnamefont {S.}~\bibnamefont
  {Sorella}},\ }\bibfield  {title} {\bibinfo {title} {\emph{Green Function
  Monte Carlo with Stochastic Reconfiguration}},\ }\href
  {https://doi.org/10.1103/PhysRevLett.80.4558} {\bibfield  {journal} {\bibinfo
   {journal} {Phys. Rev. Lett.}\ }\textbf {\bibinfo {volume} {80}},\ \bibinfo
  {pages} {4558} (\bibinfo {year} {1998})}\BibitemShut {NoStop}%
\bibitem [{\citenamefont {Sorella}\ and\ \citenamefont
  {Capriotti}(2000)}]{sorella2000}%
  \BibitemOpen
  \bibfield  {author} {\bibinfo {author} {\bibfnamefont {S.}~\bibnamefont
  {Sorella}}\ and\ \bibinfo {author} {\bibfnamefont {L.}~\bibnamefont
  {Capriotti}},\ }\bibfield  {title} {\bibinfo {title} {\emph{Green function
  Monte Carlo with stochastic reconfiguration: An effective remedy for the sign
  problem}},\ }\href {https://doi.org/10.1103/PhysRevB.61.2599} {\bibfield
  {journal} {\bibinfo  {journal} {Phys. Rev. B}\ }\textbf {\bibinfo {volume}
  {61}},\ \bibinfo {pages} {2599} (\bibinfo {year} {2000})}\BibitemShut
  {NoStop}%
\bibitem [{\citenamefont {Amari}(1996)}]{amari1996}%
  \BibitemOpen
  \bibfield  {author} {\bibinfo {author} {\bibfnamefont {S.-i.}\ \bibnamefont
  {Amari}},\ }\bibfield  {title} {\bibinfo {title} {\emph{Neural Learning in
  Structured Parameter Spaces - Natural Riemannian Gradient}},\ }in\ \href
  {https://proceedings.neurips.cc/paper_files/paper/1996/file/39e4973ba3321b80f37d9b55f63ed8b8-Paper.pdf}
  {\emph {\bibinfo {booktitle} {Advances in Neural Information Processing
  Systems}}},\ Vol.~\bibinfo {volume} {9},\ \bibinfo {editor} {edited by\
  \bibinfo {editor} {\bibfnamefont {M.}~\bibnamefont {Mozer}}, \bibinfo
  {editor} {\bibfnamefont {M.}~\bibnamefont {Jordan}},\ and\ \bibinfo {editor}
  {\bibfnamefont {T.}~\bibnamefont {Petsche}}}\ (\bibinfo  {publisher} {MIT
  Press},\ \bibinfo {year} {1996})\BibitemShut {NoStop}%
\bibitem [{\citenamefont {Amari}(1998)}]{amari1998}%
  \BibitemOpen
  \bibfield  {author} {\bibinfo {author} {\bibfnamefont {S.-i.}\ \bibnamefont
  {Amari}},\ }\bibfield  {title} {\bibinfo {title} {\emph{Natural Gradient
  Works Efficiently in Learning}},\ }\href
  {https://doi.org/10.1162/089976698300017746} {\bibfield  {journal} {\bibinfo
  {journal} {Neural Comput.}\ }\textbf {\bibinfo {volume} {10}},\ \bibinfo
  {pages} {251} (\bibinfo {year} {1998})}\BibitemShut {NoStop}%
\bibitem [{\citenamefont {Pfeuty}(1970)}]{pfeuty1970}%
  \BibitemOpen
  \bibfield  {author} {\bibinfo {author} {\bibfnamefont {P.}~\bibnamefont
  {Pfeuty}},\ }\bibfield  {title} {\bibinfo {title} {\emph{The one-dimensional
  Ising model with a transverse field}},\ }\href
  {https://doi.org/10.1016/0003-4916(70)90270-8} {\bibfield  {journal}
  {\bibinfo  {journal} {Ann. Phys.}\ }\textbf {\bibinfo {volume} {57}},\
  \bibinfo {pages} {79} (\bibinfo {year} {1970})}\BibitemShut {NoStop}%
\bibitem [{\citenamefont {Cabrera}\ and\ \citenamefont
  {Jullien}(1987)}]{cabrera1987}%
  \BibitemOpen
  \bibfield  {author} {\bibinfo {author} {\bibfnamefont {G.~G.}\ \bibnamefont
  {Cabrera}}\ and\ \bibinfo {author} {\bibfnamefont {R.}~\bibnamefont
  {Jullien}},\ }\bibfield  {title} {\bibinfo {title} {\emph{Role of boundary
  conditions in the finite-size Ising model}},\ }\href
  {https://doi.org/10.1103/PhysRevB.35.7062} {\bibfield  {journal} {\bibinfo
  {journal} {Phys. Rev. B}\ }\textbf {\bibinfo {volume} {35}},\ \bibinfo
  {pages} {7062} (\bibinfo {year} {1987})}\BibitemShut {NoStop}%
\bibitem [{\citenamefont {He}\ and\ \citenamefont {Guo}(2017)}]{he2017}%
  \BibitemOpen
  \bibfield  {author} {\bibinfo {author} {\bibfnamefont {Y.}~\bibnamefont
  {He}}\ and\ \bibinfo {author} {\bibfnamefont {H.}~\bibnamefont {Guo}},\
  }\bibfield  {title} {\bibinfo {title} {\emph{The boundary effects of
  transverse field Ising model}},\ }\href
  {https://doi.org/10.1088/1742-5468/aa85b0} {\bibfield  {journal} {\bibinfo
  {journal} {J. Stat. Mech.}\ }\textbf {\bibinfo {volume} {2017}},\ \bibinfo
  {pages} {093101} (\bibinfo {year} {2017})}\BibitemShut {NoStop}%
\bibitem [{\citenamefont {Tasaki}(2020)}]{tasaki2020}%
  \BibitemOpen
  \bibfield  {author} {\bibinfo {author} {\bibfnamefont {H.}~\bibnamefont
  {Tasaki}},\ }\href {https://doi.org/10.1007/978-3-030-41265-4} {\emph
  {\bibinfo {title} {Physics and Mathematics of Quantum Many-Body Systems}}}\
  (\bibinfo  {publisher} {Springer International Publishing},\ \bibinfo {year}
  {2020})\BibitemShut {NoStop}%
\bibitem [{\citenamefont {Katsura}(1962)}]{katsura1962}%
  \BibitemOpen
  \bibfield  {author} {\bibinfo {author} {\bibfnamefont {S.}~\bibnamefont
  {Katsura}},\ }\bibfield  {title} {\bibinfo {title} {\emph{Statistical
  Mechanics of the Anisotropic Linear Heisenberg Model}},\ }\href
  {https://doi.org/10.1103/PhysRev.127.1508} {\bibfield  {journal} {\bibinfo
  {journal} {Phys. Rev.}\ }\textbf {\bibinfo {volume} {127}},\ \bibinfo {pages}
  {1508} (\bibinfo {year} {1962})}\BibitemShut {NoStop}%
\bibitem [{\citenamefont {Ovchinnikov}\ \emph {et~al.}(2003)\citenamefont
  {Ovchinnikov}, \citenamefont {Dmitriev}, \citenamefont {Krivnov},\ and\
  \citenamefont {Cheranovskii}}]{ovchinnikov2003}%
  \BibitemOpen
  \bibfield  {author} {\bibinfo {author} {\bibfnamefont {A.~A.}\ \bibnamefont
  {Ovchinnikov}}, \bibinfo {author} {\bibfnamefont {D.~V.}\ \bibnamefont
  {Dmitriev}}, \bibinfo {author} {\bibfnamefont {V.~Y.}\ \bibnamefont
  {Krivnov}},\ and\ \bibinfo {author} {\bibfnamefont {V.~O.}\ \bibnamefont
  {Cheranovskii}},\ }\bibfield  {title} {\bibinfo {title}
  {\emph{Antiferromagnetic Ising chain in a mixed transverse and longitudinal
  magnetic field}},\ }\href {https://doi.org/10.1103/PhysRevB.68.214406}
  {\bibfield  {journal} {\bibinfo  {journal} {Phys. Rev. B}\ }\textbf {\bibinfo
  {volume} {68}},\ \bibinfo {pages} {214406} (\bibinfo {year}
  {2003})}\BibitemShut {NoStop}%
\bibitem [{\citenamefont {Bl\"ote}\ \emph {et~al.}(1986)\citenamefont
  {Bl\"ote}, \citenamefont {Cardy},\ and\ \citenamefont
  {Nightingale}}]{bloete1986}%
  \BibitemOpen
  \bibfield  {author} {\bibinfo {author} {\bibfnamefont {H.~W.~J.}\
  \bibnamefont {Bl\"ote}}, \bibinfo {author} {\bibfnamefont {J.~L.}\
  \bibnamefont {Cardy}},\ and\ \bibinfo {author} {\bibfnamefont {M.~P.}\
  \bibnamefont {Nightingale}},\ }\bibfield  {title} {\bibinfo {title}
  {\emph{Conformal invariance, the central charge, and universal finite-size
  amplitudes at criticality}},\ }\href
  {https://doi.org/10.1103/PhysRevLett.56.742} {\bibfield  {journal} {\bibinfo
  {journal} {Phys. Rev. Lett.}\ }\textbf {\bibinfo {volume} {56}},\ \bibinfo
  {pages} {742} (\bibinfo {year} {1986})}\BibitemShut {NoStop}%
\bibitem [{\citenamefont {Affleck}(1986)}]{affleck1986}%
  \BibitemOpen
  \bibfield  {author} {\bibinfo {author} {\bibfnamefont {I.}~\bibnamefont
  {Affleck}},\ }\bibfield  {title} {\bibinfo {title} {\emph{Universal term in
  the free energy at a critical point and the conformal anomaly}},\ }\href
  {https://doi.org/10.1103/PhysRevLett.56.746} {\bibfield  {journal} {\bibinfo
  {journal} {Phys. Rev. Lett.}\ }\textbf {\bibinfo {volume} {56}},\ \bibinfo
  {pages} {746} (\bibinfo {year} {1986})}\BibitemShut {NoStop}%
\bibitem [{\citenamefont {Ceperley}(1991)}]{ceperley1991}%
  \BibitemOpen
  \bibfield  {author} {\bibinfo {author} {\bibfnamefont {D.~M.}\ \bibnamefont
  {Ceperley}},\ }\bibfield  {title} {\bibinfo {title} {\emph{Fermion nodes}},\
  }\href {https://doi.org/10.1007/bf01030009} {\bibfield  {journal} {\bibinfo
  {journal} {J. Stat. Phys.}\ }\textbf {\bibinfo {volume} {63}},\ \bibinfo
  {pages} {1237} (\bibinfo {year} {1991})}\BibitemShut {NoStop}%
\bibitem [{\citenamefont {LeBlanc}\ \emph {et~al.}(2015)\citenamefont
  {LeBlanc}, \citenamefont {Antipov}, \citenamefont {Becca}, \citenamefont
  {Bulik}, \citenamefont {Chan}, \citenamefont {Chung}, \citenamefont {Deng},
  \citenamefont {Ferrero}, \citenamefont {Henderson}, \citenamefont
  {Jim\'enez-Hoyos}, \citenamefont {Kozik}, \citenamefont {Liu}, \citenamefont
  {Millis}, \citenamefont {Prokof'ev}, \citenamefont {Qin}, \citenamefont
  {Scuseria}, \citenamefont {Shi}, \citenamefont {Svistunov}, \citenamefont
  {Tocchio}, \citenamefont {Tupitsyn}, \citenamefont {White}, \citenamefont
  {Zhang}, \citenamefont {Zheng}, \citenamefont {Zhu},\ and\ \citenamefont
  {Gull}}]{leblanc2015}%
  \BibitemOpen
  \bibfield  {author} {\bibinfo {author} {\bibfnamefont {J.~P.~F.}\
  \bibnamefont {LeBlanc}}, \bibinfo {author} {\bibfnamefont {A.~E.}\
  \bibnamefont {Antipov}}, \bibinfo {author} {\bibfnamefont {F.}~\bibnamefont
  {Becca}}, \bibinfo {author} {\bibfnamefont {I.~W.}\ \bibnamefont {Bulik}},
  \bibinfo {author} {\bibfnamefont {G.~K.-L.}\ \bibnamefont {Chan}}, \bibinfo
  {author} {\bibfnamefont {C.-M.}\ \bibnamefont {Chung}}, \bibinfo {author}
  {\bibfnamefont {Y.}~\bibnamefont {Deng}}, \bibinfo {author} {\bibfnamefont
  {M.}~\bibnamefont {Ferrero}}, \bibinfo {author} {\bibfnamefont {T.~M.}\
  \bibnamefont {Henderson}}, \bibinfo {author} {\bibfnamefont {C.~A.}\
  \bibnamefont {Jim\'enez-Hoyos}}, \bibinfo {author} {\bibfnamefont
  {E.}~\bibnamefont {Kozik}}, \bibinfo {author} {\bibfnamefont {X.-W.}\
  \bibnamefont {Liu}}, \bibinfo {author} {\bibfnamefont {A.~J.}\ \bibnamefont
  {Millis}}, \bibinfo {author} {\bibfnamefont {N.~V.}\ \bibnamefont
  {Prokof'ev}}, \bibinfo {author} {\bibfnamefont {M.}~\bibnamefont {Qin}},
  \bibinfo {author} {\bibfnamefont {G.~E.}\ \bibnamefont {Scuseria}}, \bibinfo
  {author} {\bibfnamefont {H.}~\bibnamefont {Shi}}, \bibinfo {author}
  {\bibfnamefont {B.~V.}\ \bibnamefont {Svistunov}}, \bibinfo {author}
  {\bibfnamefont {L.~F.}\ \bibnamefont {Tocchio}}, \bibinfo {author}
  {\bibfnamefont {I.~S.}\ \bibnamefont {Tupitsyn}}, \bibinfo {author}
  {\bibfnamefont {S.~R.}\ \bibnamefont {White}}, \bibinfo {author}
  {\bibfnamefont {S.}~\bibnamefont {Zhang}}, \bibinfo {author} {\bibfnamefont
  {B.-X.}\ \bibnamefont {Zheng}}, \bibinfo {author} {\bibfnamefont
  {Z.}~\bibnamefont {Zhu}},\ and\ \bibinfo {author} {\bibfnamefont
  {E.}~\bibnamefont {Gull}} (\bibinfo {collaboration} {Simons Collaboration on
  the Many-Electron Problem}),\ }\bibfield  {title} {\bibinfo {title}
  {\emph{Solutions of the Two-Dimensional Hubbard Model: Benchmarks and Results
  from a Wide Range of Numerical Algorithms}},\ }\href
  {https://doi.org/10.1103/PhysRevX.5.041041} {\bibfield  {journal} {\bibinfo
  {journal} {Phys. Rev. X}\ }\textbf {\bibinfo {volume} {5}},\ \bibinfo {pages}
  {041041} (\bibinfo {year} {2015})}\BibitemShut {NoStop}%
\bibitem [{\citenamefont {Wu}\ \emph {et~al.}(2024)\citenamefont {Wu},
  \citenamefont {Rossi}, \citenamefont {Vicentini}, \citenamefont
  {Astrakhantsev}, \citenamefont {Becca}, \citenamefont {Cao}, \citenamefont
  {Carrasquilla}, \citenamefont {Ferrari}, \citenamefont {Georges},
  \citenamefont {Hibat-Allah}, \citenamefont {Imada}, \citenamefont
  {L\"{a}uchli}, \citenamefont {Mazzola}, \citenamefont {Mezzacapo},
  \citenamefont {Millis}, \citenamefont {Robledo~Moreno}, \citenamefont
  {Neupert}, \citenamefont {Nomura}, \citenamefont {Nys}, \citenamefont
  {Parcollet}, \citenamefont {Pohle}, \citenamefont {Romero}, \citenamefont
  {Schmid}, \citenamefont {Silvester}, \citenamefont {Sorella}, \citenamefont
  {Tocchio}, \citenamefont {Wang}, \citenamefont {White}, \citenamefont
  {Wietek}, \citenamefont {Yang}, \citenamefont {Yang}, \citenamefont {Zhang},\
  and\ \citenamefont {Carleo}}]{wu2024}%
  \BibitemOpen
  \bibfield  {author} {\bibinfo {author} {\bibfnamefont {D.}~\bibnamefont
  {Wu}}, \bibinfo {author} {\bibfnamefont {R.}~\bibnamefont {Rossi}}, \bibinfo
  {author} {\bibfnamefont {F.}~\bibnamefont {Vicentini}}, \bibinfo {author}
  {\bibfnamefont {N.}~\bibnamefont {Astrakhantsev}}, \bibinfo {author}
  {\bibfnamefont {F.}~\bibnamefont {Becca}}, \bibinfo {author} {\bibfnamefont
  {X.}~\bibnamefont {Cao}}, \bibinfo {author} {\bibfnamefont {J.}~\bibnamefont
  {Carrasquilla}}, \bibinfo {author} {\bibfnamefont {F.}~\bibnamefont
  {Ferrari}}, \bibinfo {author} {\bibfnamefont {A.}~\bibnamefont {Georges}},
  \bibinfo {author} {\bibfnamefont {M.}~\bibnamefont {Hibat-Allah}}, \bibinfo
  {author} {\bibfnamefont {M.}~\bibnamefont {Imada}}, \bibinfo {author}
  {\bibfnamefont {A.~M.}\ \bibnamefont {L\"{a}uchli}}, \bibinfo {author}
  {\bibfnamefont {G.}~\bibnamefont {Mazzola}}, \bibinfo {author} {\bibfnamefont
  {A.}~\bibnamefont {Mezzacapo}}, \bibinfo {author} {\bibfnamefont
  {A.}~\bibnamefont {Millis}}, \bibinfo {author} {\bibfnamefont
  {J.}~\bibnamefont {Robledo~Moreno}}, \bibinfo {author} {\bibfnamefont
  {T.}~\bibnamefont {Neupert}}, \bibinfo {author} {\bibfnamefont
  {Y.}~\bibnamefont {Nomura}}, \bibinfo {author} {\bibfnamefont
  {J.}~\bibnamefont {Nys}}, \bibinfo {author} {\bibfnamefont {O.}~\bibnamefont
  {Parcollet}}, \bibinfo {author} {\bibfnamefont {R.}~\bibnamefont {Pohle}},
  \bibinfo {author} {\bibfnamefont {I.}~\bibnamefont {Romero}}, \bibinfo
  {author} {\bibfnamefont {M.}~\bibnamefont {Schmid}}, \bibinfo {author}
  {\bibfnamefont {J.~M.}\ \bibnamefont {Silvester}}, \bibinfo {author}
  {\bibfnamefont {S.}~\bibnamefont {Sorella}}, \bibinfo {author} {\bibfnamefont
  {L.~F.}\ \bibnamefont {Tocchio}}, \bibinfo {author} {\bibfnamefont
  {L.}~\bibnamefont {Wang}}, \bibinfo {author} {\bibfnamefont {S.~R.}\
  \bibnamefont {White}}, \bibinfo {author} {\bibfnamefont {A.}~\bibnamefont
  {Wietek}}, \bibinfo {author} {\bibfnamefont {Q.}~\bibnamefont {Yang}},
  \bibinfo {author} {\bibfnamefont {Y.}~\bibnamefont {Yang}}, \bibinfo {author}
  {\bibfnamefont {S.}~\bibnamefont {Zhang}},\ and\ \bibinfo {author}
  {\bibfnamefont {G.}~\bibnamefont {Carleo}},\ }\bibfield  {title} {\bibinfo
  {title} {\emph{Variational benchmarks for quantum many-body problems}},\
  }\href {https://doi.org/10.1126/science.adg9774} {\bibfield  {journal}
  {\bibinfo  {journal} {Science}\ }\textbf {\bibinfo {volume} {386}},\ \bibinfo
  {pages} {296} (\bibinfo {year} {2024})}\BibitemShut {NoStop}%
\bibitem [{\citenamefont {Bortone}\ \emph {et~al.}(2025)\citenamefont
  {Bortone}, \citenamefont {Rath},\ and\ \citenamefont {Booth}}]{bortone2025}%
  \BibitemOpen
  \bibfield  {author} {\bibinfo {author} {\bibfnamefont {M.}~\bibnamefont
  {Bortone}}, \bibinfo {author} {\bibfnamefont {Y.}~\bibnamefont {Rath}},\ and\
  \bibinfo {author} {\bibfnamefont {G.~H.}\ \bibnamefont {Booth}},\ }\bibfield
  {title} {\bibinfo {title} {\emph{Simple Fermionic backflow states via a
  systematically improvable tensor decomposition}},\ }\href
  {https://doi.org/10.1038/s42005-025-02083-4} {\bibfield  {journal} {\bibinfo
  {journal} {Commun. Phys.}\ }\textbf {\bibinfo {volume} {8}},\ \bibinfo
  {pages} {169} (\bibinfo {year} {2025})}\BibitemShut {NoStop}%
\bibitem [{\citenamefont {Sherman}\ and\ \citenamefont
  {Kolda}(2020)}]{sherman2020}%
  \BibitemOpen
  \bibfield  {author} {\bibinfo {author} {\bibfnamefont {S.}~\bibnamefont
  {Sherman}}\ and\ \bibinfo {author} {\bibfnamefont {T.~G.}\ \bibnamefont
  {Kolda}},\ }\bibfield  {title} {\bibinfo {title} {\emph{Estimating
  Higher-Order Moments Using Symmetric Tensor Decomposition}},\ }\href
  {https://doi.org/10.1137/19m1299633} {\bibfield  {journal} {\bibinfo
  {journal} {SIAM J. Matrix Anal. Appl.}\ }\textbf {\bibinfo {volume} {41}},\
  \bibinfo {pages} {1369} (\bibinfo {year} {2020})}\BibitemShut {NoStop}%
\bibitem [{\citenamefont {Wang}\ \emph {et~al.}()\citenamefont {Wang},
  \citenamefont {Pan}, \citenamefont {Xu}, \citenamefont {Li}, \citenamefont
  {Yang}, \citenamefont {Mandic},\ and\ \citenamefont
  {Cichocki}}]{wang2023_arxiv}%
  \BibitemOpen
  \bibfield  {author} {\bibinfo {author} {\bibfnamefont {M.}~\bibnamefont
  {Wang}}, \bibinfo {author} {\bibfnamefont {Y.}~\bibnamefont {Pan}}, \bibinfo
  {author} {\bibfnamefont {Z.}~\bibnamefont {Xu}}, \bibinfo {author}
  {\bibfnamefont {G.}~\bibnamefont {Li}}, \bibinfo {author} {\bibfnamefont
  {X.}~\bibnamefont {Yang}}, \bibinfo {author} {\bibfnamefont {D.}~\bibnamefont
  {Mandic}},\ and\ \bibinfo {author} {\bibfnamefont {A.}~\bibnamefont
  {Cichocki}},\ }\href@noop {} {\bibinfo {title} {\emph{Tensor Networks Meet
  Neural Networks: A Survey and Future Perspectives}}},\ \Eprint
  {https://arxiv.org/abs/2302.09019} {arXiv:2302.09019} \BibitemShut {NoStop}%
\bibitem [{\citenamefont {Hamreras}\ \emph {et~al.}()\citenamefont {Hamreras},
  \citenamefont {Singh},\ and\ \citenamefont {Or\'{u}s}}]{hamreras2025_arxiv}%
  \BibitemOpen
  \bibfield  {author} {\bibinfo {author} {\bibfnamefont {S.}~\bibnamefont
  {Hamreras}}, \bibinfo {author} {\bibfnamefont {S.}~\bibnamefont {Singh}},\
  and\ \bibinfo {author} {\bibfnamefont {R.}~\bibnamefont {Or\'{u}s}},\
  }\href@noop {} {\bibinfo {title} {\emph{Tensorization is a powerful but
  underexplored tool for compression and interpretability of neural
  networks}}},\ \Eprint {https://arxiv.org/abs/2505.20132} {arXiv:2505.20132}
  \BibitemShut {NoStop}%
\bibitem [{\citenamefont {Sugiyama}\ \emph {et~al.}(2019)\citenamefont
  {Sugiyama}, \citenamefont {Nakahara},\ and\ \citenamefont
  {Tsuda}}]{sugiyama2019}%
  \BibitemOpen
  \bibfield  {author} {\bibinfo {author} {\bibfnamefont {M.}~\bibnamefont
  {Sugiyama}}, \bibinfo {author} {\bibfnamefont {H.}~\bibnamefont {Nakahara}},\
  and\ \bibinfo {author} {\bibfnamefont {K.}~\bibnamefont {Tsuda}},\ }\bibfield
   {title} {\bibinfo {title} {\emph{Legendre decomposition for tensors}},\
  }\href {https://doi.org/10.1088/1742-5468/ab3196} {\bibfield  {journal}
  {\bibinfo  {journal} {J. Stat. Mech.}\ }\textbf {\bibinfo {volume} {2019}},\
  \bibinfo {pages} {124017} (\bibinfo {year} {2019})}\BibitemShut {NoStop}%
\bibitem [{\citenamefont {Pang}\ \emph {et~al.}()\citenamefont {Pang},
  \citenamefont {Yi}, \citenamefont {Yin},\ and\ \citenamefont
  {Xu}}]{pang2020_arxiv}%
  \BibitemOpen
  \bibfield  {author} {\bibinfo {author} {\bibfnamefont {J.}~\bibnamefont
  {Pang}}, \bibinfo {author} {\bibfnamefont {K.}~\bibnamefont {Yi}}, \bibinfo
  {author} {\bibfnamefont {W.}~\bibnamefont {Yin}},\ and\ \bibinfo {author}
  {\bibfnamefont {M.}~\bibnamefont {Xu}},\ }\href@noop {} {\bibinfo {title}
  {\emph{Experimental Analysis of Legendre Decomposition in Machine
  Learning}}},\ \Eprint {https://arxiv.org/abs/2008.05095} {arXiv:2008.05095}
  \BibitemShut {NoStop}%
\bibitem [{dat()}]{data_availability}%
  \BibitemOpen
  \href@noop {} {\bibinfo {title} {\emph{{\rm All codes and data used in this
  manuscript are available at
  https://github.com/ryuikaneko/mps2rbm.}}}}\BibitemShut {Stop}%
\end{thebibliography}
\end{document}